\documentclass[tradiabstract,longauth]{aa}  
\usepackage[varg]{txfonts}
\usepackage{graphicx}
\usepackage[colorlinks,linkcolor=black,citecolor=blue]{hyperref}
\usepackage{natbib}

\newcommand{\pf}{JPAS-{\it Pathfinder}}

\newcommand{\jplus}{J-PLUS}
\newcommand{\splus}{S-PLUS}

\newcommand{\jp}{J-PAS}
\newcommand{\js}{J-spectra}
\newcommand{\mjp}{miniJPAS}
\newcommand{\hsc}{HSC-SSP}

\def\sext{\texttt{SExtractor}}

\def\mainurl{\url{http://archive.cefca.es/catalogues/\mjp-pdr201912/}}
\def\jphotoz{\texttt{JPHOTOZ}}

\def\magauto{\texttt{MAG\_AUTO}}
\def\auto{\texttt{AUTO}}

\def\magpsfcor{\texttt{MAG\_PSFCOR}}
\def\psfcor{\texttt{PSFCOR}}

\def\class{\texttt{CLASS\_STAR}}
\def\totalstar{\texttt{total\_prob\_star}}
\def\kron{\texttt{KRON\_RADIUS}}

\def\baysea{\texttt{BaySeAGal}} 
\def\muff{\texttt{MUFFIT}} 
\def\alstar{\texttt{AlStar}} 
\def\tgas{\texttt{TGASPEX}}
\def\dynbass{\texttt{DynBaS}}
\def\lephare{{\sc LePhare}}
\newcommand{\starlight}{{\sc starlight}}                
\newcommand{\cb}{C\&B\xspace}

\newcommand{\photoz}{photo-$z$}
\def\rb{$r_\mathrm{SDSS}$}
\def\gb{$g_\mathrm{SDSS}$}
\def\ib{$i_\mathrm{SDSS}$}
\def\uja{$u_\mathrm{JAVA}$}
\def\ujp{$u_\mathrm{JPAS}$}
\def\jred{$J1007$}
\newcommand\ageLt{\langle \log\,\mathrm{age}\rangle_\mathrm{L}}
\newcommand\ageMt{\langle \log\,\mathrm{age}\rangle_\mathrm{M}}

\newcommand\logZMt{\langle \log\,Z_\star\rangle_\mathrm{M}}

\newcommand\ageL{$\ageLt$}
\newcommand\ageM{$\ageMt$}

\newcommand\logZM{$\logZMt$}

\newcommand\logMt{\log M_\star}
\newcommand\logM{$\logMt$}

\bibpunct{(}{)}{;}{a}{}{,}
\graphicspath{{Figures/}}

\begin{document}

\title{The \mjp \ survey:}
\subtitle{Identification and characterization of galaxy populations with the \jp \ photometric system}

\authorrunning{R.~M.~Gonz\'alez Delgado et al.}
\titlerunning{Galaxy populations in the AEGIS field}

\author{
R.~M.~Gonz\'alez Delgado\inst{\ref{IAA}}
\and
L.~A.~D\'iaz-Garc\'ia\inst{\ref{IAA},\ref{ASIAA}}
\and
A.~de Amorim\inst{\ref{UFSC}}
\and
G.~Bruzual\inst{\ref{IRyA}}
\and
R.~Cid Fernandes\inst{\ref{UFSC}}
\and
E.~P\'erez\inst{\ref{IAA}}
\and
S.~Bonoli\inst{\ref{CEFCA2},\ref{DIPC},\ref{BFS}}
\and 
A.~J.~Cenarro\inst{\ref{CEFCA2}}
\and
P.~R.~T.~Coelho\inst{\ref{USP}}
\and
A.~Cortesi\inst{\ref{OV}}
\and
R.~Garc\'ia-Benito\inst{\ref{IAA}}
\and
R.~L\'opez Fern\'andez\inst{\ref{IAA}}
\and
G.~Mart\'inez-Solaeche\inst{\ref{IAA}}
\and
J.~E.~Rodr\'iguez-Mart\'in\inst{\ref{IAA}}
\and
G.~Magris\inst{\ref{CIDA}}
\and
A.~Mej\'ia-Narvaez\inst{\ref{IAUNAM}}
\and
D.~Brito-Silva\inst{\ref{UDP}}
\and
L.~R.~Abramo\inst{\ref{USP-IF}}
\and 
J.~M. Diego\inst{\ref{IFC}}
\and
R.~A.~Dupke\inst{\ref{ON},\ref{DAUM},\ref{DPA}}
\and
A.~Hern\'an-Caballero\inst{\ref{CEFCA2}}
\and
C.~Hern\'andez-Monteagudo\inst{\ref{IAC}, \ref{ULL},\ref{CEFCA2}}
\and
C.~L\'opez-Sanjuan\inst{\ref{CEFCA2}}
\and
A.~Mar\'in-Franch\inst{\ref{CEFCA2}}
\and
V.~Marra\inst{\ref{UFES},\ref{INAF}, \ref{IFPU}}
\and
M.~Moles\inst{\ref{CEFCA2},\ref{IAA}}
\and
A.~Montero-Dorta\inst{\ref{USP-IF}, \ref{Federico}}
\and
C. Queiroz\inst{\ref{USP-IF}}
\and
L.~Sodr\'e Jr.\inst{\ref{USP}}
\and 
J.~Varela\inst{\ref{CEFCA2}}
\and
H.~V\'azquez Rami\'o\inst{\ref{CEFCA2}}
\and
J.~M.~V\'ilchez\inst{\ref{IAA}}
\and
P.~O.~Baqui\inst{\ref{UFES}}
\and
N.~Ben\'\i tez\inst{\ref{IAA}}
\and
D.~Crist\'obal-Hornillos\inst{\ref{CEFCA2}}
\and
A.~Ederoclite\inst{\ref{USP}}
\and
C.~Mendes de Oliveira\inst{\ref{USP}}
\and 
T.~Civera\inst{\ref{CEFCA1}}
\and 
D.~Muniesa\inst{\ref{CEFCA1}}
\and
K.~Taylor\inst{\ref{USP}}
\and
E.~Tempel\inst{\ref{Tartu}}
\and
the J-PAS collaboration
}

\institute{
Instituto de Astrof\'{\i}sica de Andaluc\'{\i}a (CSIC), P.O.~Box 3004, 18080 Granada, Spain\newline \email{rosa@iaa.es}\label{IAA}
\and
Academia Sinica Institute of Astronomy \& Astrophysics (ASIAA), No.~1, Section 4, Roosevelt Road, Taipei 10617, Taiwan\label{ASIAA}
\and 
Departamento de F\'{\i}sica, Universidade Federal de Santa Catarina, P.O.~Box 476, 88040-900, Florian\'opolis, SC, Brazil\label{UFSC}
\and
Instituto de Radioastronom\'ia y Astrof\'isica (IRyA), Universidad Nacional Aut\'onoma de M\'exico, Campus Morelia, Michoac\'an 58089, M\'exico\label{IRyA}
\and
Centro  de  Estudios  de  F\'isica  del  Cosmos  de  Arag\'on  (CEFCA), Unidad Asociada al CSIC, Plaza San Juan 1, E-44001, Teruel, Spain\label{CEFCA2}
\and
Donostia International Physics Center (DIPC), Manuel Lardizabal Ibilbidea 4, San Sebasti\'an, Spain\label{DIPC}
\and
Ikerbasque, Basque Foundation for Science, E-48013 Bilbao, Spain\label{BFS}
\and
Universidade de S\~{a}o Paulo, Instituto de Astronomia, Geof\'isica e Ci\^encias Atmosf\'ericas, R. do Mat\~{a}o 1226, 05508-090, S\~{a}o Paulo, Brazil\label{USP}
\and
Observat\'orio do Valongo, Universidade Federal do Rio de Janeiro, 20080-090, Rio de Janeiro, RJ, Brazil\label{OV}
\and
Centro de Investigaciones de Astronom\'ia (CIDA), AP 264, M\'erida 5101-A, Venezuela\label{CIDA}
\and
Instituto de Astronom\'ia, Universidad Nacional Aut\'onoma de M\'exico, A.P.~70-264, 04510 M\'exico, CDMX, M\'exico\label{IAUNAM}
\and
N\'ucleo de Astronom\'ia, Universidad Diego Portales, Av.~Ej\'ercito Libertador 441, Santiago, Regi\'on Metropolitana, Chile\label{UDP}
\and
Instituto de F\'isica, Universidade de S\~ao Paulo, Rua do Mat\~ao 1371, CEP 05508-090, S\~ao Paulo, Brazil\label{USP-IF}
\and
Instituto de F\'isica de Cantabria (CSIC-UC). Avda. Los Castros s/n. 39005, Santander, Spain\label{IFC}
\and
Observat\'orio Nacional, Minist\'erio da Ciencia, Tecnologia, Inovaç\~ao e Comunicaç\~oes, Rua General Jos\'e Cristino, 77, S\~ao Crist\'ov\~ao, 20921-400, Rio de Janeiro, Brazil\label{ON}
\and
Department of Astronomy, University of Michigan, 311 West Hall, 1085 South University Ave., Ann Arbor, USA\label{DAUM}
\and
Department of Physics and Astronomy, University of Alabama, Box 870324, Tuscaloosa, AL, USA\label{DPA}
\and
Instituto de Astrof\'\i sica de Canarias, Calle Vía L\'actea SN, ES38205 La Laguna, Spain\label{IAC}
\and
Departamento de Astrof\'\i sica, Universidad de La Laguna, ES38205, La Laguna, Spain\label{ULL} 
\and
N\'ucleo de Astrof\'isica e Cosmologia, PPGCosmo \& Dep.~de F\'isica, Universidade Federal do Esp\'irito Santo, 29075-910, ES, Brazil\label{UFES}
\and
INAF -- Osservatorio Astronomico di Trieste, via Tiepolo 11, 34131 Trieste, Italy\label{INAF}
\and
IFPU -- Institute for Fundamental Physics of the Universe, via Beirut 2, 34151, Trieste, Italy\label{IFPU}
\and
Departamento de Astronomia, Instituto de F\'isica, Universidade Federal do Rio Grande do Sul (UFRGS), Av.~Bento Gonçalves 9500, Porto Alegre, R.S, Brazil\label{UFRGS}
\and
Centro  de  Estudios  de  F\'isica  del  Cosmos  de  Arag\'on  (CEFCA),  Plaza San Juan 1, E-44001, Teruel, Spain\label{CEFCA1}
\and
Tartu Observatory, University of Tartu, Observatooriumi~1, 61602 T\~oravere, Estonia\label{Tartu}
\and
 F\'isica, Universidad T\'ecnica Federico Santa Mar\'ia, Casilla 110-V, Avda. Espa\~na 1680, Valpara\'iso, Chile\label{Federico}
}
\date{\today}


\abstract{The Javalambre-Physics of the Accelerating Universe Astrophysical Survey (\jp) will soon start imaging thousands of square degrees of the northern sky with its unique set of $56$ filters (spectral resolution of $R\sim60$). Before the arrival of the final instrument, we observed  $1$~deg$^2$ on the AEGIS field with an interim camera with all the \jp\ filters. Taking advantage of these data, dubbed \mjp, we aim at proving the scientific potential of the \jp \ to derive the stellar population properties of galaxies via fitting codes for spectral energy distributions (SEDs), with the ultimate goal of performing galaxy evolution studies across cosmic time. One parametric (\baysea ) and three non-parametric (\muff , \alstar , and \tgas ) SED-fitting codes are used to constrain the stellar mass, age, metallicity, extinction, and rest-frame and dust-corrected $(u-r)$ colours of a complete flux-limited sample (\rb $ \le 22.5$~AB) of \mjp\ galaxies that extends up to $z=1$. We generally find consistent results on the galaxy properties  derived from the different codes, independently of the galaxy spectral type or redshift; this is remarkable considering that $25$\% of the \js\ have signal-to-noise ratios (S/N) $\sim$3. For galaxies with S/N$\geq$10, we estimate that the \jp\ photometric system will allow us to derive the stellar population properties of rest-frame $(u-r)$ colour, stellar mass, extinction, and mass-weighted age  with a precision of $0.04 \pm 0.02$ mag, $0.07 \pm 0.03$~dex, $0.2 \pm 0.09$~mag, and $0.16 \pm 0.07$~dex, respectively. This precision is equivalent to that obtained with spectroscopic surveys of similar S/N. By using the dust-corrected $(u-r)$ colour--mass diagram, a powerful proxy for characterizing galaxy populations, we find: (i) that the fraction of red and blue galaxies evolves with cosmic time, with red galaxies being $\sim38$\% and $\sim18$\% of the whole population at $z= 0.1$ and $z=0.5$, respectively, and (ii) consistent results between codes for the average intrinsic $(u-r)$ colour, stellar mass, age, and stellar metallicity of blue and red galaxies and their evolution up to $z =1$.  At all redshifts, the more massive galaxies belong to the red sequence, and these galaxies are typically older and more metal-rich than their counterparts in the blue cloud. 
Our results confirm that with \jp \ data we will be able to analyse large samples of galaxies up to $z\sim 1$, with galaxy stellar masses above $\log (M_\star/\mathrm{M}_\odot) \sim 8.9$, $9.5$, and $9.9$ at $z=0.3$, $0.5$, and $0.7$, respectively. The star formation history of a complete sub-sample of galaxies selected at $z\sim 0.1$ with $\log (M_\star/\mathrm{M}_\odot) > 8.3$ constrains the cosmic evolution of the star formation rate density up to $z\sim 3$, in good agreement with results from cosmological surveys. 
}

%




\maketitle

\section{Introduction}
\label{sec:Introduction}

Constraining the geometry and the expansion rate of the Universe is of paramount importance in cosmology since it is intimately related to the fundamental components of our Universe. Large cosmological surveys have been designed to perform complementary cosmological experiments to unveil the nature of dark matter and dark energy, the latter by precise measurements of the
expansion rate of the Universe \citep{weinberg2013}. 
Spectroscopic surveys, such as the Extended Baryon Oscillation Spectroscopic Survey \citep[eBOSS;][]{eBOSSDR2020}, the Dark Energy Spectroscopic Instrument \citep[DESI;][]{DESI2016}, and 4 metre Multi Object Spectroscopic Telescope, \citep[4MOST;][]{4MOST2012}, aim
at exquisitely characterizing the large-scale 3D clustering of galaxies by targeting large samples of pre-selected sources. Photometric surveys, instead, image large portions of the sky with a few broadband (BB) f) filters and derive cosmological information from both the galaxy angular distribution and cluster and lensing studies. Recent and ongoing efforts, to just quote a few, include
the Dark Energy Survey \citep[DES;][]{wester2005}, the Panoramic Survey Telescope and Rapid Response System 1 \citep[Pan-STARRS1;][]{chambers2016}, and the Hyper Suprime-Cam Subaru Strategic Program \citep[\hsc; ][]{aihara2018}.
The data delivered by these cosmological surveys offer the opportunity to study the evolution of the galaxy population. Spectroscopic data provide more information on individual objects, with the drawback that the selected samples suffer target-selection biases. Photometric data, instead, suffer a lack of information on individual sources as each object can be characterized by only a few photometric points.  
Photometric surveys can provide a good description of the objects detected when a large number of bands are used. Notable examples are the Classifying Objects by Medium-Band Observations \citep[COMBO-17; ][]{wolf2003}, the Cosmological Evolution Survey \citep[COSMOS; ][]{ilbert2009}, and the Advanced Large Homogeneous Area Medium Band Redshift Astronomical survey \citep[ALHAMBRA; ][]{moles2008, molino2014}. ALHAMBRA used medium-band photometry (with a full width at half-maximum of $FWHM\sim 300$~\AA), reaching a better precision in its estimations of photometric redshifts (hereafter \photoz), and opened up the possibility of characterizing the physical properties of individual objects. However, this approach has only so far been used to target a few square degrees of the sky and therefore a small volume of the Universe.

The Javalambre-Physics of the Accelerating Universe Astrophysical Survey \citep[\jp; ][]{benitez2009, benitez2014} was conceived to overcome these limitations. \jp\ is about to start scanning thousands of square degrees of the northern sky with $56$ NB filters and the Javalambre Panoramic Camera (JPCam) instrument \citep{marin-franch2017} on board the $2.5$~m telescope at the Javalambre Astrophysical Observatory \citep[Observatorio Astrof\'\i sico de Javalambre, OAJ;][]{cenarro2014}. This photometric system was designed to measure precise \photoz, with an accuracy of  up to $\Delta z = 0.003\,(1+z)$, and to perform multiple cosmological studies, including  baryon acoustic oscillation (BAO) measurements \citep[see also][]{benitez2014,bonoli2020}. Besides the potential of \jp \ for cosmology and theoretical physics, this survey is perfectly suited for galaxy evolution studies. The main reasons are related to its unique photometric system, the characteristics of the imaging camera, and the large area of the sky that will be observed over the lifetime of the survey.

The \jp\ photometric system covers the full optical spectral range, with a narrow-band (NB) filter (FWHM $\sim 145$~\AA) every $\sim 100$~\AA. This is equivalent to low-resolution spectroscopy ($R\sim60$, $\Delta v \sim$5000 km s$^{-1}$, and $\Delta \lambda \sim$ 100 \AA) for each pixel over an area similar to that covered by the Sloan Digital Sky Survey (SDSS). It allows a very good sampling of the spectral energy distribution (SED) of each source of the imaged sky. Up to intermediate redshift, \jp\ will be more competitive than other medium-band imaging surveys, such as ALHAMBRA ($R\sim20$), in obtaining the stellar population properties of galaxies thanks to a higher spectral resolution \citep{diaz-garcia2015, diaz-garcia2019a, diaz-garcia2019b} and the huge number of galaxies that will be observed. The width of the NB filters ($FWHM\sim 145$~\AA) allows for the identification of star-forming galaxies and the measurement of emission lines such as H$\alpha$ up to $z\leq0.4$, H$\beta$ and [OIII]$\lambda$5007 up to $z \leq 0.8$, and [OII]$\lambda$3727 up to $z\leq1.4$. \jp \ will be a very competitive emission line survey, highly complementary to spectroscopic surveys, such as SDSS \citep{york2000}, zCOSMOS \citep{lilly2007}, the Galaxy And Mass Assembly (GAMA) survey \citep{driver2011}, and the VIMOS VLT Deep Survey \citep[VVDS; ][]{lefevre2013}, for deriving the properties of the ionized gas in emission line galaxies \citep[ELGs; ][]{martinez-solaeche2020}. \jp \ will be complemented and supported by the Javalambre-Photometric Local Universe Survey \citep[\jplus;][]{cenarro2019} and the Southern Photometric Local Universe Survey \citep[\splus;][]{mendesdeoliveira2019}  to study star-forming galaxies in the nearby Universe ($z < 0.015$) and to retrieve the integrated \citep[][Vilella-Rojo et al.~2020 in prep.]{vilella-rojo2015}  and the radial distribution of their star formation rates \citep[SFRs;][]{logronio-garcia2019}. Furthermore, other emission line objects, such as L$\alpha$ galaxies and quasi-stellar objects (QSOs), can be detected up to high redshift. In fact,  \jp \ will increase the capability of \jplus \ and \splus \ to reveal the bright end of the L$\alpha$ luminosity function at $z = 2$--$3$ \citep{spinoso2020}.

The imaging characteristic of the survey along with the pixel scale of the camera ($0.23$\,$\arcsec$\,pixel$^{-1}$ for the BB filters and $0.4$\,$\arcsec$\,pixel$^{-1}$ for the NB filters) opens a pixel-by-pixel investigation of the spatially resolved galaxies. This makes \jp \ a competitive survey of nearby galaxies ($z\leq0.15$), similar to an integral field unit (IFU) survey, and it will complement other IFU surveys, such as the Calar Alto Legacy Integral Field Area \citep[CALIFA; ][]{sanchez2012, garcia-benito2015, sanchez2016} or the Mapping Nearby Galaxies at the Apache Point Observatory \citep[MaNGA; ][]{bundy2015, law2015}. \jp \  has the ability to retrieve the spatially resolved stellar population properties of galaxies up to distances of several effective radii, thus going farther than MaNGA, for instance.

The large volume and area of the survey ($\sim8000$~deg$^2$) will include millions of galaxies for which SEDs will be obtained. This will allow us to identify and characterize blue galaxies (BGs), and luminous red galaxies (LRGs), as well as to study the  evolution of galaxy stellar populations up to $z=1$ through SED-fitting techniques. This makes \jp \ a very competitive survey compared to other spectroscopic surveys, such as the VIMOS Public Extragalactic Survey \citep[VIPERS; ][]{guzzo2014, haines2017}, the Large Early Galaxy Astrophysics Census \citep[LEGA-C; ][]{vanderwel2016, wu2018}, and the future WHT Enhanced Area Velocity Explorer-Stellar Population at intermediate redshift Survey \citep[WEAVE-StePS; ][Iovino et al.~2020 in prep.]{costantin2019}, 
and pre-selecting sets of galaxy samples, as some spectroscopic galaxy surveys do, is not necessary. 
The continuous and non-segregated area of the \jp\ survey will allow us to explore the role of environment in galaxy formation and evolution by characterizing samples of different types of galaxies in a wide range of redshifts, halo masses, density fields, and inter-galactic medium environments. Groups and galaxy clusters at intermediate redshifts ($0.1\leq z \leq 1$) will be easily detected (Maturi et al.~2020 in prep.), and the SEDs of their galaxy members will be used to determine, in a homogeneous way, their stellar population and emission line properties according to their environmental densities. Furthermore, at low redshifts ($z < 0.1$), \jp \ will be a very competitive survey, for example as compared with the GAs Stripping Phenomena in galaxies with MUSE survey \citep[GASP; ][]{poggianti2017}, for characterizing the intra-cluster light and the ionized gas stripping, as well as its connection with the gas removal processes in galaxies.

Further examples of the scientific capabilities of \jp \ for galaxy evolution studies are presented in \citet{bonoli2020}. These results are based on the analysis of a set of \jp-like data, named \mjp,  collected in a single strip of the sky that overlaps with the All-wavelength Extended Groth Strip International survey \citep[AEGIS; ][]{davis2007}, using the \pf \ camera instead of the JPCam. The present work, based on \mjp \ galaxies at $z < 1$, has the goal of paving the way towards identifying and characterizing galaxy populations and their stellar content across cosmic time. In particular, we show the power of the \jp \ filter system to dissect the bimodal distribution of stellar populations of galaxies and its evolution up to $z\sim1$.

The well-known bimodal distribution of galaxy colours in the nearby Universe ($z <0.1$), usually referred to as the red sequence and the blue cloud, has been largely studied via colour--magnitude diagrams (CMDs), particularly through the analysis of SDSS data \citep{blanton2009}. The galaxy locus within this diagram correlates with its stellar population properties: The red sequence is populated by red, old, and metal-rich galaxies, whereas the blue cloud is mainly composed of star-forming galaxies of lower metallicities \citep{kauffmann2003a, kauffmann2003b, baldry2004, brinchmann2004, gallazzi2005, mateus2006, mateus2007}. These colour distributions also depend on the galaxy stellar mass, an important factor in the evolution of galaxies \citep[e.g.][]{perez-gonzalez2008, perez2013, gonzalez-delgado2014},
with the red sequence being populated by the most massive galaxies \citep{hernan-caballero2013, schawinski2014, diaz-garcia2019a, diaz-garcia2019b}.
The colour bimodality is also present in the colour--stellar mass diagrams and is tightly correlated with the ongoing star formation processes (or SFRs) and the stellar mass of the galaxies in the sample \citep{noeske2007, speagle2014, renzini2015, catalan-torrecilla2015, gonzalez-delgado2016, lopezfernandez18, thorne20}. 
Despite selection effects and photometric uncertainties, the colour bimodality has
been measured at intermediate redshifts from large-area surveys, such as BOSS, using 
Bayesian statistics \citep{montero-dorta2016}.  The existence of these two groups 
beyond the nearby Universe is therefore accepted, with evidence suggesting that 
they could already be in place at least at $z\simeq$4 \citep{bell2004, pozzetti2010, muzzin2013, haines2017, diaz-garcia2019a}. However, a colour--dust correction is very important for revealing the distribution of galaxies in colour--mass diagrams, as well as disentangling the real fraction of red and blue galaxies, since dusty star-forming galaxies may exhibit colours as red as the colours of red sequence galaxies \citep{williams2009, cardamone2010, whitaker2010, diaz-garcia2019a, diaz-garcia2019b}. Therefore, a clear separation between the imprint of the star formation history (SFH) and dust content is needed to identify and characterize the galaxy populations and their evolution.

The SFH of a galaxy is imprinted in its spectrum; as such, the SFH can be inferred using the fossil record encoded in the present-day stellar populations \citep{tinsley1968, tinsley1972, searle1973}. This has been extensively applied to fit the integrated spectra of nearby galaxies observed by SDSS \citep{panter2003, heavens2004, cid-fernandes2005, ocvirk2006, asari2007, tojeiro2011, koleva2011, citro2016} and GAMA \citep{bellstedt20}, as well as to spatially resolved data from IFU surveys of nearby galaxies \citep{perez2013, cid-fernandes2013, cid-fernandes2014, gonzalez-delgado2014, sanchez-blazquez2014, mcdermid2015, gonzalez-delgado2016, gonzalez-delgado2017, lopezfernandez18, garcia-benito2017, goddard2017a, goddard2017b, zibetti2017, garcia-benito2019, sanchez2019, sanchez2020}. The spectroscopic stellar continuum of a galaxy contains many absorption lines and stellar features whose intensities allow us to constrain its SFH, extinction, and the age and metallicity of the stellar population. Nevertheless, this technique suffers large uncertainties when  applied to spectra of low  signal-to-noise ratio (S/N)  
or BB photometry. However, intermediate-band photometry at optical wavelengths combined with BB near-infrared (NIR) data from the ALHAMBRA survey provided successful results in identifying star-forming and quiescent/red galaxies \citep{diaz-garcia2019a,diaz-garcia2019b,diaz-garcia2019c}. The capability of the \jp \ filter system to retrieve the stellar population properties of galaxies was first studied in \citet{benitez2014}, based on both synthetic magnitudes obtained from SDSS spectra of nearby galaxies and on mock galaxies, and later by \citet{mejia2017}. The multi-band data of \mjp, or \js, provides us with the opportunity to investigate the potential of \jp \ to identify and characterize  galaxy populations since $z\sim1$ with real data, by adapting or extending the full spectral method to multi-NB data. Some preliminary results are discussed in \citet{bonoli2020}, where we proved the potential of \js \ to retrieve the radial distribution of spatially resolved stellar population properties, such as the age up to two effective radii, with a precision similar to that obtained with MaNGA data. Here, several methodologies are explored to prove the consistency of the results in the identification and characterization of galaxy populations for a complete sample of galaxies, as well as their evolution up to $z\sim1$.

This paper is structured as follows. Section~\ref{sec:Data} briefly describes the data and properties of the sample analysed here. Section~\ref{sec:method} explains the method of analysis and the four different SED-fitting codes used to derive the properties of the galaxy stellar populations. In Sect.~\ref{sec:Results} we present the inferred stellar population properties of the galaxies in the sample, and the uncertainties in the galaxy properties are discussed in Sect.~\ref{sec:Uncertainties}. We identify and characterize blue and red galaxies and their evolution up to $z\sim1$ in Sect.~\ref{sec:Discussion}. Finally, the results are summarized in Sect.~\ref{sec:Summary}.

Throughout the paper we assume a Lambda cold dark matter ($\Lambda$CDM) cosmology in a flat Universe with $H_0=67.4$~km~s$^{-1}$ and $\Omega_\mathrm{M}=0.315$ \citep{Planck2018}. All the stellar masses in this work are quoted in solar mass units (M$_\odot$) and are scaled according to a universal \citep{chabrier2003} initial stellar mass function. All the magnitudes are in the AB system \citep{oke1983}. 

\section{Data and sample}\label{sec:Data}

This section presents a brief summary of the main instrumental characteristics and
observations contained in \mjp. A detailed and broader description of \mjp\ can be found in the \mjp\ presentation paper \citep{bonoli2020}.

\subsection{Instrumental setup}

The \mjp \  survey was carried out at the OAJ \citep{cenarro2014}, located at the Pico del Buitre in the Sierra de Javalambre in Teruel (Spain). The acquisition of the data was performed with the 2.5m Javalambre Survey Telescope (JST/T250), which is optimized to provide good image quality in the optical spectral range ($3300$--$11000$~\AA) across the $7$~deg$^2$ of the focal plane. 

The \mjp \  data were obtained with the \pf \ camera, which was the first scientific instrument installed at the JST/T250, before the arrival of the JPCam  \citep{taylor2014, marin-franch2017}. The JPCam has an effective field of view (FoV) of $4.2$~deg$^2$, and it is composed of $14$~charge coupled devices (CCDs). In contrast, the \pf \ instrument is a single CCD direct imager located at the centre of the JST/T250 FoV with a pixel scale of $0.23$\arcsec~pixel$^{-1}$, which is vignetted on its periphery, providing an effective FoV of $0.27$~deg$^2$. 

The filter system of \jp \ contains $54$ NB filters ranging from $3780$~\AA\ to $9100$~\AA\ plus two broader filters at the blue and red ends of the optical range, centred at $3479$~\AA\ (named \uja) and  $9316$~\AA\ (named \jred), respectively. The $54$ NB filters have a full width at half maximum (FWHM) of $145$~\AA\ and are equally spaced every $\sim100$~\AA, whereas the FWHM of the \uja \ band is $495$~\AA\ and \jred\ is a high-pass filter. This filter system was optimized for: (i) an accurate measurement of \photoz\ up to $z\sim1$ to carry out cosmological experiments using different tracers at different epochs; (ii) delivering low-resolution ($R\sim60$) photo-spectra (or \js) that allow us to identify and characterize the stellar populations in galaxies up to $z\sim1$; and (iii) the measurement of emission lines in galaxies and broad emission line features of QSOs and supernovae. A detailed technical description of the \jp\ filter design and characterization can be found in \citet{marin-franch2012}. 
In addition, \mjp\ includes four SDSS-like BB filters: \ujp, \gb, \rb\, and \ib.

\subsection{Data and calibration}

The \mjp \  observations  consist of four pointings in the Extended Groth Strip along a strip aligned at $45\degr$ with respect to north at ($\alpha$, $\delta$) = ($215.00\degr$, $+53.00\degr$), amounting to a total area of $\sim1$~deg$^2$. The depth achieved is fainter than $22$~mag (AB) for filters with $\lambda < 7500$~\AA\  and is $\sim22$~mag (AB) for longer wavelengths. 

The images were processed by the Data Processing and Archiving Unit group at Centro de Estudios de F\'\i sica del Cosmos de Arag\'on  
\citep[\mbox{CEFCA};][]{cristobal-hornillos2014}. All the images and catalogues are available through the CEFCA Web portal\footnote{\mainurl}, which offers advanced tools for data searches, visualizations, and data queries (Civera et al.~2020, in prep.). In this work, we only analyse the photometric data obtained with \sext\ \citep{bertin1996} in the so-called dual mode. The \rb \ filter was used as the detection band. For the rest of the filters, the \rb \  band was used as a reference to define the aperture and to build the source catalogue. Further details of the observations and the data reduction can be found in \citet{bonoli2020}.

\subsection{Selection of the sample}

The sample of galaxies analysed here is extracted from the dual-mode \mjp \ catalogue,
selected according to \rb\ magnitude, stellarity index, and \photoz. The resulting catalogue contains 64293 objects, of which $\sim$ 15000  have \photoz\ $z\leq1$ and $\class \leq 0.1$. 

We used this automatic morphological classification of \sext, as provided by the \class\  parameter, to build our galaxy sample. Other stellarity indices, such as the `stellar-galaxy locus classification' \totalstar\ parameter \citep{lopez-sanjuan2019,baqui2020}, are also listed in the \mjp \ catalogue and can be used to select extended sources.  We checked that for $\totalstar \leq 0.1$ the selected sample is very similar to that selected using \class$\leq 0.1$. 

The \photoz\ values for \mjp\ were estimated using the \jphotoz\ package developed by the \photoz\ team at CEFCA. This package is a modified version of the \lephare\ code \citep{arnouts2011} and has a new set of stellar population synthesis galaxy templates added (Hernan-Caballero et al.~2020, in prep.). As a result, the \mjp\ catalogue contains several estimates of \photoz. To select our galaxy sample, we used the \photoz\ value corresponding to the mode or the maximum of the probability density function. 

The third condition for selecting the sample galaxies is related to the quality of the detection and galaxy brightness. Only objects with a \sext\ flag-mask equal to zero in the \rb\ band were allowed. This parameter guarantees that: (i)  our sources are detected in a well-defined aperture; (ii) our sample does not contain sources close to problematic regions of the image, such as bright stars; and (iii) sources are not out of the window frame.  The \mjp \ dual-mode catalogue includes measurements for different types of aperture. We used the \magauto \ aperture as a proxy for the total galaxy magnitude. To define a flux-limited sample, we imposed the condition that the galaxies have to be brighter than $22.5$~(AB) in \rb\ according to the \magauto \ photometry. 

\citet{bonoli2020} showed that the catalogue is complete up to \rb$=23.6$~(AB) for point-like sources and up to \rb$=22.7$~(AB) for extended sources. Furthermore, the \photoz \ accuracy reaches a precision of 0.8$\%$  for most of the sources up to \rb$=22.5$~(AB), while only 15$\%$ of the sources are outliers with errors $\geq$ 5$\%$. Our selection criteria guarantee a complete sample of properly detected galaxies with a sub-percent precision at $z\leq 1$. 
Our final sample contains $\sim8500$ extended sources that obey the following criteria: 
$z \leq 1$, \rb$\leq22.5$, and \class\ $\leq 0.1$. 

\begin{figure*}
\centering
\includegraphics[width=0.75\textwidth,clip=True]{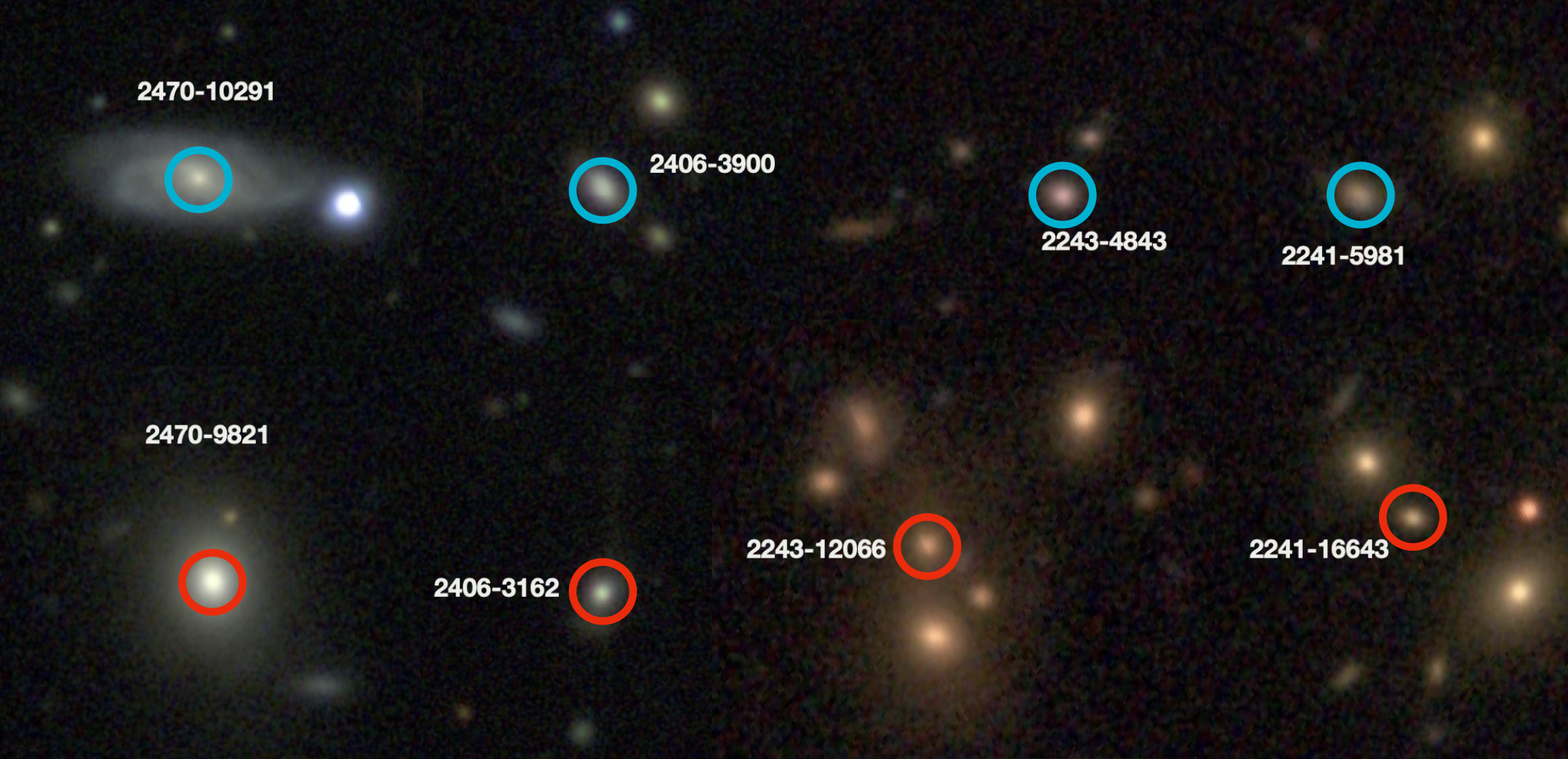}

\includegraphics[width=0.95\textwidth,clip=True]{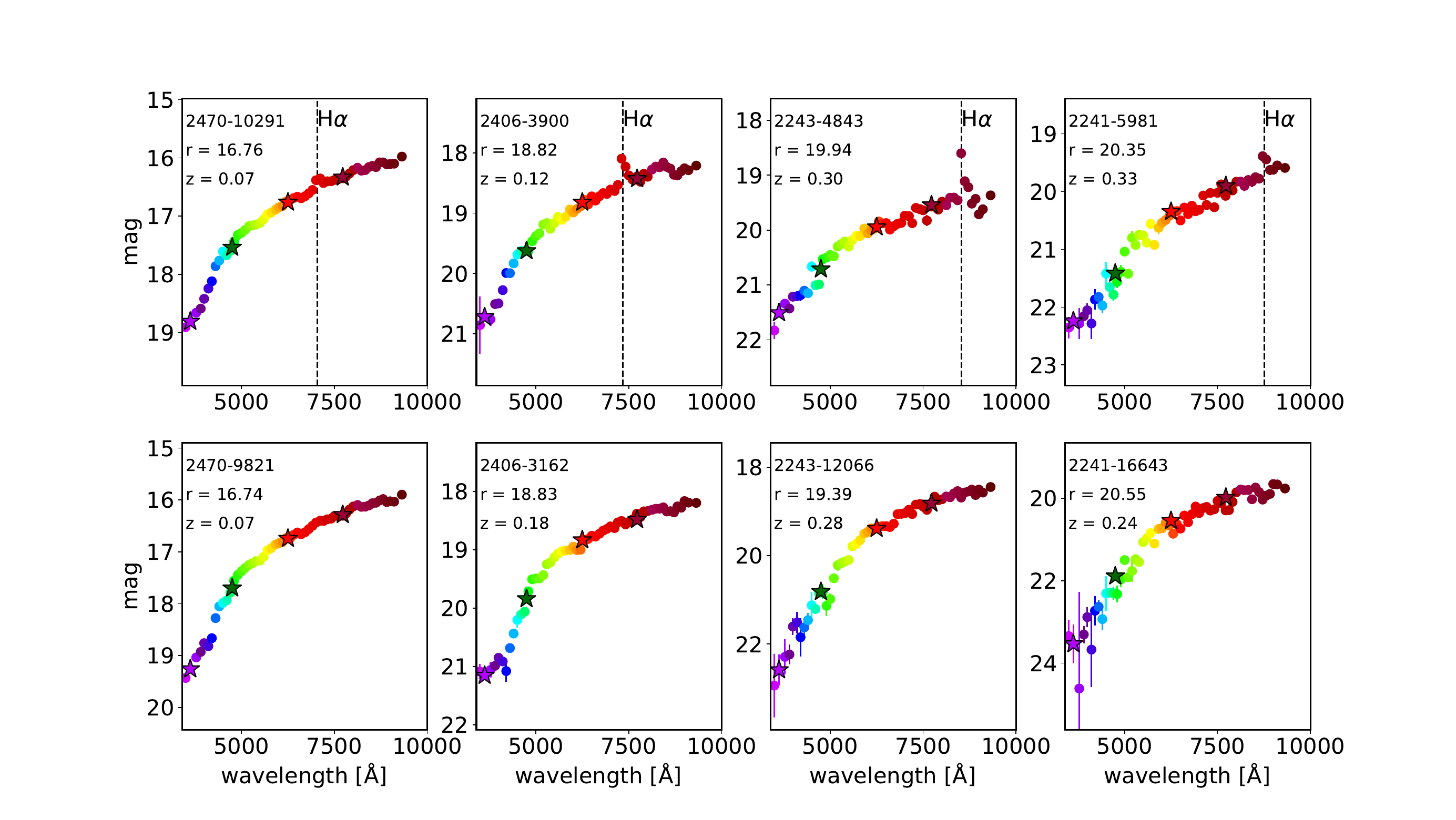}
\caption{Images and \js \ of several galaxies in the AEGIS strip.  {Top panels}: Mosaic image of red and blue galaxies.   \textit{Bottom panels}: Corresponding \js\ (coloured points) in \magpsfcor\  for the galaxies above. We note the H$\alpha$ emission in the BGs. The star-shaped markers are the BB magnitudes (\uja, \gb, \rb, and \ib). }
\label{fig:data}
\end{figure*}

It is worth noting that our sample contains galaxies with a wide variety of morphologies, colours, and environments (see Fig.~\ref{fig:data}). Even though the bulk of the sample is at $z \sim 0.3$, it is also representative of nearby galaxies ($z \sim 0.1$). We can distinguish spiral structures (e.g. ~2470$-$10291 in Fig.~\ref{fig:data}), spheroidal morphologies (e.g.~2470$-$9821 at $z=0.07$ in Fig.~\ref{fig:data}), and even radio-jet structures at higher redshifts (e.g.~2406$-$3162 at $z=0.18$ in Fig.~\ref{fig:data}). Galaxies in groups are also well identified (e.g.~2243$-$12066 and 2241$-$16643 in Fig.~\ref{fig:data} at $z=0.28$ and $0.24$, respectively), as well as galaxies in less dense environments. The \js\ of these galaxies (lower panels of  Fig.~\ref{fig:data}) show the power of \jp\ to identify blue and star-forming galaxies, to detect the recombination nebular lines H$\alpha$ up to $z\sim0.4$ and H$\beta$ up to $z\,\sim$\,0.8, and to detect other nebular collisional lines, such as [OIII]$\lambda$5007 up to $z \sim$0.8 and  [OII]$\lambda$3727 up to $z \leq$1.4. Luminous red galaxies are also very well identified in the survey (see four examples of this type of galaxy in Fig.~\ref{fig:data}). These galaxies populate the upper limit of the brightness distribution of the sample, but fainter and more distant galaxies are also properly detected (see Sect.~\ref{sec:method}).


\subsection{Observational properties of the sample}

\begin{figure}
\centering
\includegraphics[width=0.5\textwidth]{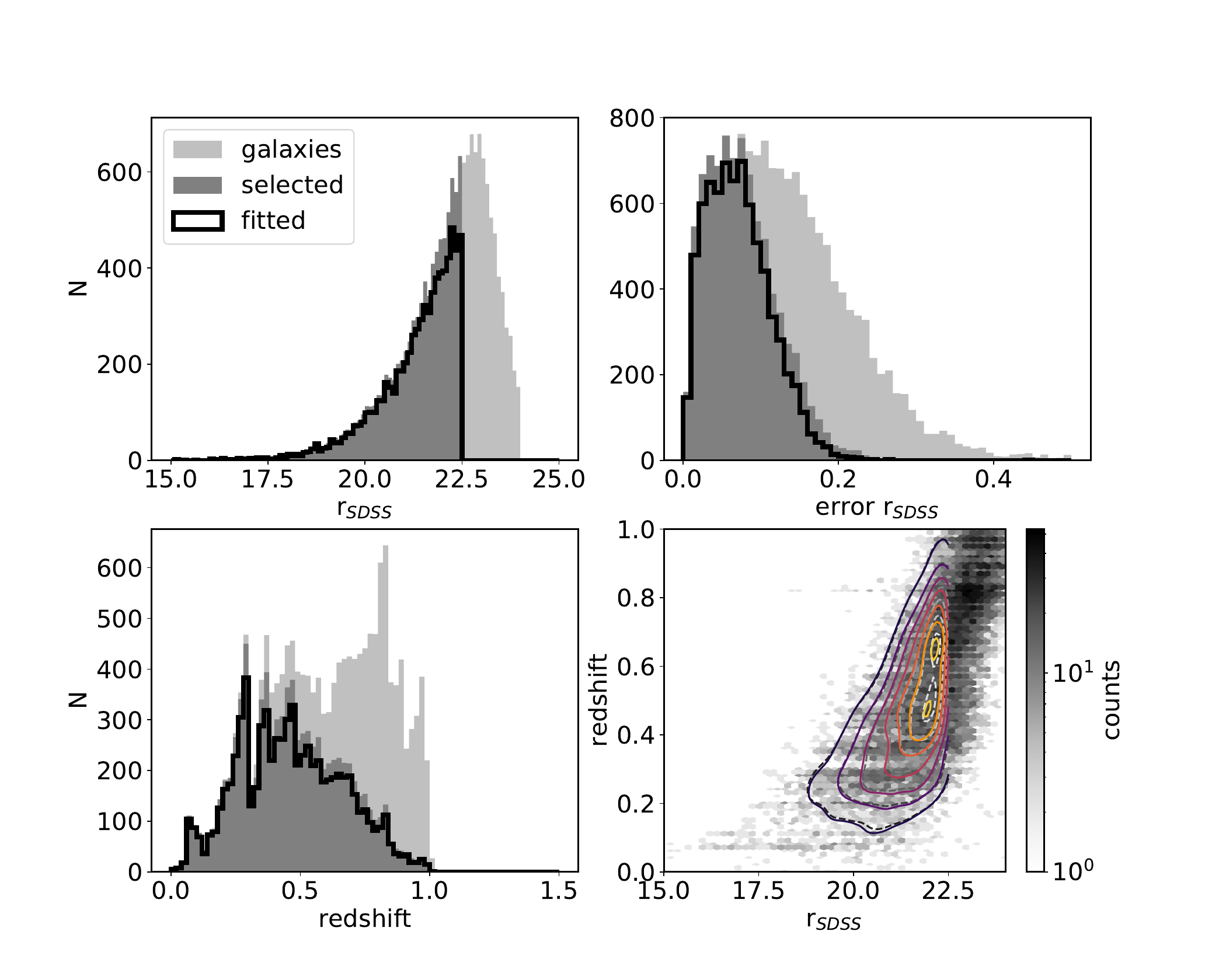}
\caption{Distributions of several observational properties. \textit{Top-left, top-right, and bottom-left panels}: Distribution of  \rb\ magnitude and errors, and redshift of galaxies with \photoz$\leq 1$ identified in \mjp \ (light grey) and galaxies with \rb$\leq 22.5$ and  \photoz$\leq 1$ (dark grey).\ The black line indicates the galaxies that were fitted. All magnitudes and related errors correspond to the \magauto \ photometry. \textit{Bottom-right panel}: Density map of the number count of galaxies with $z\leq 1$, shown with the density contours of the selected sample (dashed lines) and the galaxies that were fitted (solid lines).
}
\label{fig:histredshift}
\end{figure}

Figure~\ref{fig:histredshift} compares the distributions of redshift, brightness, and error in the \rb\  band of the final sample with those of the original sample of \mjp\ galaxies up to $z=1$. A large fraction of galaxies in the sample have \photoz  \ between $0.2$ and $0.6$. Few galaxies with \photoz\ between $0.8$ and $1$ were selected due to the cut in magnitude at $22.5$~(AB) in \rb. Most of the extended objects at $z \geq 0.8$ are fainter than $22.5$~(AB) and are barely detected. In comparison, the mean error in the \rb\ band for the selected sample is $0.1$~(AB), whereas for the full sample it is $0.2$~(AB). As a consequence, the galaxies of the final sample are detected with an average S/N of $10$,  which is higher than the average S/N of the whole sample in the dual-mode catalogue. 

We can fit $\sim95$\% of the $\sim8500$ \js\ in the selected sample with the four SED-fitting codes used in our analysis (Sect.~\ref{sec:method}). The unfitted spectra do not fulfil quality requirements imposed by the SED-fitting codes, such as the minimum number of bands or the minimum S/N in each band, which may differ from code to code. Figure~\ref{fig:histredshift} shows that the selected sample and the whole sample have similar distributions in redshift, \rb\ band brightness, and error, and, in terms of detection, both datasets have similar S/N distributions.

\subsection{The sample in the \auto\ and \psfcor\ photometry}\label{sec:sample_auto_psf}

In addition to \magauto, the dual catalogue also includes the \magpsfcor\ photometry.  This photometric method was developed for the analysis of the ALHAMBRA data \citep{molino2014} and adapted for \mjp. For each object it performs corrections 
that take into account the differences in the point spread functions (PSFs) between the different bands. This process guarantees a good determination of the galaxy colour and the shape of the continuum, which is crucial for retrieving the properties of the galaxy stellar populations via the full spectral fitting techniques applied here. 
Instead of the total magnitude that estimates the flux within an elliptical aperture determined by the \kron\ (such as \magauto), \magpsfcor\ is measured for a smaller aperture (i.e. it does not provide the total flux).  

\begin{figure}
\centering
\includegraphics[width=0.5\textwidth]{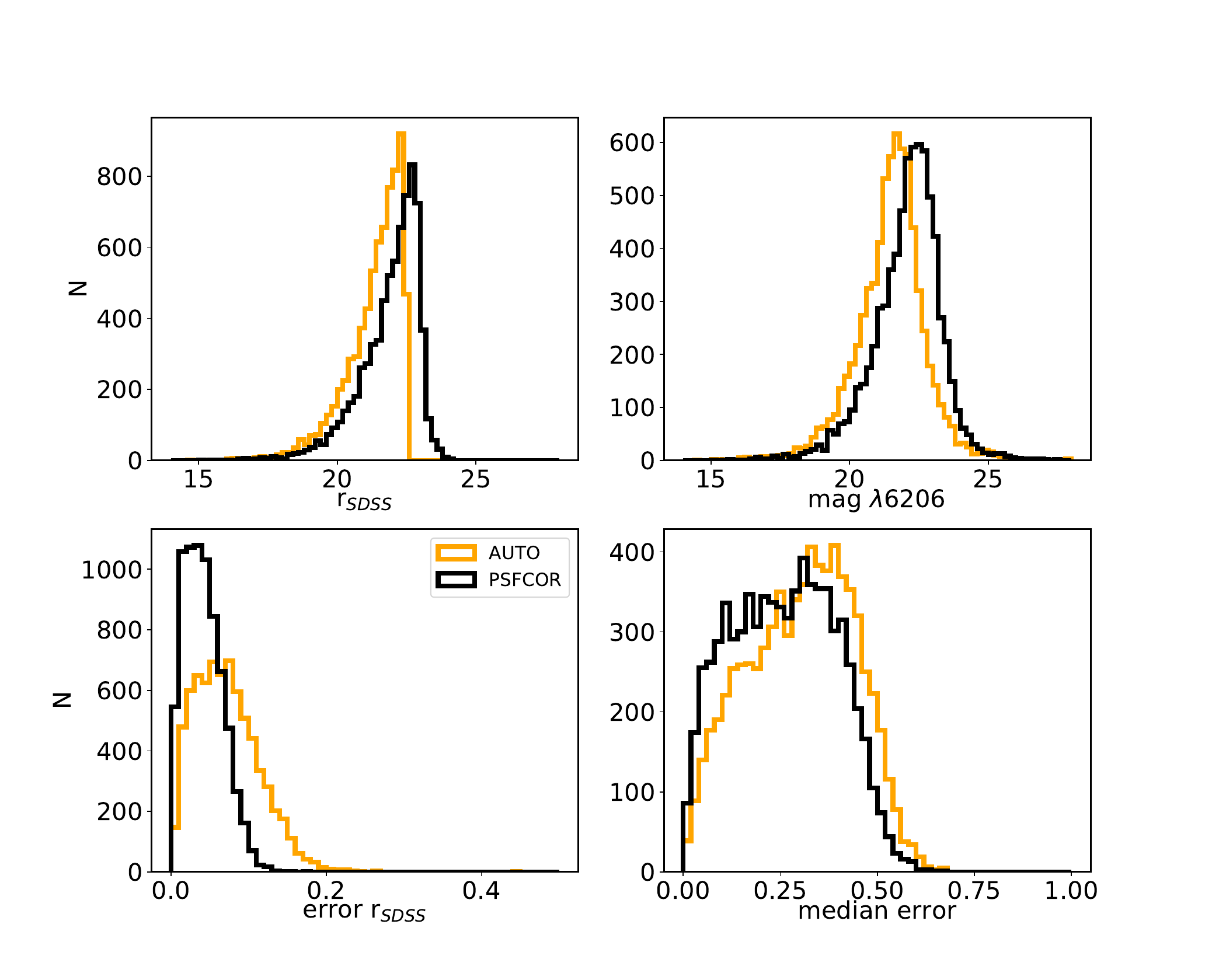}
\caption{Distribution of galaxy magnitudes for the \rb\ BB filter and the NB filter        at $\lambda6206$~\AA\ (\textit{top panels}), and the distribution of the magnitude errors in the \rb\ BB and the median error of the NB filters for each galaxy (\textit{bottom panels}). The coloured and black lines correspond to the values obtained by the \magauto \ and \magpsfcor\ photometry, respectively. 
}
\label{fig:hist_mag_err}
\end{figure}

In this work we performed the analysis of the sample using both the \magpsfcor\ and the \magauto\ photometry, which may differ (Fig.~\ref{fig:hist_mag_err}). 
The magnitude distributions in the \rb\ BB filter and in the NB filter with a central wavelength at $\lambda6206$~\AA\ ($NB6206$) indicate that galaxies are fainter in \magpsfcor\ by $0.55$~mag because the aperture is typically smaller than for \magauto. Flux uncertainties in \rb\ for \magpsfcor\ are slightly smaller, by $0.03$~mag, than for
\magauto.
The same is true for the NB filters. In particular, the error in flux for the $NB6206$ filter is smaller by $0.05$~mag for \magpsfcor\  and, in general, by $\sim0.04$~mag for the rest of the NB filters; the S/N distribution for \magpsfcor\ is slightly better than for \magauto.

\begin{figure}
\centering
\includegraphics[width=0.5\textwidth]{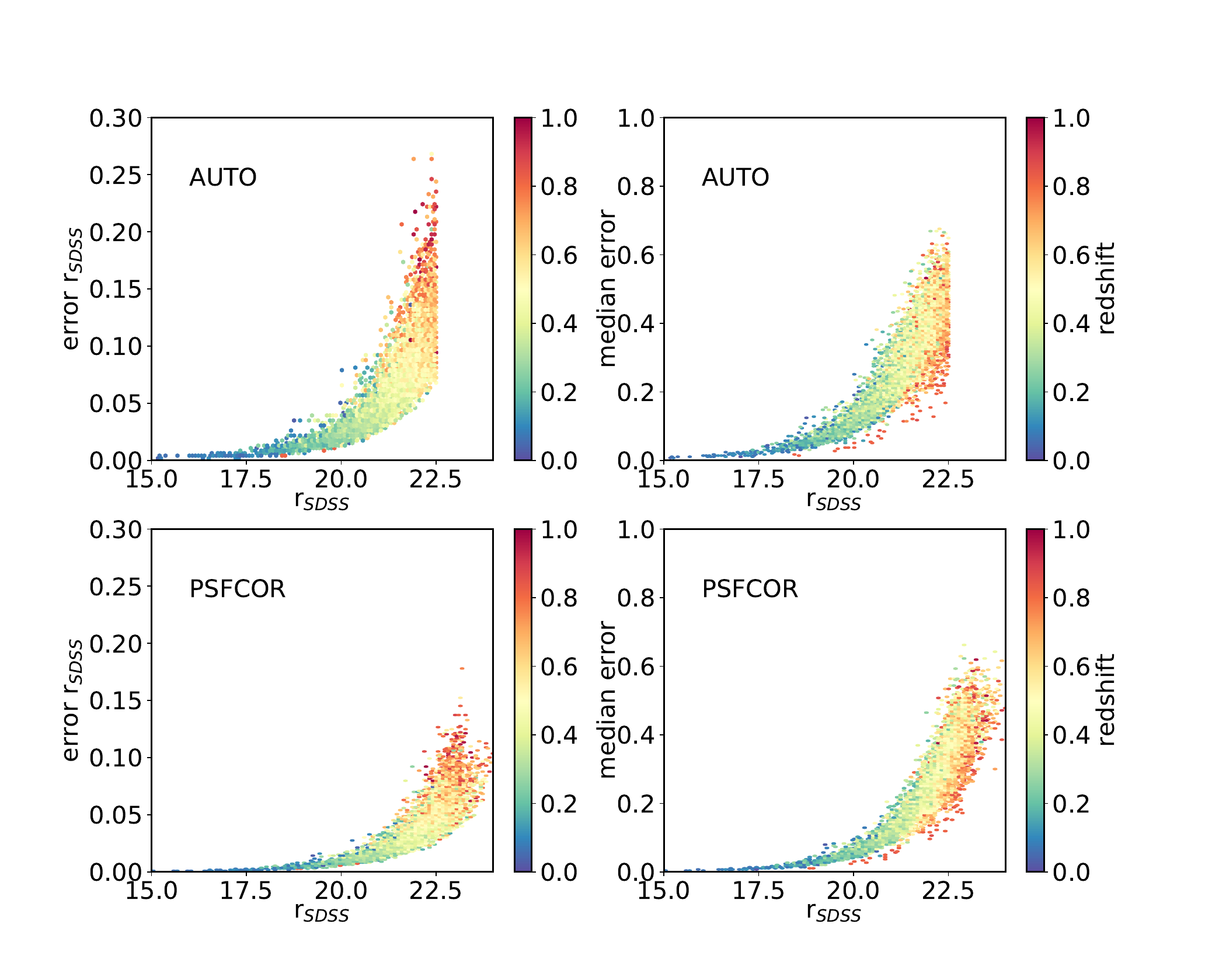}
\caption{Errors of the \rb\ band (left) and NB (right) filters as a function of the \rb\ magnitude (detection band). The \textit{top and bottom panels} show the \magauto\ and \magpsfcor \ apertures, respectively. The colour bar illustrates the redshift of each galaxy.
 }
\label{fig:mag_err}
\end{figure}

As expected, there is a clear correlation between the error in the magnitude and the magnitude of the object (Fig.~\ref{fig:mag_err}). Errors in the \rb\ and NB filters increase exponentially at faint magnitudes. The slope is steeper for the NB filters than for the BB \rb\ filter. For \magauto\ magnitudes, an \rb$=20$~(AB) galaxy has an error $<0.05$ mag, while the median error for the NB filters is $<0.1$ mag. For an \rb$=22$~(AB) galaxy, the respective errors are $<0.15$ and $<0.5$ mag. 
Thus, while faint sample galaxies are properly detected in the \rb\ band, the median average S/N is low: $\rm S/N\sim1$--$3$ at \rb$=22.5$~(AB). 
The uncertainties for \magpsfcor\ are smaller than for \magauto, but both show similar behaviours. For this reason, we performed the spectral fits to the \js\ using the \magpsfcor\ magnitudes, and we used them as reference values despite the fact that the sample was defined according to \magauto. 

\begin{figure}
\centering
\includegraphics[width=0.45\textwidth]{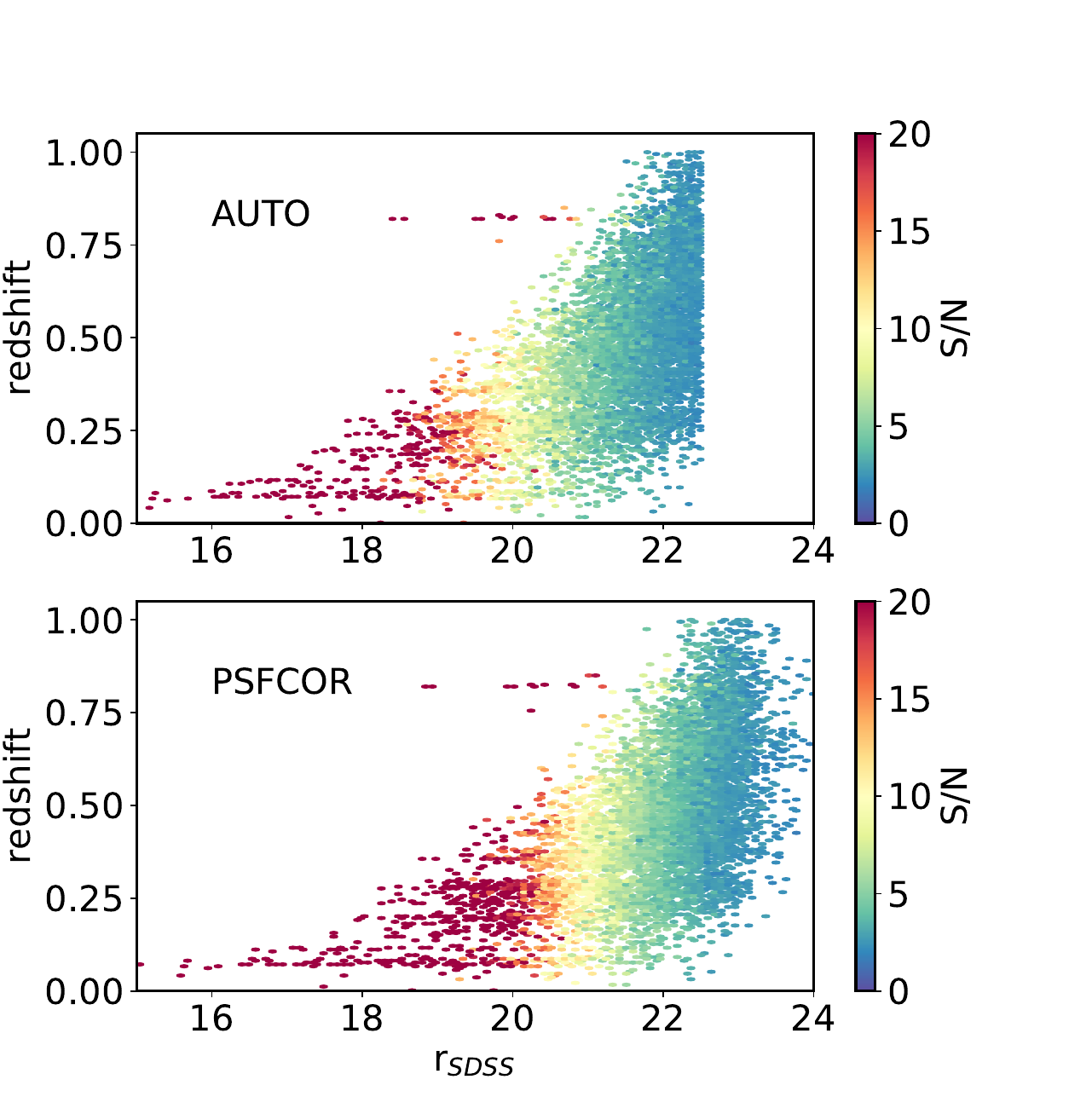}
\caption{Magnitudes observed in the \rb\  band as a function of redshift for \magauto \ and \magpsfcor\ apertures. The colour bar shows the median S/N in the NB filters.}
\label{fig:mag_redshift}
\end{figure}

The magnitude in the \rb\ band shows a clear dependence on redshift (Fig.~\ref{fig:mag_redshift}). Galaxies with \rb\,$\leq 20$ (\magauto) are typically found at $z<0.5$, while fainter galaxies are at any distance. The S/N is also a clear function of the brightness of the galaxy, as indicated by the average median errors of the \rb\ and the NB filters (Fig.~\ref{fig:mag_err}). The increase in the S/N  for \magpsfcor \ with respect  to \magauto \ is also clear in Fig.~\ref{fig:mag_redshift}, although this increase is small. For instance, galaxies with \rb\,$\sim 22$~(AB) have a S/N of $\sim5$ for \magpsfcor\ apertures, while for \magauto\ this value is $\sim 3$.

Our selection criteria exclude most of the bright sources (\rb\ $<$ 21) at $z\sim0.82$ in the \mjp\ catalogue, which are probably red dwarf stars that do not have well-estimated \photoz. Of these sources, 14 with \class\ $<0.1$ are still present in our sample, appearing as a horizontal red cluster at $z\sim0.82$ in Fig.~\ref{fig:mag_redshift}; eight of these sources have \totalstar\ $>0.9$, as expected for red dwarf stars. The fraction of these sources in our sample is quite small, less than $0.2$\% of the fitted \js , and their inclusion in the sample will not have any statistical impact on our results.

\section{Method of analysis}
\label{sec:method}

There is a long list of SED-fitting codes developed to fit data ranging from the far ultraviolet to the far infrared (FIR), such as PROSPECT \citep{robotham20}, CIGALE \citep{boquien19}, PROSPECTOR \citep{leja19}, BAGPIPES \citep{carnall18}, BEAGLE \citep{chevallard16}, and MAGPHYS \citep{dacunha08}. Their ingredients, input, and output are different. 
They include parametric and/or non-parametric SFHs, diverse stellar initial mass functions (IMFs), and a variety of dust attenuation laws and dust emission models to fit the FIR, as well as state-of-the-art stellar population templates. 
In this work we used our own SED-fitting codes, specifically developed or adapted to fit the galaxy SEDs as traced by the 56 \jp\ bands. The objective is to estimate the SFH of each galaxy in order to derive various properties of the galaxy stellar population. One of these codes (\baysea) is a parametric code, that is, the SFH is described by an analytical model. In contrast,  the other codes (\muff, \alstar, and \tgas) are non-parametric, that is, the SFH is expressed as an arbitrary superposition of different simple stellar populations (SSPs). This section describes the common features shared by these codes, as well as their specific aspects that may cause differences in the results.

\subsection{General aspects common to the SED-fitting codes}

\subsubsection{Common inputs and assumptions}

{\it Stellar population models:} In order to minimize discrepancies in the results due to the methodologies embedded in the codes, the four fitting codes used the latest versions of the \citet{bruzual2003} stellar population synthesis models
\citep[][hereafter \cb]{plat2019}. The \cb models follow the PARSEC evolutionary tracks \citep{marigo2013,chen2015} and use the Miles \citep{sanchez-blazquez2006,falcon-barroso2011,prugniel2011} and IndoUS \citep{valdes2004,sharma2016} stellar libraries in the spectral range covered by the \jp \ data.
These models are available for different metallicities ranging from $\log \left(Z_\star/Z_\odot \right) = -2.23$ to $0.55$~dex and run in age from $\sim$\,0 to $14$\,Gyr. 
For this paper we used the \cb SSP models computed for the \citet{chabrier2003} IMF. 
Specifically, we used a common set of seven metallicities,  $\log \left(Z_\star/Z_\odot \right) =  -2.23$, $-1.53$, $-0.63$, $-0.33$, $0$, $0.25$, and $0.55$, and 
all the ages to build the SFH of the parametric models in \baysea, the `two-burst' composite stellar population (CSP) models in \muff, and the `square burst' CSP models in \alstar\ and \tgas.

{\it Dust attenuation law:} Attenuation by dust is a key ingredient for a proper interpretation of galaxy colours. We assumed a common attenuation law for all four codes. Model spectra were attenuated by a factor formally expressed as  $e^{-q_{\lambda} \tau_V}$,  where $\tau_V$ is the dust attenuation parameter in the $V$ band and $q_\lambda \equiv \tau_\lambda / \tau_V$ denotes the reddening law. For the present work we chose the attenuation law proposed by \citet{calzetti2000}, which we added as a unique foreground screen with a fixed ratio of $R_V=3.1$ (the average value for the Milky Way).

{\it Emission lines:} The presence of emission lines from either young star-forming regions or an active galactic nucleus (AGN) component may strongly increase the flux in certain \jp\ bands. Since emission lines are not included in SSP models, the NBs affected by strong emission lines (mainly H$\alpha$; H$\beta$; [NII]$\lambda$ 6584, 6548; [OIII]$\lambda$ 5007, 4959; and [OII]$\lambda$3727) were removed from the analysis at the redshift of each galaxy to ensure that the stellar population properties reported by the four codes came exclusively from the stellar continuum.

{\it Maximum age of the stellar population and redshift:} In the non-parametric codes, the age span of the SSP base used to fit a given galaxy SED is truncated at $t_\mathrm{max}$, the age of the Universe at the galaxy redshift. In \baysea, the lookback time for the onset of star formation is limited to $0.99 \times t_\mathrm{max}$. The redshift of each galaxy is fixed to the mode of the \photoz\ value provided in the \mjp\ catalogues (Hern\'an-Caballero et al.~2020, in preparation).

\subsubsection{Merit function and confidence intervals}

In \alstar, \tgas, and \muff,\ the best-fitting solution is obtained by computing the non-negative values of the coefficients $x_{tZ}$ that minimize the merit function

\begin{equation}
\chi^2=\sum_\lambda \frac{[F^\mathrm{obs}_\lambda - \sum_{t,Z} x_{tZ}\  f_{\lambda, tZ}(\tau_V)]^2}{\sigma_\lambda^2},
\label{chi2}
\end{equation}

\noindent which is used to measure the goodness-of-fit. In Eq.~(\ref{chi2}), $F^\mathrm{obs}_\lambda$ is the observed flux in each of the \mjp\ bands, $f_{\lambda,tZ}$ are the model photo-spectra, and $\sigma_l^2$ is the corresponding uncertainty in each band. The sum is calculated over all the filters (index $\lambda$) and models (indices $t$ and $Z$). Both \alstar\ and \tgas\ use the non-negative least squares (NNLS) algorithm \citep{lawson1974} to find the vector $x_{tZ}$ that minimizes $\chi^2$, using an outer loop to minimize it by the dust attenuation $\tau_V$.
\muff , on the other hand, uses an algebraic solution to compute the values of the coefficients, with the dust attenuation embedded in the pre-computed models. In contrast, \baysea\ follows a Markov chain Monte Carlo (MCMC) approach, using the same figure-of-merit function to compute the probability of each step, and $\tau_V$ is one of the model parameters.

Given the nature of the photometric uncertainties and the known correlations and degeneracies amongst colours and stellar population properties, it is essential to perform a statistical analysis based on the observed photon-noise and its impact on the results. \muff, \alstar, and \tgas\ follow the frequentist approach of Monte Carlo-ing the input (by adding Gaussian noise with observationally defined amplitudes) and repeating the fit many times, assuming that the errors in the different bands are uncorrelated. A probability distribution function (PDF) for each stellar population property is built by weighting the results from each iteration by the likelihood $\mathcal{L} \propto \exp(-\chi^2/2)$. In \baysea\ the problem is treated in a Bayesian way, and the posterior PDF for each parameter results naturally from the MCMC algorithm.

For the four codes, the inferred value for each galaxy property is obtained directly from the corresponding marginalized PDF. 
Each property is thus characterized by the mean, the median, and the percentiles that define the confidence interval in the distribution. 

\subsubsection{Stellar population properties}
 
The four codes use the same definitions of the three stellar population properties to characterize the AEGIS galaxies. We briefly describe them here.\ 

First, the galaxy stellar mass ($M_\star$) is the stellar mass of a galaxy at present. It is calculated from the mass converted into stars according to the SFH of the galaxy, taking into account the mass loss of the SSP owing to stellar evolution. We usually plot $\log_{10} M_\star [M_\odot]$.

Second, for the age of the stellar population, we define the mass-weighted logarithmic age (hereafter the mass-weighted age) following Eq.\ (9) from \citet[][]{cid-fernandes2013} as
\begin{equation} 
\label{eq:at_mass}
\langle \log\ \mathrm{age} \rangle_\mathrm{M} = \sum_{t,Z} \mu_{tZ} \times \log t,
\end{equation}
\noindent where $\mu_{tZ}$ is the fraction of the mass of the base element, with age $t$ and metallicity $Z$. Similarly, the light-weighted age is defined as
\begin{equation} 
\label{eq:at_light}
\langle \log\ \mathrm{age} \rangle_\mathrm{L} = \sum_{t,Z} x_{tZ} \times \log t,
\end{equation}
\noindent where $x_{tZ}$ is the fraction of light at the normalization wavelength ($5635$~\AA) corresponding to the base element with age $t$ and metallicity $Z$.

Finally, for the intrinsic and rest-frame $(u-r)$ colour, the intrinsic $(u-r)$ colour is calculated by convolving the resulting fitted synthetic spectrum at rest-frame with the \uja\ and \rb\ filter transmission functions.  The rest-frame $(u-r)$ is calculated in a similar manner but also including the reddening effects in the synthetic SED (i.e.~$e^{-q_{\lambda} \tau_V}$).
 
\subsection{Specific aspects of each SED-fitting code}

\subsubsection{\baysea}\label{SFR}
 
\baysea\ is based on the method developed by \citet{lopezfernandez18} to fit the GALEX and CALIFA data for a sample of nearby galaxies. This methodology has been modified to fit \js \ (de Amorim et al. 2020, in prep.), instead of the GALEX colours, as well as the stellar features measured in the CALIFA spectra. The code assumes an SFH $=$ SFH$(t; \Theta)$, where $t$ is the lookback time and $\Theta$ is a parameter vector that includes the stellar metallicity ($Z_\star$), a dust attenuation parameter ($\tau_V$), and parameters ($k$,\,$t_0$,\,$\tau$) that control the temporal behaviour of the SFR, $\psi(t)$.

The synthetic spectrum for a given $\psi$ and $\Theta$ is computed from
\begin{equation}
\label{eq:ModelSpectrum}
L_\lambda(\Theta) =  e^{-q_{\lambda} \tau_V} \int \mathrm{SSP}_\lambda(t,Z) \, \psi(t; \Theta) \,  dt,
\end{equation}
where SSP$_\lambda(t,Z)$ is the spectrum at age $t$ of an SSP of metallicity $Z$ and initial mass 1\,$\mathrm{M}_\odot$, and
$\Theta = (k, t_0, \tau, Z_\star, \tau_V)$.
In this study we used the exponential-$\tau$ and the delayed-$\tau$ SFR laws, which are frequently used in the literature.
For the exponential-$\tau$ SFR, $\psi(t) = \psi_{0}e^{-(t_{0} - t)/\tau}$, and for the 
delayed-$\tau$ SFR, $\psi(t) = k (t_0-t)/\tau e^{-(t_{0} - t)/\tau}$. In both cases, $t$ is the lookback time,
$\tau$ is the SFR e-folding time, and $t_0$ is the lookback time for the onset of star formation.
The $\psi_0$ and $k$ are normalization constants related to the mass formed in stars.

The code explores parameter space and constrains the parameters $\Theta$ that fit our data using an MCMC method. The advantage of this Bayesian analysis is that we can marginalize over the parameters. The three main characteristics of the code are as follows.\ First, given a set of 56 NB magnitudes, the MCMC algorithm draws a set of values of $\Theta$ in parameter space according to some probabilistic rules. These values are a representation of the PDF over the entire parameter space. Second, the code determines the corresponding total stellar mass formed by maximizing the likelihood of the scale-dependent observables (the photometric magnitudes, in our case). Third, using the SFH library, the code derives the PDF for each of the stellar population properties (mass, stellar age, dust attenuation, stellar metallicity, and colours). The PDFs represent the complete solution to the inference problem. 
For each galaxy property, we used the median and sigma of the corresponding PDF as its inferred value and error. 

\subsubsection{\muff }\label{sec:muffit}

The code \muff\ \citep[MUlti-Filter FITting for stellar population diagnostics, extensively detailed in][]{diaz-garcia2015} is a generic SED-fitting tool specifically designed for the detailed analysis of the SEDs of stars and galaxies. Namely, \muff\ is optimized to deal with multi-band photometric data and allows one to determine the stellar population properties of the miniJPAS galaxies: age, metallicity, stellar mass, rest-frame luminosities, extinction, \photoz, etc. To constrain these fundamental parameters, the code performs an error-weighted $\chi^2$-test assuming two-burst composite models of stellar populations.
\muff\ has proven to be a reliable and powerful tool with capabilities for tracing the evolution of the stellar content of galaxies in agreement with spectroscopic studies at intermediate redshifts \citep[see e.g.][]{san-roman2018,diaz-garcia2019c}, as well as with mock galaxy samples and SDSS data based on spectroscopic indices \citep[][]{diaz-garcia2015}. 

\subsubsection{\alstar }\label{sec:alstar}

Instead of postulating an a priori $\psi(t; \Theta)$ function, one can describe galaxy evolution in terms of an arbitrary positive combination of $N$ stellar populations of different ages and metallicities. This non-parametric approach is the one implemented in the \starlight\ code \citep{cid-fernandes2005},  which has been employed in numerous spectroscopic studies based on SDSS, CALIFA, MaNGA, and other datasets. Its advantage with respect to parametric codes is that the more flexible description of the SFH in general leads to better spectral fits, particularly in objects that undergo multiple star formation episodes (parametric codes have a hard time dealing with multi-modal SFHs). The disadvantage is that the huge parameter space ($N \sim 100$) makes it hard to implement conventional Bayesian sampling schemes to evaluate the probability distribution function of the parameters. 
In this work we used an algebraic version of \starlight, henceforth dubbed  \alstar. The code performs the same spectral decomposition in terms of a base of $N$ model spectra, but solving the NNLS problem instead of the 
MCMC-like sampling scheme implemented in \starlight. The non-linear effects of dust attenuation are dealt with separately, solving the problem for different values of $\tau_V$ and finding the best solution. The specific base used in this paper comprises populations of 16 ages and seven metallicities, as in \citet{werle2019}, except that here each age corresponds to a `square burst' (a period of constant SFR between adjacent age bins). 

\subsubsection{\tgas }\label{sec:tgas}

\tgas\ is a non-parametric spectral fitting code that can be used to estimate the SFH of a stellar population. The possibility of determining the physical properties of galaxies from the J-PAS NB filters was studied by \cite{mejia2017} for a variety of SFHs using the \tgas\ and \dynbass\ spectral fitting codes described in detail by \cite{magris2015}. These authors compared results obtained from SDSS galaxy spectra and J-PAS synthetic photometry, concluding that the J-PAS filter system yields the same trend in the age--metallicity relation as SDSS spectroscopy for typical S/N values. They also quantified the biases, correlations, and degeneracies that affect the retrieved parameters using mock galaxy samples.
Similar to \alstar , \tgas\ derives the SFH by using the NNLS algorithm.
The best-fit values of the physical properties of the galaxy are determined directly from the best-fit solution, computed for several independent values of $\tau_V$.
For the implementation of \tgas\ used in this paper, we used the nine metallicities available in the \cb\ models described above.
A maximum of $100$ time steps ranging in age from $0.5$~Myr to $t_\mathrm{max}$ were considered for each metallicity.
The best solution rarely comprised more than ten spectra of different ages and metallicities.


\begin{figure*}
\centering
\includegraphics[width=\textwidth]{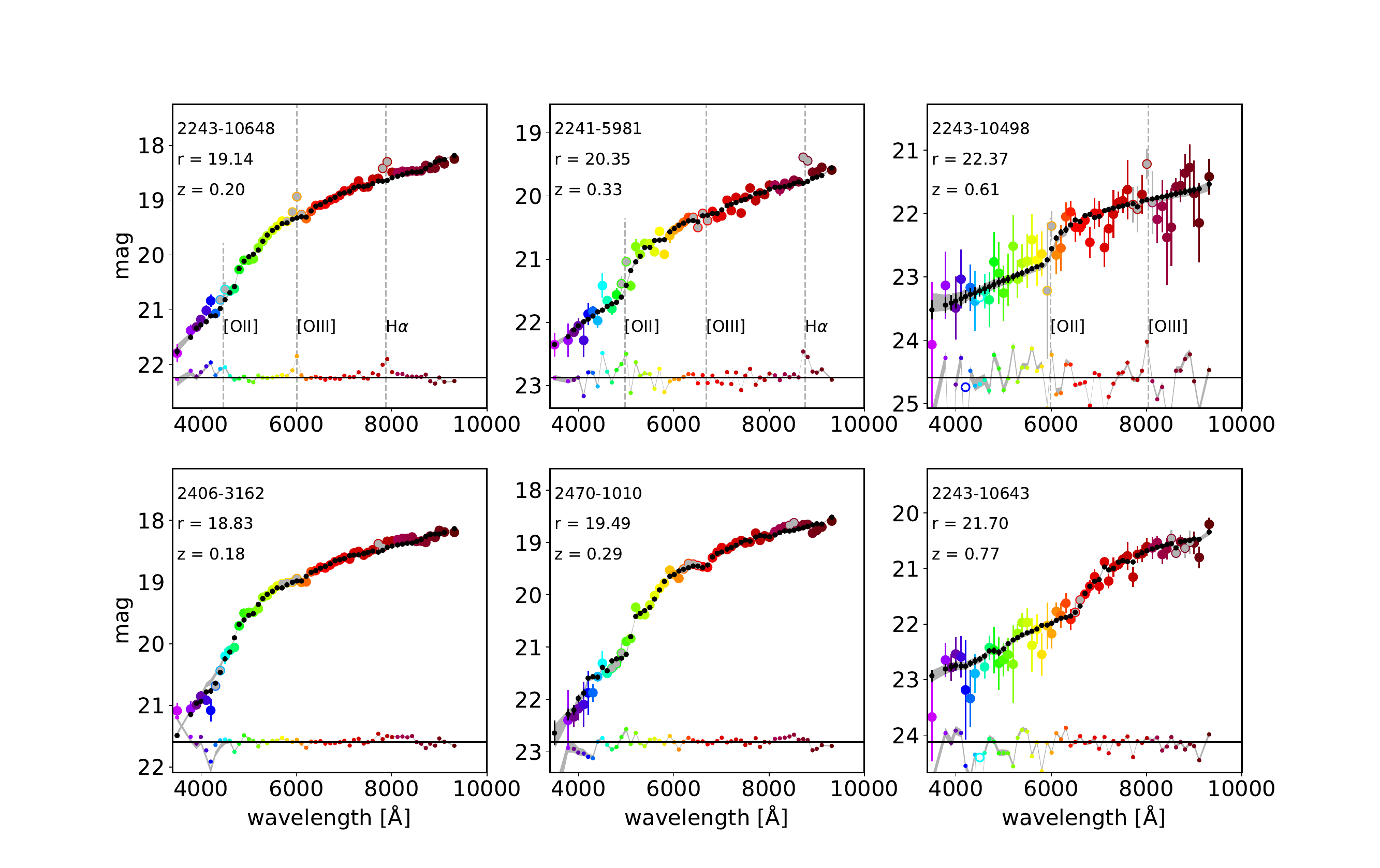}
\caption{
\magpsfcor\ \js\ of different galaxies with redshift  $z = 0.18$--$0.77$ and brightness $19.1<$\rb$<22.4$ (\magauto \ in \rb ). Masked filters (white coloured circles) and filters overlapping with H$\alpha$, [NII], [OIII], H$\beta$, or [OII] lines (grey coloured circles) are not used in the fit. The best model fitted to the continuum with \baysea \ is plotted as black points, and the grey band shows the magnitudes of the mean model at the $\pm$1$\sigma$ uncertainty level. The difference between the observed and the best model fitted magnitudes are plotted as small coloured points around the black horizontal line, which represents a null difference between the observed and the best-fitted model. The grey bands around these dots show the difference  $\pm$1$\sigma$ variation. 
}
\label{fig:J-fits}
\end{figure*}

\section{The stellar population properties of the sample}\label{sec:Results}

This section presents the general picture of the distributions of the stellar population properties obtained with our four SED-fitting codes for the \mjp\ photometry. We explore the differences and the impact on the results introduced by the methodologies embedded in \baysea, \muff, \alstar, and \tgas.

\subsection{Fits and quality assessment }

As mentioned above, SED-fitting solutions reproduce the 56 \mjp\ NB magnitudes of galaxies of different types (e.g. ELGs or LRGs)  quite well within the uncertainties and independently of the redshift and brightness range (see Fig.~\ref{fig:J-fits}). Indeed, spectral features of ELGs, such as the H$\alpha$, [OIII] $\lambda$5007, and [OII] $\lambda$3727 emission lines, are easily detected (see the top panels of Fig.~\ref{fig:J-fits}), whereas other features in different bands, such as the noticeable $4000$~\AA-break in LRGs, are also well reproduced (see bottom panels of Fig.~\ref{fig:J-fits}). The residuals obtained from the fits, defined as the difference between the observed and the best-fitting magnitudes, are higher for fainter galaxies (see the right-hand panels of Fig.~\ref{fig:J-fits}) and typically increase towards the bluer bands (see the middle panels of Fig.~\ref{fig:J-fits}). This behaviour is related to the higher uncertainties in the data, which increase at fainter magnitudes. It is also worth noting that the uncertainty in the red filters ($\lambda \geq 8000$~\AA) can be significantly high owing to fringing effects, which have a larger effect on fainter galaxies (e.g. $2243$--$10498$).

The \js\ are well reproduced by the four codes.
The quality of the fits for the whole sample can be assessed through different estimators or parameters, such as the reduced $\chi^2$ value ($\chi^2_\mathrm{reduced}$) and the normalized residual. 
A discussion of these parameters as derived by \baysea\ is provided in Appendix A.
The quality assessment of the fits for the four codes is very similar to Fig.~\ref{fig:merit}, yielding equivalent conclusions about the fitting errors.


\subsection{Results from \baysea \ }

In this section, we present the results of the stellar population properties obtained by \baysea \  using  the \magpsfcor \ and \magauto\ photometry.

\subsubsection{Distributions of stellar population properties}

\begin{figure*}
\centering
\includegraphics[width=\textwidth]{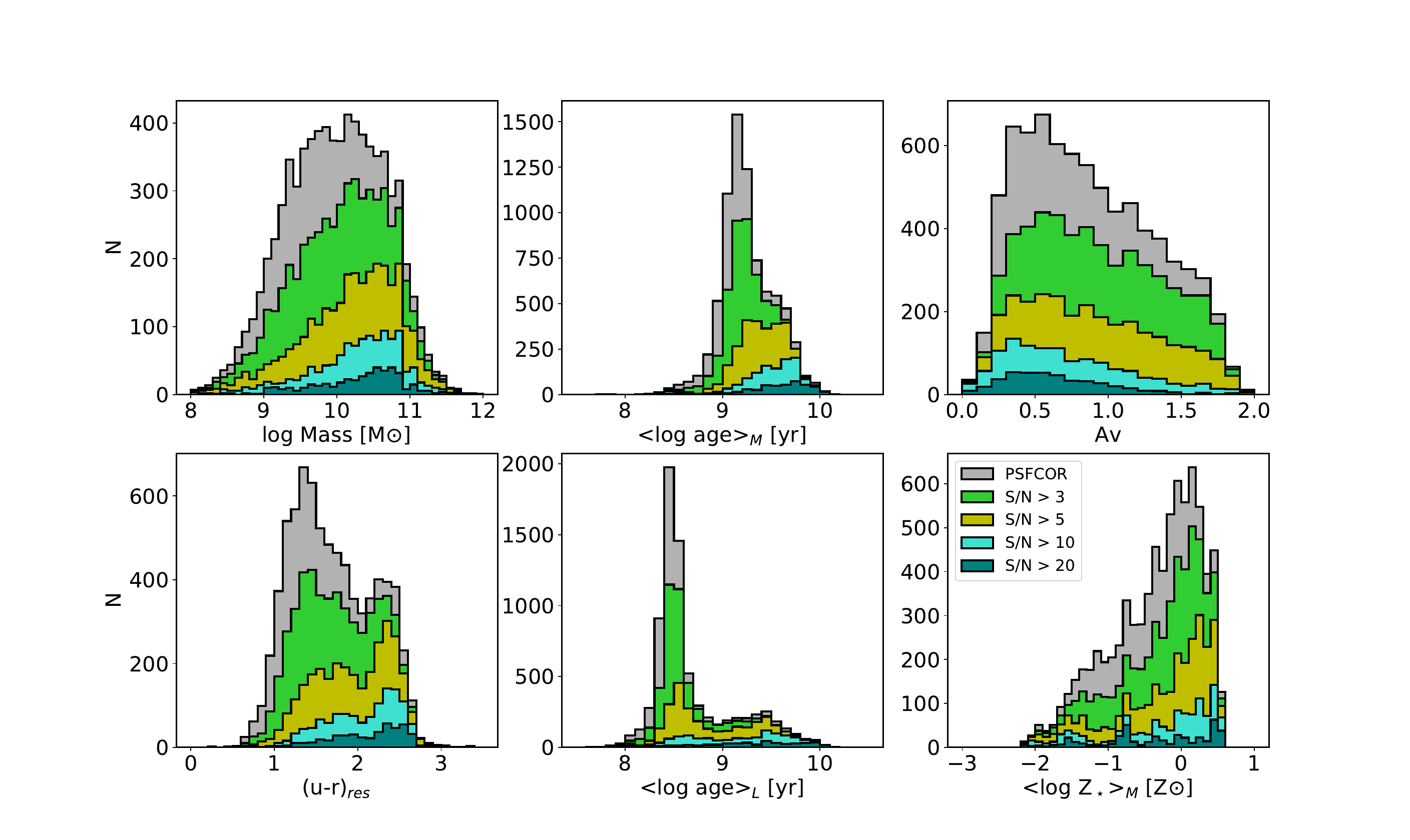}
\caption{Stellar population properties from the full spectral fitting of the \js \ with \magpsfcor \ obtained with \baysea . \textit{From left to right, and from top to bottom}: Stellar mass, mean mass-weighted age, intrinsic extinction, rest frame $(u-r)$ colour, mean light-weighted age (at $5630$~\AA), and stellar metallicity. The different coloured histograms correspond to different values of the median S/N in the NB filters.
}
\label{fig:distributionSP}
\end{figure*}

The grey histograms in Fig.~\ref{fig:distributionSP} show the distributions of mass, age, extinction, rest-frame colour, and metallicity (\logM , \ageM\,and \ageL, $A_V$, $(u-r)_\mathrm{res}$, and $\langle \log\ Z \rangle_\mathrm{M} $, respectively)
obtained with \baysea\ for the galaxies in the AEGIS field.
The distribution of stellar mass ranges from $8$ to $\sim12$~dex, with a plateau from $9.4$ to $\sim 10.6$. Extinction is distributed between $0$ and $2$~mag, peaking at $\sim0.5$~mag. The\ $\langle \log\ Z \rangle_\mathrm{M} $ covers most of the metallicity range of the SSPs, but the distribution peaks at around solar and super-solar metallicity. The colour $(u-r)_\mathrm{res}$ shows the well-known bimodal distribution of galaxies, where the maximum density is at $(u-r)_\mathrm{res}\sim1.5$ for the blue cloud galaxies and at $(u-r)_\mathrm{res} \sim2.5$ for the red sequence. 
A bimodal distribution is also clearly detected in \ageL\  with peaks at \ageL\,[yr] $\sim8.5$ and $\sim9.5$, while the whole distribution ranges from \ageL\,[yr] $8$ to $10$. The \ageM\ is shifted towards older ages with respect to \ageL. This is expected because small amounts of mass in recent star formation episodes translate into a high contribution to the luminosity of the galaxy, decreasing the light-weighted age. The distribution in \ageM\,[yr] peaks at $\sim9.1$, with an extended wing to older ages reaching significant numbers up to \ageM\,[yr] $\sim9.7$.

The coloured histograms in Fig.~\ref{fig:distributionSP} show these distributions binned according to S/N.
The $(u-r)_\mathrm{res}$ distribution clearly shows that the fraction of BGs detected with high S/N  is lower than the fraction of red galaxies. Similarly, the fraction of massive galaxies with high S/N\index{N}  is higher than for the less massive galaxies (which mainly populate the blue cloud), and the fraction of old galaxies with high S/N  is also higher than that of the younger galaxies. In summary, the red, old, and massive galaxies are detected at higher S/N  than the blue, young, and less massive ones.

\subsubsection{Stellar population properties: \magauto \ versus \magpsfcor}

\begin{figure*}
\centering
\includegraphics[width=\textwidth]{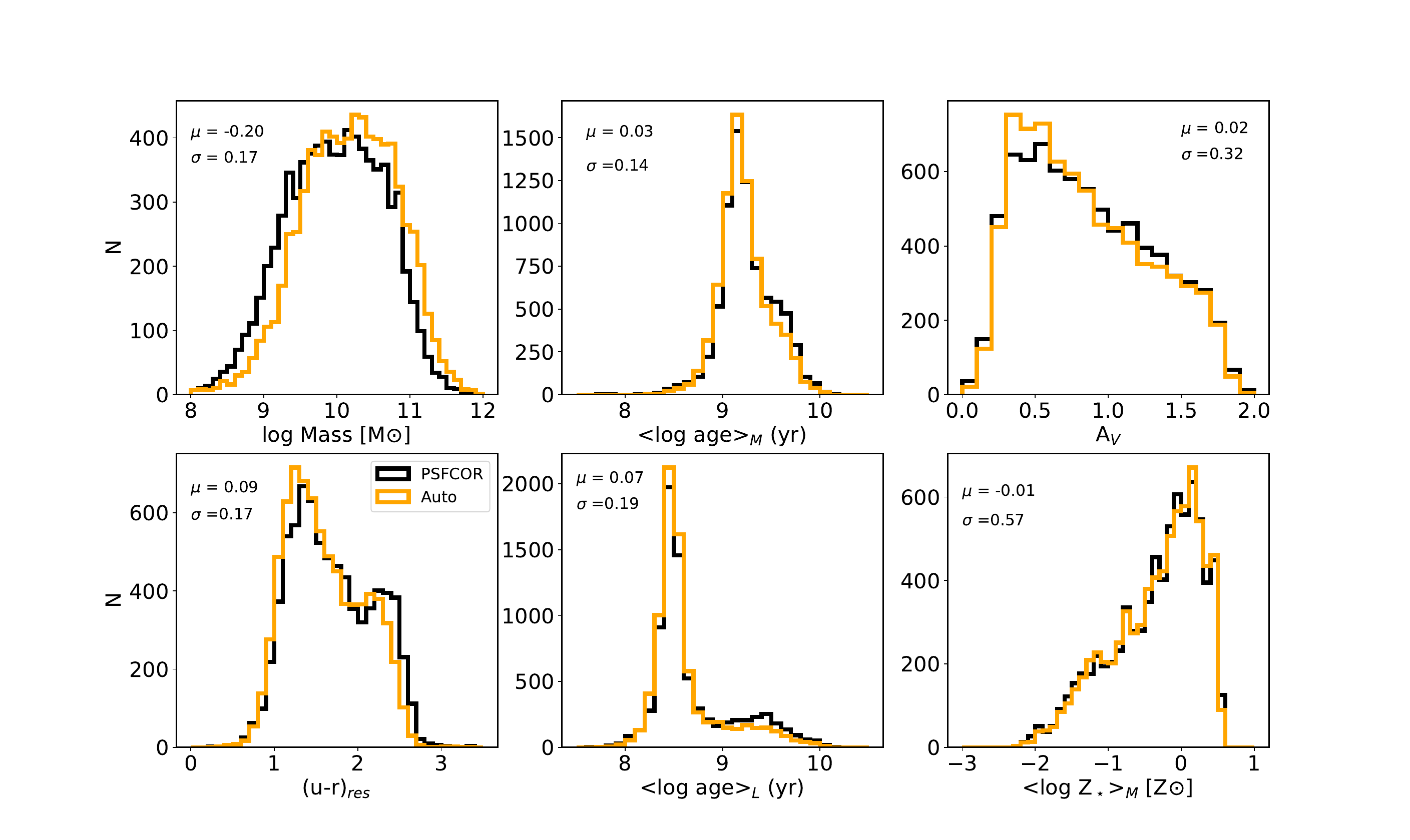}
\caption{Comparison of the distributions of the stellar population properties derived with \baysea \  by spectral fitting using \magpsfcor \ (black) and \magauto\ (orange) magnitudes for stellar mass (present and initial), rest-frame intrinsic $(u-r)$ colour, mass- and light-weighted ages, stellar extinction, and metallicity. In each panel, $\mu$ and $\sigma$  indicate the mean and standard deviation of the difference between the value of each galaxy property derived with \magpsfcor \ and \magauto .
}
\label{fig:PsfcorAuto}
\end{figure*}

We fitted the \js\ using both  \magpsfcor\  and \magauto\ magnitudes. In this way we were able to assess the impact of the photometric choice on the stellar population properties of the \mjp\ galaxies. As mentioned in Sect. \ref{sec:sample_auto_psf}, the flux is integrated through a larger aperture in \magauto \ than in \magpsfcor, the former being closer to the total flux. Therefore, the fitting to the \js\ performed using \magauto \ should provide a more representative estimate of the integrated stellar population properties than those obtained using \magpsfcor. 

From Fig.~\ref{fig:PsfcorAuto} we see that the main differences appear in the \logM \ and $(u-r)_\mathrm{res}$ distributions.  \magauto\ results are shifted to larger masses by $0.2$~dex ($\sigma=0.17$), and to bluer colours by $-0.09$ ($\sigma = 0.17$). These shifts are compatible with the larger aperture used by \magauto\ with respect to \magpsfcor. 
The small shift to younger ages by \ageL\,$\sim-0.07$ ($\sigma=0.07$) can be interpreted similarly. The \ageM \ is not affected by the larger aperture because this property is biased towards the older stellar populations, which are concentrated in the inner regions of the galaxies. The $A_V$ and \logZM\  are not affected by the aperture, although the small decrease in \logZM\ is possibly due to the negative gradient of  $\log Z_\star$ in galaxies \citep{gonzalez-delgado2015}.

\subsection{Results from \muff, \alstar, and \tgas}
\label{sec:codesdistributions}

\begin{figure*}
\centering
\includegraphics[width=\textwidth]{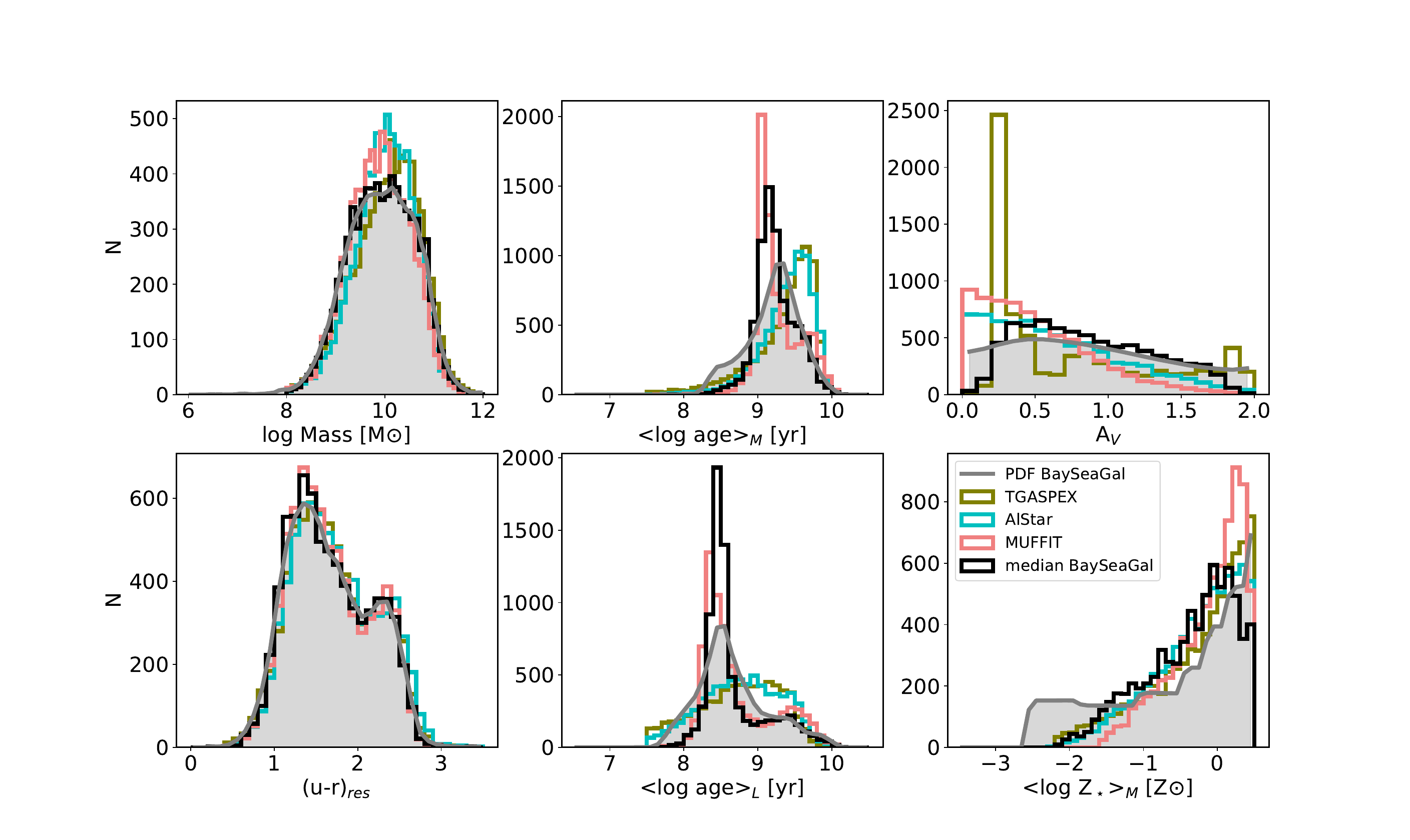}
\caption{Distributions of mean values of stellar population properties obtained by our SED-fitting codes \baysea, \muff, \alstar, and \tgas\ for the \mjp\ galaxies (black, red, blue, and green lines, respectively). The grey curves illustrate the posterior probability distribution functions (PDF) obtained by \baysea\ assuming a delayed-$\tau$ SFH. All the results were obtained using the \psfcor\ photometry.}
\label{fig:hist_SP_methods}
\end{figure*}

Next, the \js\ (in \magpsfcor) were fitted using the non-parametric codes \muff, \alstar, and \tgas, which are\ described in Sects.~\ref{sec:muffit} to \ref{sec:tgas}. The two main differences with respect to \baysea\ are: (i) These three codes use combinations of SSPs instead of a parametric $\psi(t)$. (ii) \baysea\ builds the full probability distribution function (PDF) of the properties via an MCMC process, whereas the other codes use a Monte Carlo approach. These codes provide a characterization of the stellar populations in the \mjp\ galaxies, in agreement with the \baysea\ results. The figures of merit for the three codes are comparable to Fig.~\ref{fig:merit}. 
The $\chi^2$ distributions indicate fits similar in quality to the \baysea\ ones. 

In Fig.~\ref{fig:hist_SP_methods} we compare the distributions of galaxy properties obtained with the four codes.
The distributions of $(u-r)_\mathrm{res}$ are remarkably similar, indicating that the four codes reach similar quality fits.
The distributions of \logM\ are quite compatible, although \alstar\ and \tgas\ show a mild excess of galaxies with stellar mass
10 $\leq$ \logM $<$ 11 [M$_\odot$] with respect to \muff \ and \baysea. On average, the \alstar\ and \tgas\ stellar masses are $0.06$ and $0.11$~dex larger, respectively, than the \baysea\ values.
\muff \ stellar masses are $0.05$~dex lower than those of \baysea.
The distributions of \logZM \ are very similar for the four codes, although \muff\ predicts a larger fraction of galaxies with super-solar metallicity. It is worth mentioning that the \baysea\ PDF is more extended at the low-metallicity end than those of the other codes. On average, \muff , \alstar , and \tgas \ galaxies are $0.24$, $0.1$, and $0.1$~dex more metal-rich than those of \baysea, respectively. However, these differences are small considering the coarse distribution of metallicity in the input models.
The distributions of $A_V$ from \baysea, \muff, and \alstar\ are similar, although the \baysea\ distribution shows a flatter slope and a cutoff at $A_V \leq 0.3$. The distribution of $A_V$ from \tgas\  is significantly different from those of the other three codes. 
On average, $A_V$ results from \muff, \alstar, and \tgas\  are larger by 0.35, 0.2, and 0.1, respectively, than those from \baysea.
We find that there are more discrepancies in the distributions of age than for the other properties. 
On the one hand, the distributions of \ageL\ and \ageM\ obtained by \baysea\ and \muff\ are almost identical, showing a double-peaked distribution. On the other hand, the \alstar\ and \tgas\ \ageL\ distributions show one maximum, shifted towards older ages with respect to the \baysea\ and \muff\ results. 
The \ageM\ results from \alstar\ and \tgas\ are typically $0.15$~dex and $0.1$~dex older than those from the other two codes. The differences in the mean value and dispersion of the galaxy properties measured with the non-parametric codes and \baysea\ are listed in Table~\ref{tab:SPcodes}. 

\begin{table*}
    \centering
    \caption{Mean and standard deviation of the difference between the value of each galaxy property derived with \muff , \alstar , and \tgas \  with respect to \baysea \ and the delayed-$\tau$ model.}
    \begin{tabular}{l r r r}
    \hline
    \hline
        SP & \muff \  &  \alstar \  &  \tgas \  \\
     \hline
        $(u-r)_\mathrm{res}$ & $0.004 \pm 0.11$  & $0.07 \pm 0.16 $ & $0.04 \pm 0.18$ \\
        \logM \       & $-0.06 \pm 0.14$ & $0.06  \pm 0.27$ & $0.11 \pm 0.28$ \\
        $A_V$         & $-0.35 \pm 0.33$  & $-0.20  \pm 0.50$ & $-0.10 \pm 0.65$ \\
        \logZM \      &  $0.24 \pm 0.52$  & $0.10   \pm 0.50$ & $0.12  \pm 0.74$ \\
        \ageM \       &  $0.02 \pm 0.21$  & $0.15  \pm 0.30$ & $0.10   \pm 0.40$ \\
        \ageL \       & $-0.04 \pm 0.26$  & $0.04  \pm 0.45$ & $-0.04 \pm 0.52$ \\
        \hline
    \end{tabular}
    
    \label{tab:SPcodes}
\end{table*}

\subsection{Mass--colour diagram}

\begin{figure*}
\centering
\includegraphics[width=\textwidth]{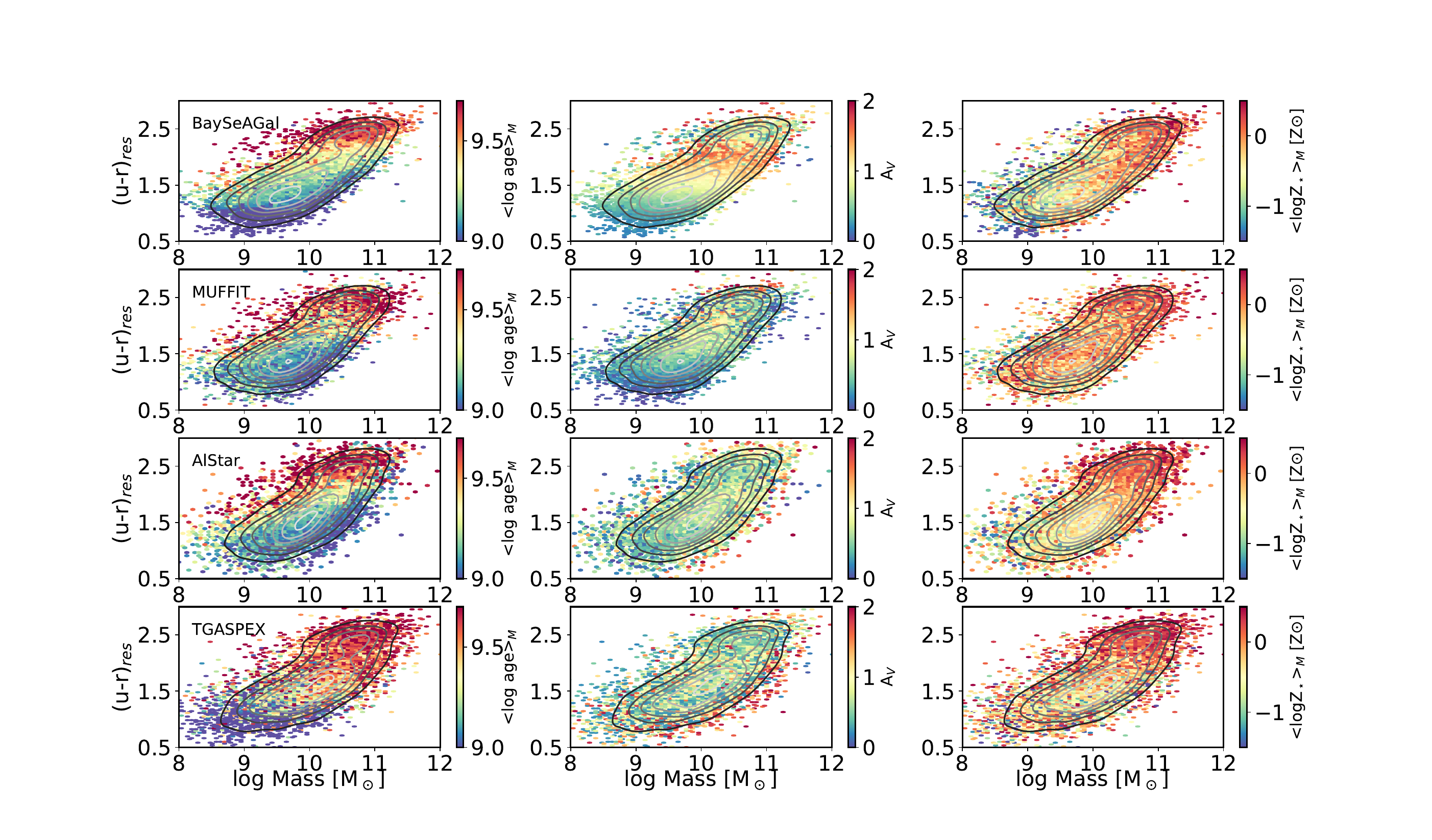}
\caption{Mass--colour relation, with the coloured bar  showing the stellar population properties of the galaxies (age, extinction, and metallicity). Properties were derived using the \psfcor\ photometry. \textit{From top to bottom}: Results from \baysea, \muff, \alstar, and \tgas.}
\label{fig:MassColor}
\end{figure*}

The $(u-r)_\mathrm{res}$ colour of \mjp\ galaxies shows a bimodal distribution that indicates the presence of two galaxy populations, which we refer to as red and blue galaxies. One of the diagrams frequently used to study the differences between these two populations is the galaxy stellar mass--colour diagram \citep[see e.g.][and references therein]{moresco2013,schawinski2014,diaz-garcia2019a}.
If we focus on the distribution of galaxies in the \logM--$(u-r)_\mathrm{res}$  diagram (Fig.~\ref{fig:MassColor}), the blue cloud and red sequence of galaxies are clearly identified. Galaxies that belong to the red sequence are typically old and metal-rich. Regarding age, we find that \ageM\,[yr] is typically above $9.6$ in the red sequence, while blue cloud galaxies are younger, with \ageM\,[yr] down to $9$. Most of the galaxies in the red sequence have solar and super-solar \logZM\ metallicity, while for the blue cloud it is below solar values. Overall, the extinction is similar in the blue cloud and in the red sequence, $A_V < 1$, while the larger $A_V$ values are located at different positions in the diagram, depending on the code. For example, \baysea\ galaxies with $A_V > 1.5$ are located at the intermediate part of the diagram, which is usually referred to as the green valley. This finding is in agreement with previous results that suggest that green valley galaxies are indeed star-forming galaxies from the blue cloud; they are strongly reddened by dust, producing colours much redder than the intrinsic ones \citep{diaz-garcia2019a}.
In contrast, the reddest galaxies from \tgas \ are located in the right ridge of the histogram, where the high redshift galaxies of the sample are located; these galaxies have larger uncertainties in the observed magnitudes. These differences in the properties of the galaxies between the four fitting codes illustrate the degeneracy between the outputs and the dependence of the results on the uncertainties in the observed magnitudes.


\section{Uncertainties in the stellar population properties}
\label{sec:Uncertainties}

\subsection{Uncertainties of the observed magnitudes and redshifts}

\begin{figure*}
\centering
\includegraphics[width=\textwidth]{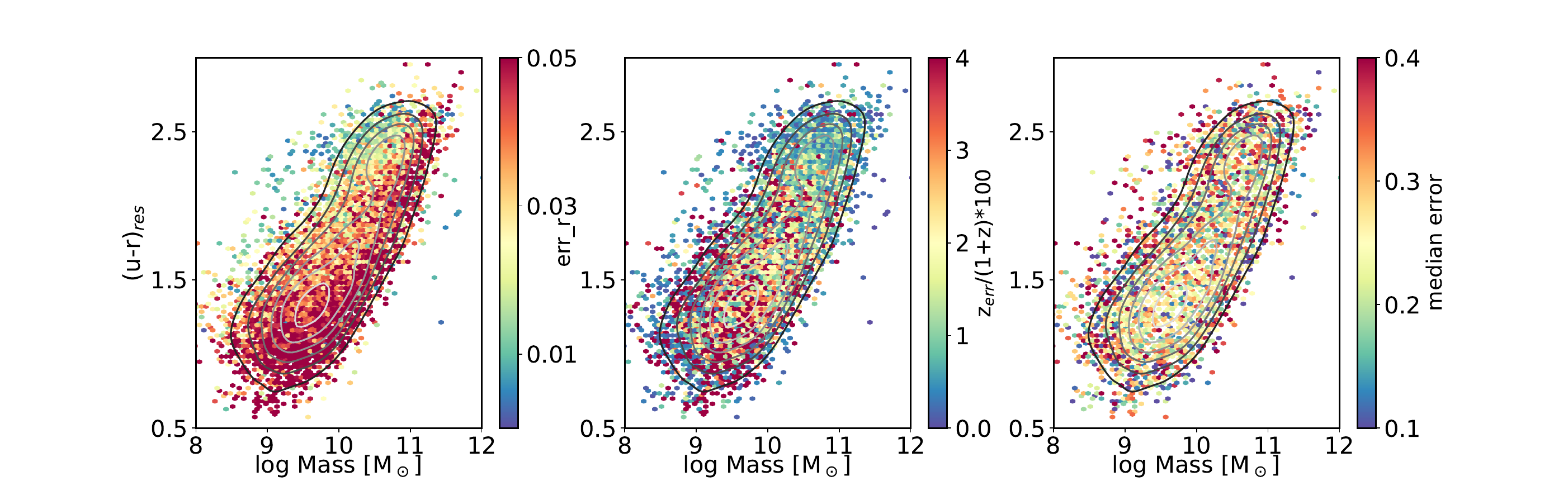}
\caption{Mass--colour relation derived by \baysea \ and using the \psfcor \ photometry, with the coloured bar showing the error in the magnitude in the \rb\  band, the error in redshift, and the median S/N in the NB filters.}
\label{fig:MassColor_uncer}
\end{figure*}

One of our aims is to explore how the uncertainties in the observed magnitudes and redshift affect the locus of galaxies in the $(u-r)_\mathrm{res}$ diagram, as well as their spectral classification as red or blue galaxies (see Fig.~\ref{fig:MassColor_uncer}). Faint galaxies are mainly located in the right ridge of the  \logM\ --  $(u-r)_\mathrm{res}$ diagram; this  reflects the position of the galaxies with higher \photoz, which are also fainter. The precision in the estimation of the \photoz\ also affects the fits and the values of \logM\ and $(u-r)_\mathrm{res}$. If we measure the precision in \photoz\ by the relative error, 
$\sigma_z /(1 + z)$\,$\times$\,$100$ (where $z$\,=\,\photoz ), then galaxies in the red sequence exhibit lower \photoz\ uncertainty than galaxies in the blue cloud. The $4000$~\AA\ break, more prominent in LRGs than in BGs, controls
this behaviour. This stellar continuum feature helps us obtain more precise values of \photoz, as well as more precise properties of the galaxy, since it is correlated with the age of the stellar population \citep{gonzalez-delgado2005}. Nonetheless, the identification of red and blue galaxies in the  \logM\ --  $(u-r)_\mathrm{res}$  diagram does not show a strong limitation, according to the quality of the \js. Galaxies with similar S/N, as traced by the median error in the NB magnitudes, are equally distributed across the red sequence and blue cloud, but most of the galaxies with low S/N fall near the border of the area
filled by the bulk of the galaxies in the \logM\ --  $(u-r)_\mathrm{res}$ diagram in Fig.~\ref{fig:MassColor_uncer}.

\begin{figure*}
\centering
\includegraphics[width=\textwidth]{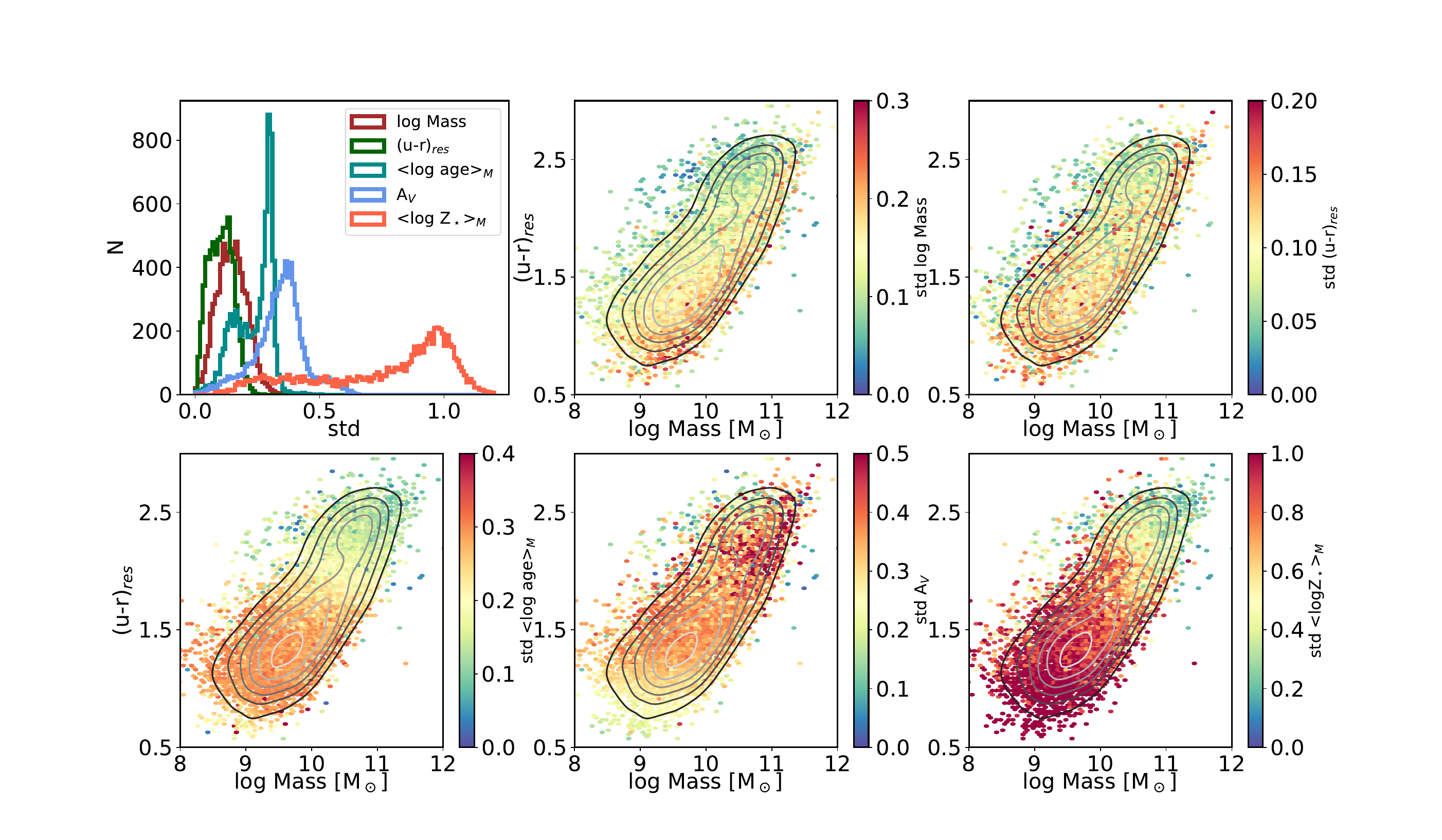}
\caption{Distribution of standard deviations of galaxy properties (\textit{top-left panel}) and (from left to right and top to bottom) the colour--mass plane coloured by the standard deviation of the stellar population properties: \logM, $(u-r)_\mathrm{res}$, \ageM, $A_V$, and \logZM, respectively.}
\label{fig:MassColorSPerrors}
\end{figure*}

\subsection{Precision of the stellar population properties}

With \baysea\ we can easily evaluate the uncertainties on the galaxy SFH and the inferred stellar population properties thanks to the MCMC process and the use of a parametric SFR. For a given $\psi(t)$ we obtain the PDF and from it the mean and the standard deviation ($\sigma$) for each galaxy property ($t_0$,\,$\tau$,\,$A_V$,\,$\log Z$,\,\logM).
Overall, $(u-r)_\mathrm{res}$ and \logM \ are the properties with the lowest uncertainties (see the upper-left panel of Fig.~\ref{fig:MassColorSPerrors}) with $\langle \sigma ((u-r)_\mathrm{res}) \rangle = 0.1 \pm 0.05$ mag and $\langle \sigma (\logMt ) \rangle = 0.15 \pm 0.06$ dex. The property with the largest uncertainty is \logZM , with $\langle \sigma (\logZMt ) \rangle = 0.76 \pm 0.28$ dex. This is expected from the coarse sampling of metallicity in the SSP models. The stellar extinction $A_V$ is significantly better estimated than \logZM , with $\langle \sigma (A_V) \rangle = 0.34 \pm 0.1$ mag (see the upper-left panel of Fig.~\ref{fig:MassColorSPerrors}). The \ageM \ is estimated better than $A_V$, but globally worse than $(u-r)_\mathrm{res}$.
The $\sigma(\ageMt )$ presents a bimodal distribution with peaks at $\sim0.3$ and $\sim0.15$. These peaks are related to the bimodal distribution of the galaxy population. The ages for the BGs are less constrained than for the red galaxies, with $\langle \sigma (\ageMt ) \rangle  = 0.25 \pm 0.06$ dex and $0.14\pm0.05$ dex, respectively, for $(u-r)_\mathrm{res}$ smaller or larger than $1.8$.

The precision depends on the S/N. The mean S/N of our sample is\,$\sim$\,$8$. In particular, if we consider a sub-sample of BGs with S/N\,$\geq$\,$10$, which is the mean S/N of spectroscopic samples in future surveys such as WEAVE/STePs \citep{costantin2019}, the precision of the inferred galaxy properties improves considerably. For this sub-sample of \mjp\ blue cloud galaxies, the stellar metallicity is estimated with $\langle \sigma (\logZMt ) \rangle = 0.42 \pm 0.25$ dex and $\langle \sigma (\ageMt ) \rangle = 0.16 \pm 0.07$ dex. Higher  precision is obtained for S/N\,$\geq$\,$20$\,--\,$30$, approaching IFU surveys of nearby galaxies, such as CALIFA \citep[See Table ~\ref{tab:precision};][]{cid-fernandes2014, gonzalez-delgado2015}. 

Exploring the distribution of the uncertainty across the \logM -- $(u-r)_\mathrm{res}$ diagram (see also Fig.~\ref{fig:MassColorSPerrors}), we find that these properties are not estimated with equal precision for the different galaxy populations; as such, the precision is not evenly distributed across the \logM -- $(u-r)_\mathrm{res}$ diagram. 
The bimodality of the distribution is clearly seen in $\sigma$ (\ageM) (bottom-left panel of Fig.~\ref{fig:MassColorSPerrors}). 
For red sequence galaxies, \ageM\ is estimated better ($\sigma (\ageMt) \leq 0.15$ dex) than for galaxies in the blue cloud ($\sigma (\ageMt) \sim 0.3$ dex). Uncertainties in \logM\ (top-central panel in Fig.~\ref{fig:MassColorSPerrors}) are also smaller in the red sequence than in the blue cloud, although the larger uncertainties in \logM\ correspond to galaxies in the right ridge of the $\logMt$ -- $(u-r)_\mathrm{res}$ diagram, as discussed above for Fig.~\ref{fig:MassColor}. In addition, $(u-r)_\mathrm{res}$ is constrained slightly better in the red sequence than in the blue cloud (top-right panel of Fig.~\ref{fig:MassColor}). The uncertainties in $A_V$ (bottom-central panel of Fig.~\ref{fig:MassColor}) are similar throughout the blue cloud (though somewhat lower for the less massive galaxies), while the higher $\sigma (A_V)$ correspond to the more massive galaxies in the red sequence. The precision in \logZM\ (bottom-right panel of Fig.~\ref{fig:MassColor}) shows a clear distribution in the $\logMt$ -- $(u-r)_\mathrm{res}$ diagram; it is less well constrained for galaxies in the blue cloud, with $\sigma$(\logZM) up to $1$ dex. However, for galaxies in the red sequence, $\sigma$(\logZM) can be as low as $0.2$ dex. Thus, the solar or super-solar metallicity of red sequence galaxies is well determined.

\begin{table*}[]
    \centering
    \caption{ Precision (mean standard deviation) of the stellar population properties of the \mjp \ sub-sample with S/N$\geq$30, 20, 10, 5, and 3. The units of $\langle \sigma ((u-r)_\mathrm{res}) \rangle$ and  $\langle \sigma (A_V) \rangle$ are in magnitudes, and $\langle \sigma (\logMt) \rangle$, $\langle \sigma (\ageMt) \rangle$ ($\langle \sigma (\ageLt)), \rangle$, and $\langle \sigma (\logZMt) \rangle$ are in dex of [M$_\odot$], [yr], and [Z$_\odot$].
    The number of galaxies included in each sub-sample is indicated.  }
    \begin{tabular}{l  r r r r r}
    \hline
    \hline
        SP &  S/N$\geq$30  &  S/N$\geq$20  &  S/N$\geq$10  & S/N$\geq$5 & S/N$\geq$3 \\
     \hline  
      $\langle \sigma ((u-r)_\mathrm{res}) \rangle$  &  $0.03 \pm 0.02$ & $0.03 \pm 0.02$  & $0.04 \pm  0.02$ & $0.06 \pm 0.03$ & $0.09 \pm 0.04$ \\
      $\langle \sigma (\logMt) \rangle$              &  $0.04 \pm 0.03$ & $0.05 \pm 0.03$  & $0.07 \pm  0.03$ & $0.09 \pm 0.03$ & $0.12 \pm 0.05$ \\
      $\langle \sigma (A_V) \rangle$                 &  $0.10 \pm 0.05$ & $0.12 \pm 0.06$  & $0.20 \pm  0.09$ & $0.29 \pm 0.12$ & $0.33 \pm 0.11$ \\
      $\langle \sigma (\ageMt) \rangle$              &  $0.10 \pm 0.07$ & $0.12 \pm 0.07$  & $0.16 \pm  0.07$ & $0.19 \pm 0.08$ & $0.22 \pm 0.08$ \\
      $\langle \sigma (\ageLt) \rangle$              &  $0.12 \pm 0.07$ & $0.14 \pm 0.07$  & $0.19 \pm  0.08$ & $0.23 \pm 0.08$ & $0.26 \pm 0.07$ \\
      $\langle \sigma (\logZMt) \rangle$             &  $0.30 \pm 0.20$ & $0.34 \pm 0.22$  & $0.42 \pm  0.25$ & $0.56 \pm 0.28$ & $0.69 \pm 0.28$ \\
      \hline
      Number of galaxies  & $229$ & $444$ & $1211$ & $2954$ & $5463$ \\
      \hline
    \end{tabular}
    \label{tab:precision}
\end{table*}

\subsection{Uncertainties associated with the star formation law}
\label{sec:laws}

\begin{figure*}
\centering
\includegraphics[width=\textwidth]{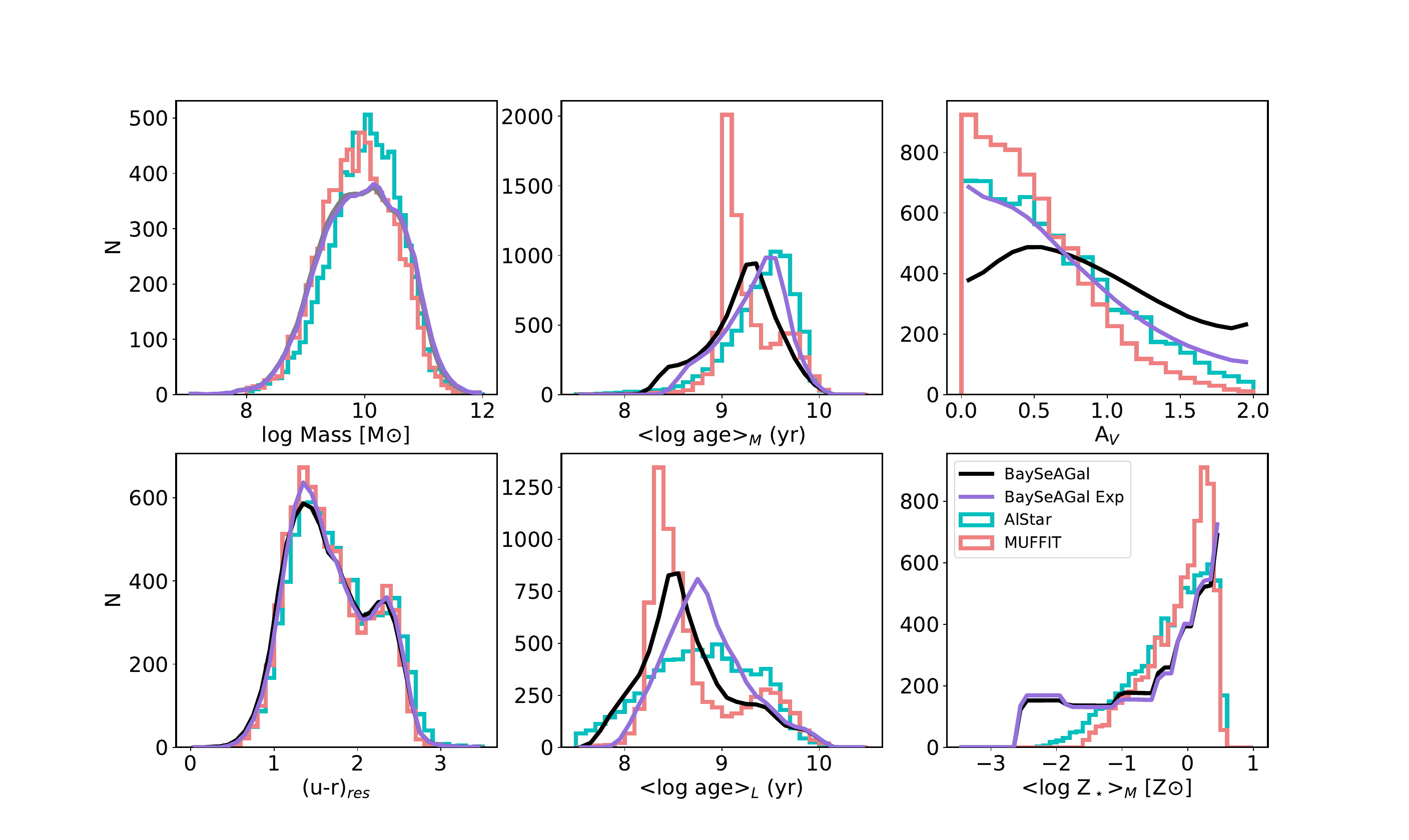}
\caption{Distributions of stellar population properties obtained by \baysea\ for the \mjp\ galaxies using the \psfcor\ photometry and assuming different star formation laws: delayed (black) and  exponential (violet) star formation laws. Results obtained with \muff \ and \alstar \ are also shown (see inset).
}
\label{fig:hist_SP_laws}
\end{figure*}

Here we explore the influence of $\psi(t)$ on the inferred stellar population properties using \baysea\ and the two SFR laws defined in Sect.\,\ref{SFR}. 
The results in Fig.\,\ref{fig:hist_SP_laws} indicate that the PDF for \logM, $(u-r)_\mathrm{res}$ and \logZM\ are very similar for the two cases. The exponential-$\tau$ SFR produces values of $A_V$ that are lower, on average, by $0.09 \pm 0.65$ magnitudes than the delayed-$\tau$ SFR. The peaks of the \ageM\ and \ageL\ PDFs for the exponential-$\tau$ SFR are shifted to older ages with respect to the delayed-$\tau$ SFR by \ageM\,=\,$0.11 \pm 0.09$ dex and \ageL\,=\,$0.17 \pm 0.12$ dex.
This `aging' of the galaxies is related to the decrease in $A_V$ in the delayed-$\tau$ SFR model with respect to the exponential-$\tau$ SFR model. 
This degeneracy between age and extinction is well known in stellar population synthesis \citep[e.g.~][]{cid-fernandes2005}. 

Table~\ref{tab:Expcodes} lists the average differences between the properties from the four codes and the exponential-$\tau$ SFR model. Figure~\ref{fig:hist_SP_laws} compares the distributions of the properties from \baysea \ with the results from the exponential and delayed-$\tau$ models as well as the results for the non-parametric codes. We notice that the distributions of $A_V$ and \ageM\ from the exponential-$\tau$ SFR are closer to the results from \alstar\ and \tgas\ than the results from the delayed-$\tau$ SFR model. On average, the difference between galaxy properties inferred with \alstar\ and the exponential-$\tau$ SFR  model are $-0.02 \pm 0.5$ mag and $0.04 \pm 0.29$ dex for $A_V$ and \ageM, respectively. Thus, we conclude that the results from the non-parametric codes, in particular \alstar\ and \muff, are represented better by an exponential-$\tau$ than by a delayed-$\tau$ SFR. However, it is known that exponential SFR models produce an early and rapid growth of the galaxy stellar mass that is not compatible with the peak in the cosmic density of star formation at $z\sim2$ \citep{chiosi2017,lopezfernandez18}.  

\begin{table*}[]
    \centering
    \caption{Mean and standard deviation of the difference between the value of each galaxy property derived with \muff , \alstar , and \tgas \  with respect to the  \baysea \ exponential model. The last column shows the differences between the results with \baysea \ and the delayed-$\tau$ and exponential models.}
    \begin{tabular}{l  r r r r }
    \hline
    \hline
        SP & \muff \  &  \alstar \  &  \tgas \  & \baysea \\
     \hline  
        $(u-r)_\mathrm{res}$ &  $-0.006 \pm 0.10$ & $0.06 \pm  0.16$   &  $0.03 \pm 0.18$ & $-0.01 \pm 0.03$ \\
        \logM \       &  $-0.08  \pm 0.13$ & $0.035 \pm 0.26$   &  $0.09 \pm 0.27$ & $-0.02 \pm 0.05$ \\
        $A_V$         &  $-0.15  \pm 0.33$ & $-0.02 \pm 0.50$   &  $0.09 \pm 0.65$ & $0.09 \pm 0.65$  \\
        \logZM \      &  $0.25   \pm 0.56$ & $0.11  \pm 0.59$   &  $0.14 \pm 0.79$ & $0.015 \pm 0.22$  \\
        \ageM \       &  $-0.09  \pm 0.22$ & $0.04  \pm 0.29$   & $-0.015 \pm 0.40$ & $-0.11 \pm 0.09$ \\
        \ageL \       &  $-0.13 \pm 0.27$  & $-0.12 \pm 0.45$   & $-0.21 \pm 0.52$ & $-0.17 \pm 0.12$  \\
        \hline
    \end{tabular}
    
    \label{tab:Expcodes}
\end{table*}


\section{Discussion}
\label{sec:Discussion}

The \photoz \ precision makes \jp\ an excellent survey for studying the properties of galaxies across cosmic time. The \mjp\ survey comprises only $1$~deg$^2$ of the sky, and the number of galaxies per redshift bin is somehow limited. Nevertheless, it allows us to explore how galaxy stellar properties evolve with cosmic time. Furthermore, it allows us to identify and characterize blue and red galaxies, as well as their cosmic evolution.

In this work we did not carry out a comprehensive and complete study of the evolution and formation of galaxies with redshift. Rather, we roughly explored the global properties of the galaxies available in the \mjp\ survey, and we compared the results from different SED-fitting codes. For a proper interpretation of the results, we should take into account that flux-limited samples are affected by the so-called Malmquist bias, meaning that galaxies below a certain luminosity limit are not observed at a given redshift. In this regard, less massive galaxies are typically affected by this bias and it is convenient to define a stellar mass completeness limit for a more detailed and comprehensive study. The definition of these limits -- as well as other aspects and properties, such as co-moving number densities as a function of mass and redshift or stellar mass and luminosity functions -- are beyond the scope of this work, and they will be tackled in future works (e.g.~D\'iaz-Garc\'ia et al.~2021, in prep.). For galaxy evolution studies extended in redshift, as presented in the present work, a consequence of the Malmquist bias is that less massive galaxies are only imaged at lower redshifts, where red or quiescent galaxies are even more strongly affected by this effect owing to a higher mass--luminosity relation. For instance, quiescent galaxies of $\log M_\star \sim 9.5$~dex in \mjp\ are only observed at $z<0.35$, whereas star-forming galaxies at this mass range can be easily imaged at $z>0.6$ (see Fig.~\ref{fig:MassColorEvolution}). 

\subsection{Evolution of galaxy populations in the mass--age diagram} 

\begin{figure*}
\centering
\includegraphics[width=\textwidth]{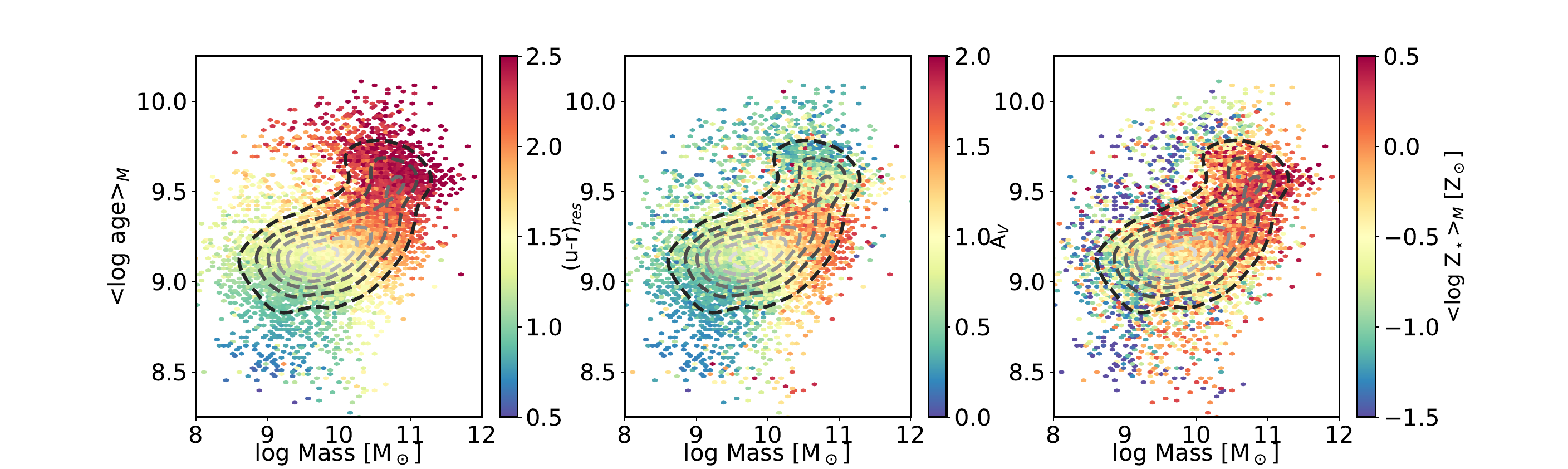}
\includegraphics[width=\textwidth]{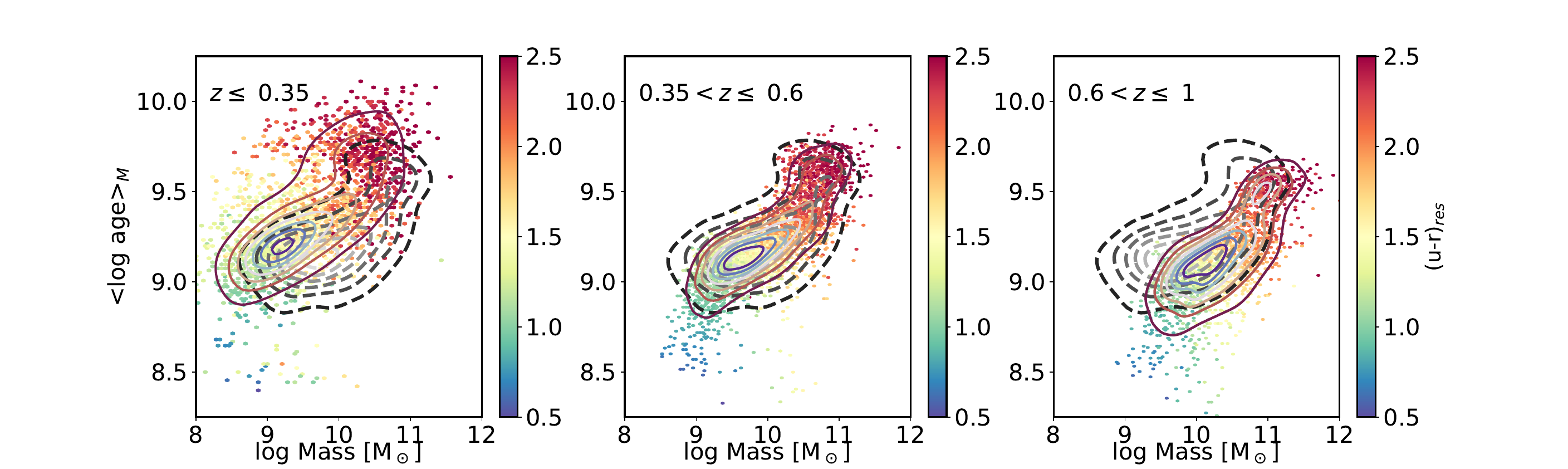}
\caption{Distributions of $(u-r)_\mathrm{res}$, $A_V$, and \logZM\  across the mass--age relation. The values of all the parameters are coded according to the colour bars. The \textit{bottom panels} illustrate the points and contour distribution of $(u-r)_\mathrm{res}$ in the mass--age relation at $z \leq 0.35$, $0.35 < z \leq0.6$, and $0.6 < z \leq$1. The dashed contours show the distribution of galaxies with $z \leq$1.}
\label{fig:MassAge}
\end{figure*}

The bimodal distribution of galaxies is also reflected in the stellar mass--age diagram (\logM \ -- \ageM ). Figure~\ref{fig:MassAge} (which only shows the results from \baysea \ with the delayed-$\tau$ model) depicts the bimodal distribution of the galaxy populations in AEGIS. Old galaxies are typically redder and more metal-rich than young galaxies. Galaxies in the green valley have larger $A_V$ than both red and blue galaxies. 
The \logM\ -- \ageM\ diagram also shows a bimodal correlation between the mass and the stellar population age. Late-type galaxies, which mainly populate the blue cloud in the nearby Universe, show a linear correlation with the stellar mass, which is steeper than for early-type galaxies or red sequence galaxies \citep{kauffmann2003a, gonzalez-delgado2014}. 

These results are revealed more clearly when only galaxies of similar cosmic epochs are plotted (lower panels of Fig.~\ref{fig:MassAge}). For this reason, we split our sample into three different redshift bins, where the width of each subset was defined to include a sufficient number of galaxies: $z\leq 0.35$, $0.35< z \leq 0.6$, and $0.6< z \leq 1$. At low redshift, the mass--age relation shows a remarkable change in slope for galaxies more massive than $\sim 10.5$~dex \citep{kauffmann2003a, gonzalez-delgado2014}, for which the relation \logM\ -- \ageM\ is almost flat. This mass limit marks the green valley, the transition from the blue cloud to the red sequence. However, red galaxies in the AEGIS \logM\ -- \ageM\ diagram show a decrease in \ageM\ with decreasing \logM, demonstrating that less massive galaxies were assembled during more recent cosmic epochs. This behaviour is reproduced at any redshift $z\sim1$.

\subsection{Evolution of galaxy populations in the mass--colour diagram, corrected for extinction}

\begin{figure*}
\centering
\includegraphics[width=\textwidth]{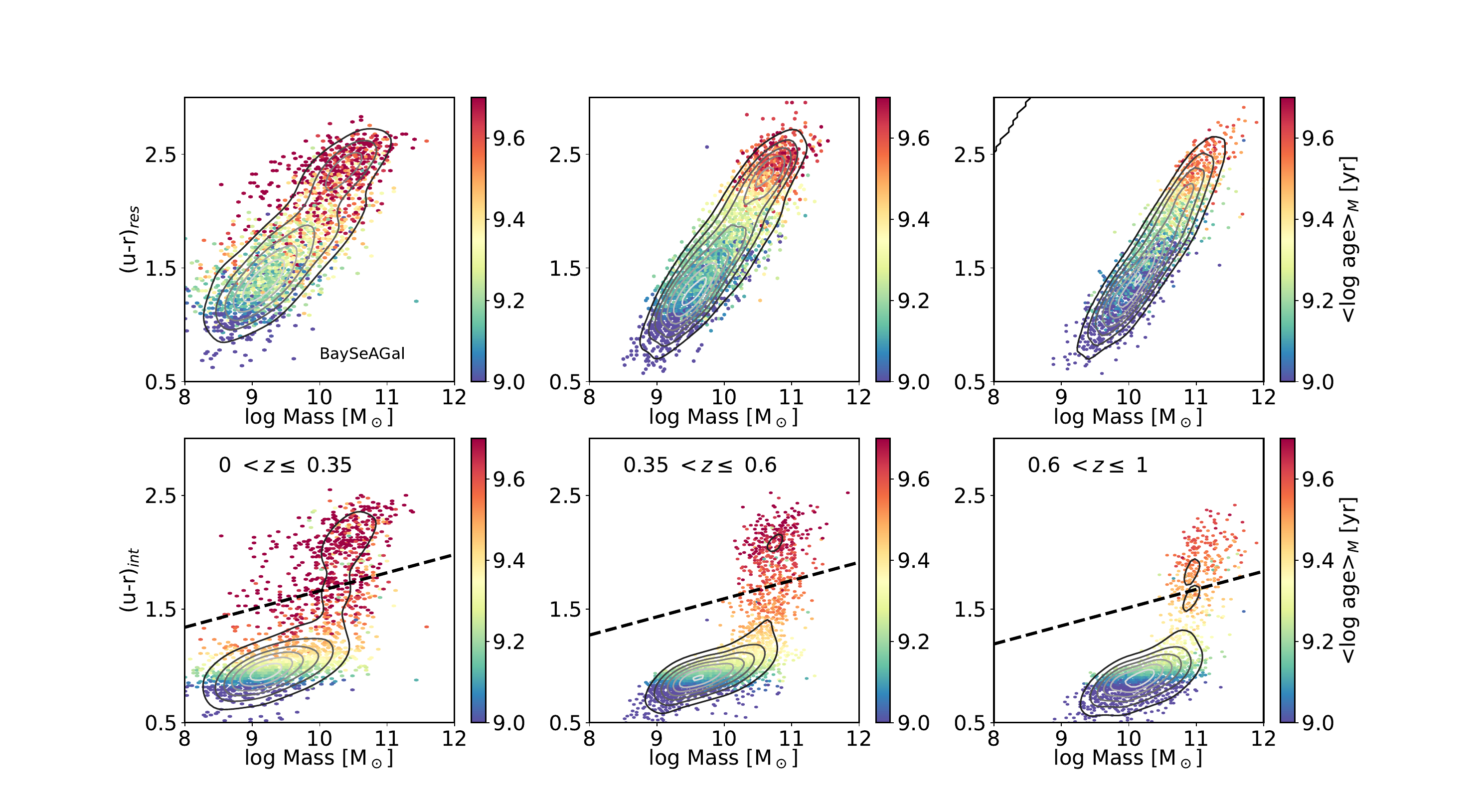}
\caption{Rest-frame (\textit{top panels}) and intrinsic (\textit{bottom panels}) colour $(u-r)$ vs. stellar mass for the redshift bins $z \leq 0.35$, $0.35 < z \leq 0.6$, and $0.6 < z \leq 1$ (\textit{from left to right}) for the results determined by  \baysea \ and the delayed-$\tau$ model. The dashed line shows the $(u-r)_\mathrm{int}^\mathrm{lim}$ for the mean redshift in each bin (details in the text).
}
\label{fig:MassColorEvolution}
\end{figure*}

Here we explore how the bimodal distribution of galaxy populations changes with cosmic time in the mass--colour diagram. We discuss the differences in the distribution of galaxy populations when the dust-corrected colour (referred to as intrinsic colour)  $(u-r)_\mathrm{int}$ is considered instead of the $(u-r)_\mathrm{res}$. Certainly, dust extinction reddens galaxies, and to correct colours for extinction the demographics of galaxies in the blue cloud and red sequence are relevant \citep[e.g.][]{hernan-caballero2013, schawinski2014, diaz-garcia2019a}. 

The mass--$(u-r)_\mathrm{int}$ diagram shows that the distribution of galaxies across the mass-colour diagram is remarkably different after colours are corrected for extinction effects. This fact is due to the presence of a non-negligible fraction of dusty star-forming galaxies populating the so-called green valley. 
Many of the galaxies move to bluer colours after $(u-r)_\mathrm{res}$ is corrected for extinction, and the fraction of galaxies that de-populate the red sequence is larger, increasing towards higher redshifts. 
A significant fraction of the galaxies in the green valley \citep[$30$--$65$\%;][]{diaz-garcia2019a} are actually obscured star-forming galaxies, whose fraction depends on redshift and stellar mass. In the nearby Universe ($z \leq 0.1$), and as traced by the SDSS colours, the fraction of galaxies that de-populate the red sequence is not negligible \citep{schawinski2014}, although it is smaller than at higher redshifts.

Figure~\ref{fig:MassColorEvolution} shows the results from \baysea \ and the delayed-$\tau$ models. A similar diagram is derived with \muff. However, some differences are found with respect to the results from \alstar \ and \tgas, which impact the estimation of the red and blue galaxies of the sample. In the appendix, we discuss the similarities and discrepancies of the distributions of \logM  and $(u-r)_\mathrm{int}$ in the results from the four codes. In particular, we focus on the distributions of \logM and $(u-r)_\mathrm{int}$ at the redshift bins of $z\leq0.35$, $0.35< z \leq 0.6$, and $0.6 < z \leq 1$ (Fig.~\ref{fig:hist_SP_methods_evolution}).

\subsection{Identification of blue and red galaxies}

\label{sec:Identification}

\begin{figure*}
\centering
\includegraphics[width=\textwidth]{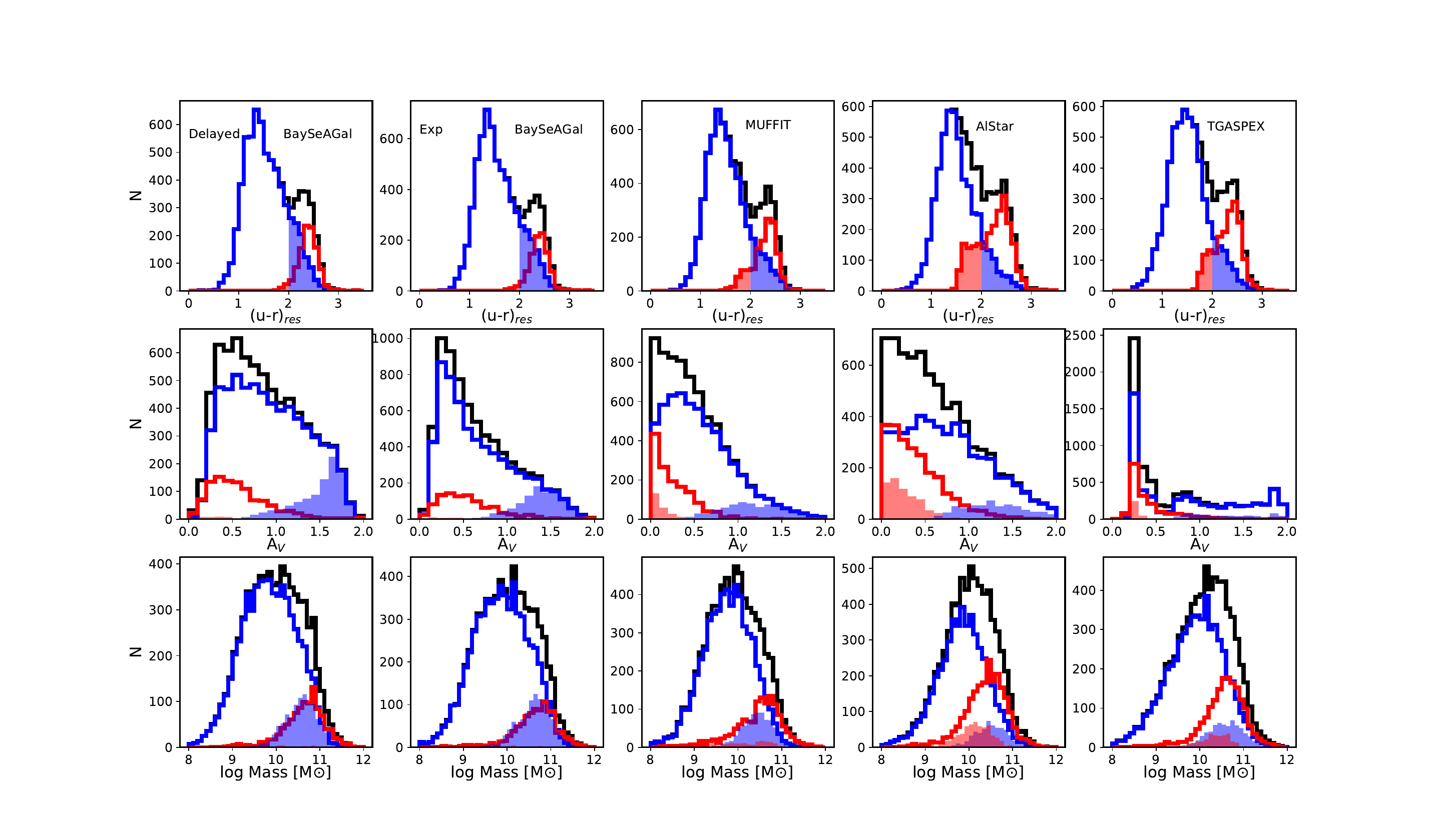}
\caption{Distributions of $(u-r)_{res}$ (upper panels), extinction (middle panels), and  \logM\ (lower panels) of
red (red lines) and blue (blue lines) galaxies of the AEGIS sample identified by \baysea \ with (from left to right) a delayed-$\tau$ model, an exponential model, \muff, \alstar, and \tgas . Filled-blue histograms are the distributions of BGs with  $(u-r)_{res} > 2$, and filled-red histograms are the distributions of red galaxies with $1.5 < (u-r)_{res} < 2$. 
}
\label{fig:histBlueRed}
\end{figure*}

\citet{diaz-garcia2019a} developed a method to discern between red and blue galaxies using a sample of galaxies from the ALHAMBRA survey at redshifts $0.1 < z < 1.1$. The authors also used \muff\ to perform an SED-fitting analysis of the 20 intermediate-band filters covering the optical range and the $J$, $H$, and $K_\mathrm{s}$ NIR bands. They obtained the stellar mass, rest-frame colour, and extinction of each galaxy from ALHAMBRA and subsequently performed a classification to identify quiescent and star-forming galaxies. They determined the fraction of dusty star-forming galaxies in the green valley through the intrinsic $(m_{F365} - m_{F551})$ and  $(m_{F551} - J)$ colours, as well as the contamination in quiescent galaxy samples defined via classical colour--colour diagrams owing to these obscured star-forming galaxies. As a result, \citet{diaz-garcia2019a} concluded that the $\logMt - (u-r)_\mathrm{int}$ diagram can reduce the contamination of the red sample by a fraction of $20$\% with respect to previous colour--colour diagrams, also without any bias at the low stellar mass regime. 

We followed a similar method to that in \citet{diaz-garcia2019a} and  classified galaxies as red or blue (quiescent or star-forming) according to their intrinsic colour, stellar mass, and redshift. The limiting intrinsic colour originally computed by \citet[][see Eq.~3]{diaz-garcia2019a} was adapted to match the \mjp\ photometric system. As a result, \mjp\ galaxies are labelled as quiescent when they exhibit intrinsic colours redder than the limiting value of $(u - r)_\mathrm{int}^\mathrm{lim}$, which is formally expressed as
\begin{equation}
\label{eq:redbluegalaxies}
(u - r)_\mathrm{int}^\mathrm{lim} =  0.16 \times (\logMt  \ - 10.) - 0.3 \times (z - 0.1) + 1.7
\end{equation}

\noindent \citep[Eq.~3 of][]{diaz-garcia2019a}, where $z$ is the \photoz\ of the galaxy and $\logMt$ is its stellar mass. Otherwise, the galaxy is labelled as star-forming. After a visual inspection, the colour limit set by Eq.~(\ref{eq:redbluegalaxies}) clearly separates the blue cloud from the red sequence at any redshift (see the dashed line in Fig.~\ref{fig:MassColorEvolution}). 

\begin{figure}
\centering
\includegraphics[width=\columnwidth]{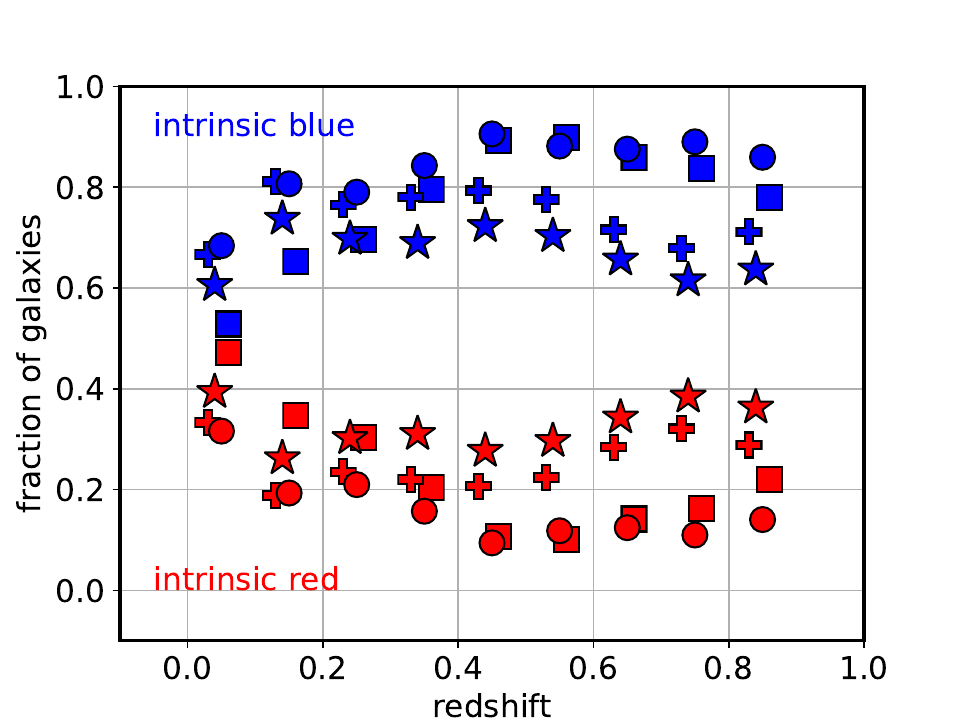}
\caption{Fraction of red and blue galaxies at different redshifts obtained by \baysea \ for a delayed-$\tau$ SFH (circles), \muff \ (squares), \alstar \ (stars), and \tgas \ (crosses).
}
\label{fig:fractionRedBlue}
\end{figure}

With this classification, we were able to estimate the fraction of red and blue galaxies in the AEGIS field that are brighter than \rb$=22.5$ and at $z \leq$1. We find that 85\% (\baysea), 81\% (\muff), 69\% (\alstar), and 76\% (\tgas) of the galaxies in the sample are in the blue cloud. In general, red galaxies have intrinsic red colours and lower extinction values than BGs (see Fig.~\ref{fig:histBlueRed}). There is, however, a fraction of galaxies that have intrinsic blue colours but are reddened by dust. For example, 13\% (\baysea), 11\% (\muff), 8\% (\alstar), and 9\% (\tgas) of galaxies have $(u-r)_\mathrm{res} > 2$ and $(u-r)_\mathrm{int}$ below the $(u - r)_\mathrm{int}^\mathrm{lim}$. The four codes identify these galaxies as dusty galaxies and as the most massive ones of the whole sample (Fig.~\ref{fig:fractionRedBlue}). We notice that the differences between codes in the fraction of galaxies in the blue cloud cannot be only due to differences in the estimation of BGs with red colours, $(u-r)_\mathrm{res} > 2$. While the maximum difference in the percentage of BGs between the results from the codes is $\sim 16$\% for \baysea \ and  \alstar, the difference in the percentage of BGs with $(u-r)_\mathrm{res} > 2$ is only $\sim 4$\% between these two codes. 
However, differences between codes in the estimations of $A_V$ of galaxies with intermediate colours, $1.5 < (u-r)_\mathrm{res} < 2$ (see the upper panels of  Fig.~\ref{fig:histBlueRed}), contribute to explain the differences in the fraction of BGs in the sample. Around 1\% (\baysea), 3\% (\muff), 9\% (\alstar), and 4\% (\tgas) of galaxies with $(u-r)_\mathrm{int}$ close to the $(u - r)_\mathrm{int}^\mathrm{lim}$  are identified as red galaxies. They have lower extinctions than most of the red but intrinsically  blue galaxies, and they are less massive than the reddest BGs (filled red and blue histograms in the middle panels of Fig.~\ref{fig:histBlueRed}). The mean extinction of this 9\% of \alstar \ red galaxies with $ 1.5 < (u-r)_\mathrm{res} < 2$  is $A_V = 0.26 \pm 0.18$, while for these galaxies the mean extinction is $0.38 \pm 0.29$ (\tgas), $0.48 \pm 0.29$ (\muff), and $1.07 \pm 0.35$ (\baysea). The four codes obtain a similar \logM \ ($\sim 10.1 \pm 0.5$) and stellar metallicity (\logZM \ $\sim -0.1 \pm 4$) for these galaxies. The 9\% of \alstar \ red galaxies have $z = 0.53 \pm 0.20$ and a median S/N of only $3.7$.  

We also explored the evolution of the fraction of blue and red galaxies in the sample (see Fig.~\ref{fig:fractionRedBlue}). Overall, the fraction of red galaxies decreases with increasing redshift (\magauto\ photometry), although the results from the four codes show significant differences in the highest redshift bins. For example, 
at $z\sim0.1$ and at $z=0.5,$ the percentage of red galaxies is $\sim38$\% and  $\sim 18$\%, respectively. At the highest redshift bin, the fraction of red galaxies  is not consistent between the four different codes and it ranges from values of $\sim10$--$15$\% with \baysea  \ and \muff \ to  $\sim30$\%  with \alstar. A percentage of red to blue galaxies of $20$\% has been reported in the analysis of ALHAMBRA data \citep{diaz-garcia2019a,diaz-garcia2019b}. 

A further study based on the detection of emission lines in these galaxies will be necessary to properly discriminate between red and blue galaxies with $1.5 < (u-r)_\mathrm{res} < 2$ and redshift above $0.5$. However, a S/N above $5$ will also help to classify them with better agreement between the different codes.

\subsection{Characterization of blue and red galaxies}

\begin{figure*}
\centering
\includegraphics[width=0.49\textwidth]{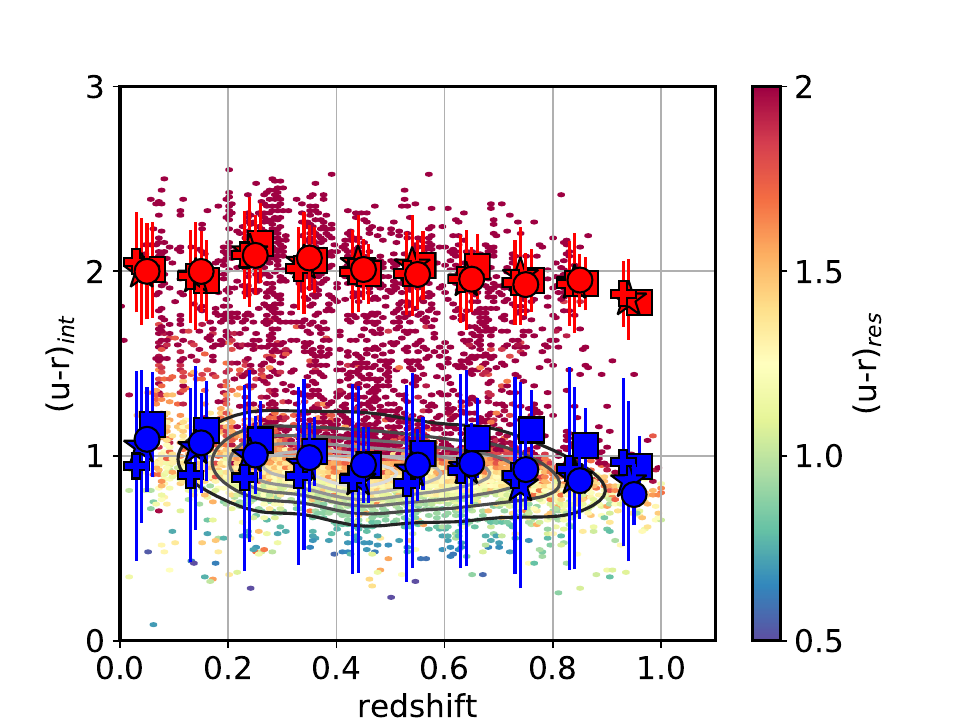}
\includegraphics[width=0.49\textwidth]{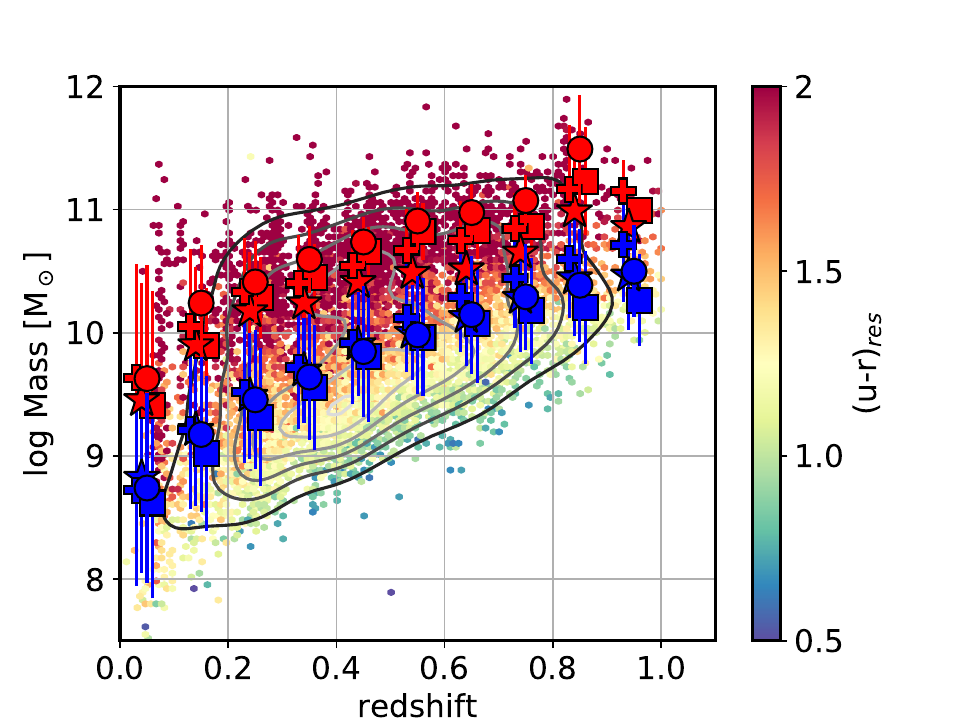}

\includegraphics[width=0.49\textwidth]{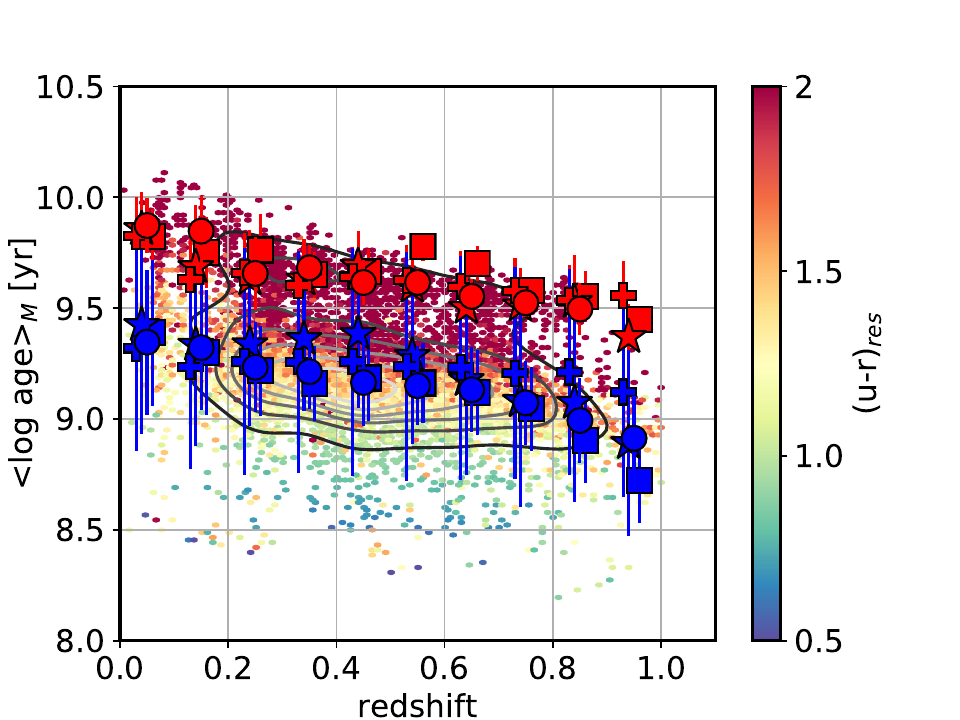}
\includegraphics[width=0.49\textwidth]{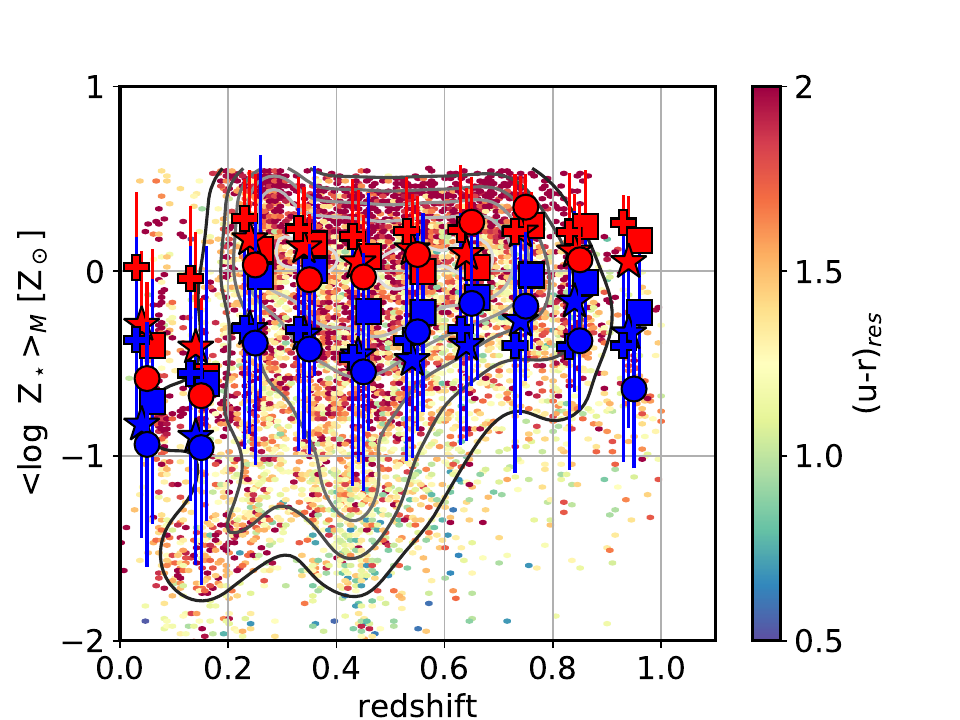}

\caption{Evolution of the intrinsic colour, stellar mass, age, and  metallicity of galaxies obtained by \baysea \ for a delayed-$\tau$ SFH. Dots represent the average values of each property in each redshift bin derived by \baysea \ (circles), \muff \ (squares), \alstar \ (stars), and \tgas \ (crosses). The dispersions with respect to the average values are shown as error bars. Blue and red dots correspond to blue cloud and red sequence galaxies, respectively.
}
\label{fig:SPEvolution}
\end{figure*}

The colour segregation of galaxy populations may be interpreted in terms of the differences of either their stellar content or their evolutionary pathways. Here we discuss the stellar population properties of red and blue galaxies as a function of redshift.

In particular, we explore the evolution of $(u-r)_\mathrm{int}$, \logM, \ageM, and \logZM\ for blue and red galaxies (see also Fig.~\ref{fig:SPEvolution}). We average the value of each galaxy property at any redshift bin. We notice that red and blue galaxies are properly distinguished according to their stellar content, whereas the properties of red galaxies are better constrained than for BGs. 

The colour $(u-r)_\mathrm{int}$ of blue and red galaxies is bluer at higher redshifts. For red galaxies, the two bins at lower redshifts show slightly bluer colours than galaxies at intermediate redshift. As a reference, a colour $(u-r)_\mathrm{int} = 2$ is equivalent to an SSP model of 2~Gyr of age and solar metallicity, while  $(u-r)_\mathrm{int} = 1$ is equivalent to an SSP of several $100$~Myr and half solar metallicity.

Blue galaxies are typically on average less massive than red galaxies by $\sim0.7$~dex, the difference being smaller at $z = 1$ than at $z = 0.1$. The masses of both blue and red galaxies are typically larger at higher redshifts. This is a  consequence of the incompleteness of the sample because faint and/or less massive galaxies are not imaged at high redshift. In addition, the larger number of massive galaxies at higher redshifts is just a consequence of the larger volume observed. 
Blue and red galaxies are very segregated in terms of their values of \ageM. At any redshift, red galaxies are older by $\sim0.5$~dex. This is probably a consequence of different SFHs and/or formation epochs in the blue cloud and in the red sequence, at least at $z = 1$. However, the \ageM\ of both blue and red galaxies decreases with increasing redshift, indicating ongoing star formation and/or reflecting a biased sample for the low-mass galaxies at higher redshifts. 

Blue and red galaxies also differ in \logZM. While red galaxies mostly show solar metallicity and above, BGs are always below solar values. However, the metallicity values obtained from the different SED-fitting codes are more uncertain, which makes the distinction of a separated \logZM\ relation between the two galaxy populations more difficult. Given the large uncertainty of the results, there is no evidence of a metallicity evolution with redshift. However, there is a  drop of $\sim0.5$~dex for metallicity at $z\leq0.2$ for both populations, which is likely produced by the drop in mass (see the top-left panel of Fig.~\ref{fig:SPEvolution}) due to detection bias. In general, the detection bias and the behaviours of mass and metallicity (top-right and bottom-right panels of Fig.~\ref{fig:SPEvolution}, respectively) may account for the lack of evidence in metallicity evolution.

\subsection{Volume incompleteness of the sample }

\begin{figure}
\centering
\includegraphics[width=0.49\textwidth]{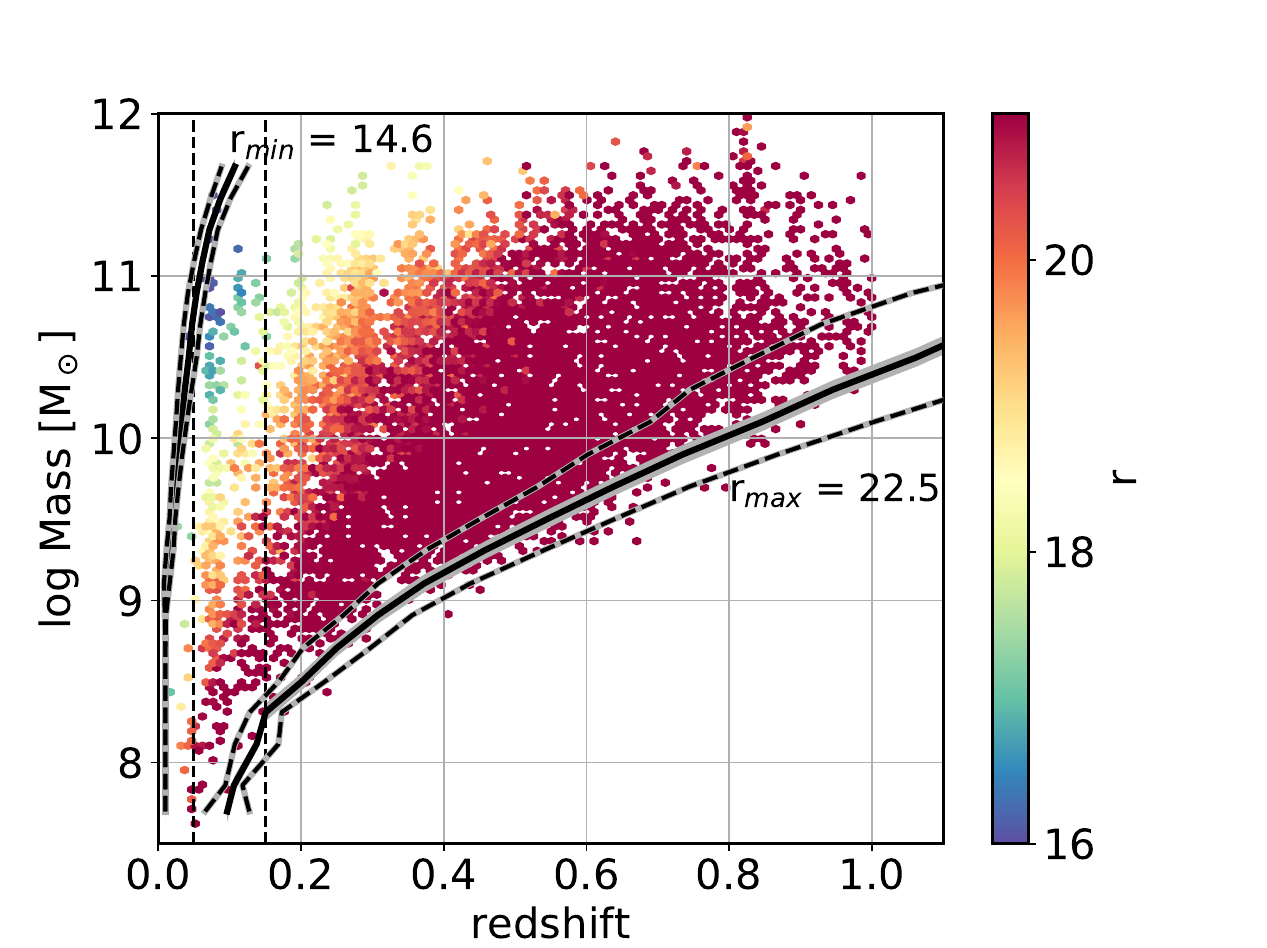}
\caption{Redshifts and stellar masses obtained by \baysea\ using the \magauto\ photometry for each of the galaxies in our sample. The black lines show the $z_\mathrm{max}$ and $z_\mathrm{min}$ values that correspond to the limiting magnitudes of the \mjp \ galaxy sample selection analysed here ($14.6\leq r \leq22.5$). The shaded regions show the dispersion of this limiting magnitude in the y-axis, and the black-grey lines the dispersion in the x-axis. The dashed lines illustrate the sub-sample of galaxies at $0.05 \le z \le 0.15$ used to explore the evolution of the SFRD ($\rho_\star$). All the points are colour-coded according to the galaxy magnitudes in the \rb\ band and \magauto \ apertures.}
\label{fig:MassEvolution}
\end{figure}

Before we can estimate the star formation rate density (SFRD) of the Universe from our data, we need to characterize the volume incompleteness of our sample and the stellar mass limits as a function of redshift. For this particular case  we used the stellar masses  obtained using the \magauto \ photometry as they  provide a better estimate of  the total galaxy stellar mass than those derived using \magpsfcor. Firstly, we needed to set the minimum and maximum redshifts ($z_\mathrm{min}$ and $z_\mathrm{max}$, respectively) at which every galaxy can be detected in our sample owing to the \mjp\ detection limits. We estimated the $z_\mathrm{min}$ and $z_\mathrm{max}$ values (see the black lines in Fig.~\ref{fig:MassEvolution}) by using the SFH and stellar population properties of each galaxy and calculating at which redshift the observed magnitude in the $r$ band would be equal to $14.6$ and $22.5$, respectively. Then, the average $z_\mathrm{min}$ and $z_\mathrm{max}$ and their 1$\sigma$ dispersions were determined after binning the sample in stellar mass bins of  $\Delta \logMt = 0.2$~dex. It is also of note that our predictions of $z_\mathrm{min}$ and $z_\mathrm{max}$ are based on the SEDs obtained from the SED-fitting analysis of the \js\ fits rather than other traditional methods in which the $k$-correction is based on a predefined set of galaxy templates.

As a result, galaxies with a stellar mass of $\log (M_\star/\mathrm{M}_\odot)\sim 10$ can be detected up to $z= 0.8$, and low-mass galaxies of $\log (M_\star/\mathrm{M}_\odot)\sim 8.3$ up to $z= 0.15$ (see Fig.~\ref{fig:MassEvolution}). Our results show that \jp \ will be able to study samples of galaxies with stellar masses above $\log (M_\star/\mathrm{M}_\odot) \sim 8.9$, $9.5$, and $9.9$ at $z=0.3$, $0.5$, and $0.7$, respectively. These limits are $\sim 1.3$--$1.5$~dex below the lower limit of the stellar mass that will be covered by future spectroscopic surveys, such as WEAVE/StePS.

\begin{figure*}
\centering
\includegraphics[width=\textwidth]{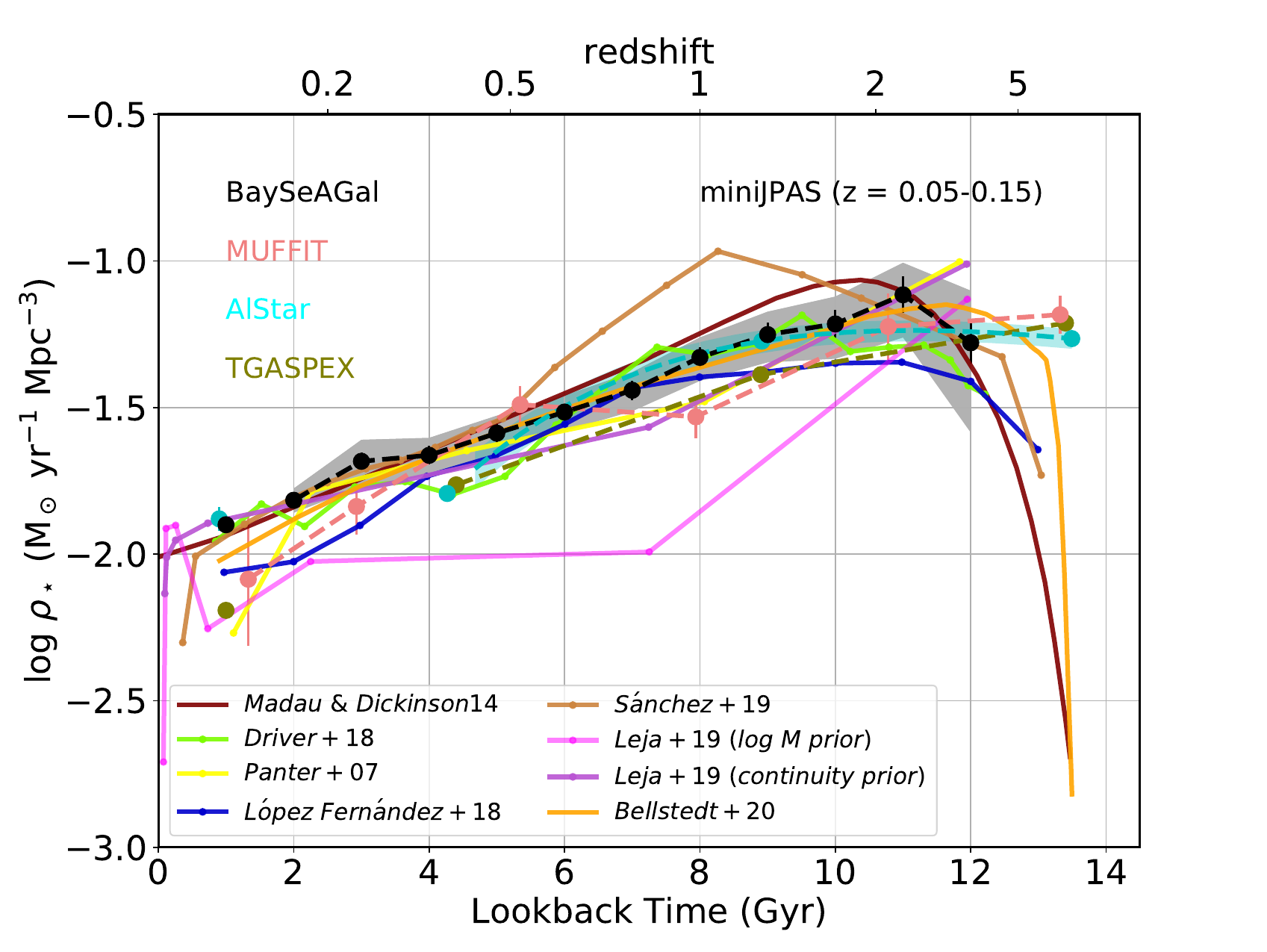}
\caption{Cosmic evolution of the SFRD ($\rho_\star$) obtained from the SED-fitting results of \baysea \ (black dots), \muff \ (coral dots), \alstar \ (cyan dots), and \tgas \ (olive dots), with the nearby galaxies ($0.05\leq z \leq0.15$). Grey (\baysea) and cyan (\alstar) shaded regions represent the uncertainties associated with the results. The different lines represent the SFRDs obtained in other works (see inset) and recently compiled by \citet{bellstedt20}. 
}
\label{fig:SFRD}
\end{figure*}
\bigskip
\subsection{Cosmic evolution of star formation rate density}

One of the most significant observational results obtained from galaxy redshift surveys is the cosmic evolution of the SFR of the Universe. It is well known that the SFRD ($\rho_\star$) peaks at $\sim3.5$~Gyr after the Big Bang, $z\sim2$, and then decreases until the present day \citep{lilly96, madau98, hopkinsbeacom06, fardal07, gunawardhana13, madau14, driver18}. This result has also been obtained by low-$z$ galaxy surveys using fossil records of nearby galaxies  \citep{heavens04, panter07, lopezfernandez18, sanchez19, bellstedt20}. Here, we test the capability of \jp\ data, along with our methods, to derive the SFH of galaxies and their properties, as well as to compare the cosmic evolution of the $\rho_\star$ of the Universe from a subset of nearby galaxies in our sample. 

Our results (see Fig.~\ref{fig:MassEvolution}) suggest that galaxies in a redshift bin centred at $z\sim0.1$ can be particularly useful for retrieving the cosmic evolution of $\rho_\star$ using the SFH of nearby galaxies ($0.05 \leq z \leq 0.15$). At $z=0.15,$  \mjp \ includes galaxies with  stellar masses above $2\times10^8$ $M_\odot$. While galaxies with stellar masses below this limit present a minor contribution to the total stellar mass density, they may significantly contribute to the SFR of the Universe in the last $4$ Gyr. This is because low-mass galaxies experienced their main star formation processes during recent epochs \citep{asari2007, bellstedt20}. However, we detected a small fraction of these galaxies in our sample, and it is difficult to account for the mass incompleteness below 2$\times$10$^8$ $M_\odot$. 

To get $\rho_\star$ and take the volume incompleteness effect into account, we divided the SFR of each galaxy at $0.05 \leq z \leq 0.15$ by its maximum co-moving volume, also known as~$V_\mathrm{max}$, over which the galaxy could be observed. However, we note that only a small fraction ($\sim4$~\%) of the galaxies in this redshift range would not be observed at $z=0.15$ (i.e.~$z_\mathrm{max} < 0.15$). Therefore, for most of the galaxies in our nearby sub-sample of galaxies, $V_\mathrm{max}$ is equal to the co-moving volume ($V_\mathrm{c}$) at this redshift range, meaning that $V_\mathrm{max} = \Delta V = V_\mathrm{c}(z=0.15) - V_\mathrm{c}(z=0.05)$. 

For \baysea, the SFRs were obtained from the parametric SFHs for a $1$~Gyr resampling, while for the non-parametric codes we used a larger interval in lookback time of $\sim3$--$4$~Gyr. This is because our redshift bin $0.05\leq z\leq0.15$ includes a small number of galaxies ($360$ galaxies) and the ages of the components in the non-parametric codes are discretized according to the ages of the SSP model set. This is not an issue for parametric codes because the SFH is described by a smoother and continuous composition of stellar population models.
The error of $\log \rho_\star$ in each epoch is obtained by propagating the dispersion in the stellar mass formed in each bin, although these errors are typically lower than $0.05$~dex for most of the epochs. The results obtained from the four codes (see Fig.~\ref{fig:SFRD}) demonstrate that $\rho_\star$ increases with redshift, up to $z\sim2.5$ for \baysea \ and to higher redshifts for the parametric codes. It is worth mentioning that fossil record methods hardly distinguish between stellar populations of ages between $11$ and $13$~Gyr. Thereby, it is hard to distinguish between populations formed above $z >3$.

We also compared the cosmic evolution of $\rho_\star$ obtained in this work with previous results in the literature \citep[][; see Fig.~\ref{fig:SFRD}]{panter07,madau14,driver18,lopezfernandez18,sanchez19,leja19,bellstedt20}. These results are derived from different datasets and using different approaches to the analysis. For instance, the works by \citet{madau14} and \citet{driver18} were performed with data from galaxy redshift surveys. The remaining studies are based on nearby galaxies and fossil record methods applied to stellar populations, where the cosmic SFR was constrained by using non-parametric \citep{panter07, leja19, sanchez19} and parametric \citep{lopezfernandez18, bellstedt20} SFH models. There are many similarities and discrepancies between the results that are mainly related to the properties of the samples, the quality of the data, and the methodologies, as well as the ability to correct for cosmic variance, AGN contribution, and dust effects. Overall, we conclude that the analysis of  \mjp \ data  provides successful results that are in good agreement with cosmological surveys \citep{madau14, driver18} as well as the fossil record analysis of SDSS \citep{panter07}, IFS CALIFA \citep{lopezfernandez18}, and GAMA data \citep{bellstedt20}. In this regard, we have proven that our data and analysis techniques exhibit a remarkable potential to predict the evolution of both the SFR and stellar mass density of the Universe with cosmic time.

\bigskip
\section{Summary and conclusions}
\label{sec:Summary}

 In this paper we have presented an analysis of  \mjp \ data using the full \jp \ filter system to evaluate the potential of  \jp\  for galaxy evolution studies. Our primary aim is to identify and characterize the stellar population properties of galaxies and their evolution up to $z=1$. Using the fossil record method for stellar populations, we analysed the observed optical SEDs (the \js) of $\sim 8000$ galaxies selected from the general \mjp\ catalogue with \rb$\leq 22.5$ (\magauto\ magnitudes), $z \leq 1$, and \class $\leq 0.1$. The \js\ of these galaxies were fitted with different codes to constrain the stellar mass, rest-frame and intrinsic $(u-r)$ colours, extinction, age, and metallicity of their stellar populations. The bimodal distribution of galaxies was identified  by using the stellar mass--colour diagram corrected for extinction, \logM -- $(u-r)_\mathrm{int}$, and its evolution was explored across cosmic time up to $z=1$. The impact on the results of the photon-noise errors in the \js\ photometry and the \photoz\ uncertainties were explored, together with the uncertainties in the estimation of the different stellar population properties across the \logM--$(u-r)_\mathrm{int}$ plane.
 
One parametric (\baysea) and three non-parametric SED-fitting codes (\muff, \alstar, and \tgas) were used to obtain the distribution of the stellar population properties of the sample, as well as to check the consistency of the results between codes. We used a common set of SSP models from an updated version of the \citet{bruzual2003} synthesis models, with the attenuation law by \citet{calzetti2000} and assuming the IMF of \citet{chabrier2003}. The three main differences between the codes are: (i) \baysea \ is a Bayesian approach that derives the PDF for each of the stellar population parameters by assuming a delayed-$\tau$ or an exponential SFR. (ii) \muff \ combines two-burst SSP models. (iii) \alstar \ and \tgas \  use an arbitrary combination of SSP models to solve the NNLS problem.
\muff, \alstar, and \tgas \ do not perform a fully Bayesian evaluation of the PDF; instead, they follow a Monte Carlo approach by adding Gaussian noise to the observed fluxes to iteratively repeat the SED-fitting process. 
The median and mean values of the properties obtained for each galaxy by the four SED-fitting codes were compared. The full PDF from \baysea \  was also compared to the median/mean distributions.

 From the results obtained by \baysea, we can draw some conclusions that can be extended to the other codes. The galaxy stellar mass, extinction, metallicity, mass- and luminosity-weighted ages, and rest-frame and dust-corrected colours can be estimated from the fits of \js \ of galaxies brighter than $22.5$~(AB) in the \rb\ band that have a mean of the median signal-to-noise in the NB filters of $\sim8$. The precisions (the standard deviations of the distribution for each galaxy) in  the stellar population parameters are: $0.1\pm0.05$, $0.15\pm0.06$~dex, and $0.34\pm0.1$ for $(u-r)_\mathrm{res}$ colour, stellar mass, and extinction, respectively. The precision of the mass-weighted age of the stellar populations is $0.14\pm0.05$ and $0.25\pm0.06$ for red and blue galaxies, respectively. However, stellar metallicity is less precisely constrained, in particular for galaxies in the blue cloud. We also find that these results are independent of the SFR law adopted by \baysea . The precision in the results is remarkable considering that $\sim25$\% of the sample have S/N $\leq 3$. Better precision is obtained when only galaxies with S/N $\geq 10$ are selected. In this case, the precisions for $(u-r)_\mathrm{res}$ colour, stellar mass, extinction,  mass-weighted age, and stellar metallicity are: $0.04\pm0.02$, $0.07\pm0.03$~dex, $0.20\pm0.09$, $0.16\pm0.07$, and $0.42\pm0.25$~dex, respectively. This precision is only slightly below what will be obtained in spectroscopic surveys of similar S/N, such as WEAVE/StePS, which will get precision in the mass-weighted age of $\sim0.1$~dex for galaxies at $0.3 < z < 0.7$ and $I_{AB} < 20.5$.

 The main conclusions after comparing the stellar population properties obtained by the four SED-fitting codes are:
 
 \begin{itemize}
 
 \item{The distributions of galaxy properties, such stellar mass, rest-frame colours, extinction, and metallicity, are very similar. However, the distributions of the stellar ages show differences. The distributions of age from \baysea \ (assuming a delayed-$\tau$ SFR model) and \muff \ are similar, and they are shifted towards younger ages with respect to the distributions obtained with \alstar \ and \tgas. This is probably a consequence of the earlier formation epoch and  rapid mass growth in the SFH of BGs found by \alstar \ and \tgas \ with respect to \muff \ and \baysea . However, the consistency in the results (e.g. between \baysea \ + delayed-$\tau$ and \alstar, which are the two models with larger discrepancies) is within  $0.06$ in \logM, $0.2$~mag in $A_V$, $0.1$ in \logZM, $0.15$ in \ageM, and $0.04$  in \ageL. }

 \item{Red and blue galaxies are easily identified in the stellar mass and dust-corrected $(u-r)$ diagram. However, the fraction of red and blue galaxies varies according to the SED-fitting code, especially in the highest redshift bins considered here.
 We have estimated that the percentage of galaxies in the blue cloud varies between $85$\% (\baysea ), $81$\% (\muff ), $69$\% (\alstar ), and $76$\% (\tgas ); however, the percentages of `dusty star-forming galaxies' (BGs with $(u-r)_\mathrm{res} > 2$ and $A_V$ >1 ) are $13$\% (\baysea ), $11$\% (\muff ), $8$\% (\alstar ), and $9$\% (\tgas ), therefore consistent within a few percent. 
 }

 \item{Red and blue galaxy populations can be characterized by the correlations between the stellar population properties and the position of each galaxy in the stellar mass and dust-corrected $(u-r)$ diagram. All the SED-fitting codes provided consistent results for the stellar population properties and their evolution with redshift.
 Red and blue galaxies are well separated by their $(u-r)_\mathrm{int}$ colour, with mean values equal to $\sim$2 and $\sim$1, respectively.  
 In each redshift bin, red galaxies are more massive and have older and more metal-rich stellar populations than BGs. }

\item{In terms of redshift evolution, 
blue and red galaxies are older at the present day than at $z\sim1$. The mean stellar mass of galaxies increases with redshift, as expected from selection effects of flux-limited samples; at high redshift,  only  the brightest galaxies are detected. An increase in the metallicity since $z\sim 1$ is not observed, suggesting the lack of a significant chemical enrichment since $z\sim 1$; however, this is  probably a consequence of the observed bias in stellar mass.}
  
 \end{itemize}

The \mjp \ survey and its novel \jp \ filter system have proven their capability to identify and characterize galaxy populations and their evolution across cosmic time up to $z = 1$. The \jp \ survey will cover several thousand times the sky area observed by \mjp, and it will provide complete and statistically significant samples of galaxies with \js \ with S/N above 3 to retrieve the stellar population properties as a function of redshift and environment. A similar precision regarding the analysis of the spectroscopic datasets will be derived by analysing the sub-sample of galaxies with S/N  $\geq$10.  In addition,  \jp \ will be able to study samples of galaxies with stellar masses above  $\log (M_\star/\mathrm{M}_\odot) \sim 8.9$, $9.5$, and $9.9$ at $z=0.3$, $0.5$, and $0.7$, respectively. These limits are $> 1$~dex below the lower limit of the stellar mass that will be covered by future spectroscopic surveys, such as WEAVE/StePS.
Analysis of the \mjp \  galaxies with $\log (M_\star/\mathrm{M}_\odot) > 8.3$  at $z\sim0.1$ is able to retrieve the evolution of $\rho_\star$  up to $z\sim 3$ in good agreement with results from other cosmological surveys. These results show the strong potential of \jp \ to constrain the evolution of the SFR and stellar mass density with cosmic time.
\begin{acknowledgements} 

R.G.D., L.A.D.G., R.G.B., G.M.S., J.R.M., and E.P. acknowledge financial support from the State Agency for Research of the Spanish MCIU through the "Center of Excellence Severo Ochoa" award to the Instituto de Astrof\'\i sica de Andaluc\'\i a (SEV-2017-0709), and to the AYA2016-77846-P and PID2019-109067-GB100.

L.A.D.G.~also acknowledges financial support by the Ministry of Science and Technology of Taiwan (grant MOST 106-2628-M-001-003-MY3) and by the Academia Sinica (grant AS-IA-107-M01).

G.B. acknowledges financial support from the National Autonomous University of M\'exico (UNAM) through grant DGAPA/PAPIIT IG100319 and from CONACyT through grant CB2015-252364.

SB  acknowledges PGC2018-097585- B-C22, MINECO/FEDER, UE of the Spanish Ministerio de Economia, Industria y Competitividad. 

L.S.J. acknowledges support from Brazilian agencies FAPESP (2019/10923-5) and CNPq (304819/201794).

P.O.B. acknowledges support from  the Coordenação de Aperfeiçoamento de Pessoal de Nível Superior – Brasil (CAPES) – Finance Code 001.

P.R.T.C. acknowledges financial support from Funda\c{c}\~{a}o de Amparo \`{a} Pesquisa do Estado de S\~{a}o Paulo (FAPESP) process number 2018/05392-8 and Conselho Nacional de Desenvolvimento Cient\'ifico e Tecnol\'ogico (CNPq) process number  310041/2018-0. 

V.M. thanks CNPq (Brazil) for partial financial support. This project has received funding from the European Union’s Horizon 2020 research and innovation programme under the Marie Skłodowska-Curie grant agreement No 888258.

E.T. acknowledges support by ETAg grant PRG1006 and by EU through the ERDF CoE grant TK133.

Based on observations made with the JST/T250 telescope and PathFinder camera for the miniJPAS project at the Observatorio Astrof\'{\i}sico de Javalambre (OAJ), in Teruel, owned, managed, and operated by the Centro de Estudios de F\'{\i}sica del  Cosmos de Arag\'on (CEFCA). We acknowledge the OAJ Data Processing and Archiving Unit (UPAD) for reducing and calibrating the OAJ data used in this work.

Funding for OAJ, UPAD, and CEFCA has been provided by the Governments of Spain and Arag\'on through the Fondo de Inversiones de Teruel; the Arag\'on Government through the Research Groups E96, E103, and E16\_17R; the Spanish Ministry of Science, Innovation and Universities (MCIU/AEI/FEDER, UE) with grant PGC2018-097585-B-C21; the Spanish Ministry of Economy and Competitiveness (MINECO/FEDER, UE) under AYA2015-66211-C2-1-P, AYA2015-66211-C2-2, AYA2012-30789, and ICTS-2009-14; and European FEDER funding (FCDD10-4E-867, FCDD13-4E-2685).

\end{acknowledgements}



\bibliographystyle{aa}
\bibliography{GalEvo_AEGIS}

\begin{appendix}

\section{Fits and quality assessment with \baysea }

\begin{figure*}
\centering
\includegraphics[width=\textwidth]{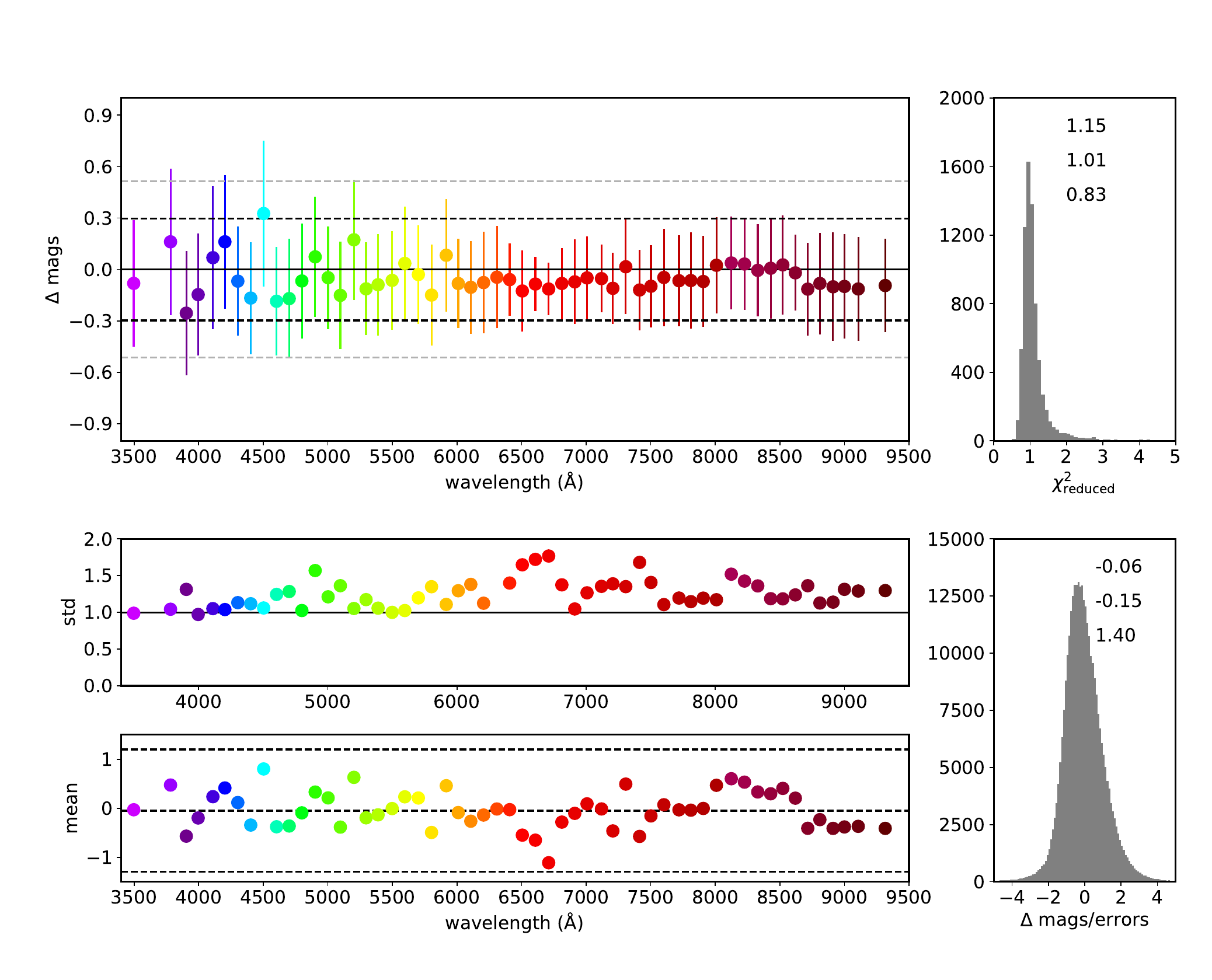}
\caption{Residuals and figures of merit of the \magpsfcor\ \js\ fits. \textit{Top-right panel}: Distribution of  $\chi^2$. The mean, median, and standard deviation of the $\chi^2$ distribution are indicated. \textit{Top-left panel}: Mean value of the difference between the observed and the best-fitting model. The error bar is the 1~sigma uncertainty level for each filter, while the dashed lines are the global averages for the 1 and 2~sigma uncertainty levels. \textit{Bottom-right panel}: Distribution of the ratio between $\Delta$mag and the error for all the filters and galaxies. The mean, median, and standard deviation of this distribution are indicated. \textit{Middle- and bottom-left panels}: Variation of the mean and standard deviation for each filter. \textit{Upper panels} are for \magpsfcor,\ and \textit{bottom panels} are for \magauto \ apertures.
}
\label{fig:merit}
\end{figure*}

The quality of \js \ fits for the whole sample was confronted in three ways: firstly, by its reduced $\chi^2$ value ($\chi^2_\mathrm{reduced}$; see the top-right panel of Fig.~\ref{fig:merit}), which is defined as the $\chi^2$ value divided by the number of available bands used during the SED-fitting analysis; secondly, by the residuals, $\Delta$mag, calculated as the difference between the observed and the model magnitudes for each NB filter (see the top-left panel of Fig.~\ref{fig:merit}) -- model magnitudes correspond to the SFH with the minimum $\chi^2$ value; and finally, by the normalized residual, meaning the residuals divided by the photon-noise uncertainties of each band (see the bottom-right, middle-left, and bottom-left panels of  Fig.~\ref{fig:merit}).

The distribution of $\chi^2_\mathrm{reduced}$ values has a mean equal to $1.15$ and a median equal to $1.01$, where the standard deviation of the distribution is $\sigma=0.83$. It is also worth mentioning that the distribution of the normalized residuals is properly centred with mean $-0.06$, median $-0.15$, and $\sigma = 1.4$. These two distributions indicate that the \js \ are properly fitted within the errors. The residuals and the normalized residuals change across the spectrum, although always within the errors. The $\Delta$mag and the dispersion are smaller for the red than for the blue bands ($\lambda < 5000$~\AA). Nonetheless, for some filters, errors in the data seem to be slightly underestimated, in particular for filters at intermediate wavelengths, because the dispersion of the normalized residuals is larger than unity. Although these values were derived for the \magpsfcor\ \js, similar results were obtained using the \magauto \ photometry. For this case and for all the filters, the normalized residual (i.e.~$\Delta$mag/error) is close to unity, indicating that the errors of  \magauto\  are not underestimated.

\section{Similarities and discrepancies between the codes for the evolution of galaxy populations in the mass--colour diagram }

\begin{figure*}
\centering
\includegraphics[width=\textwidth]{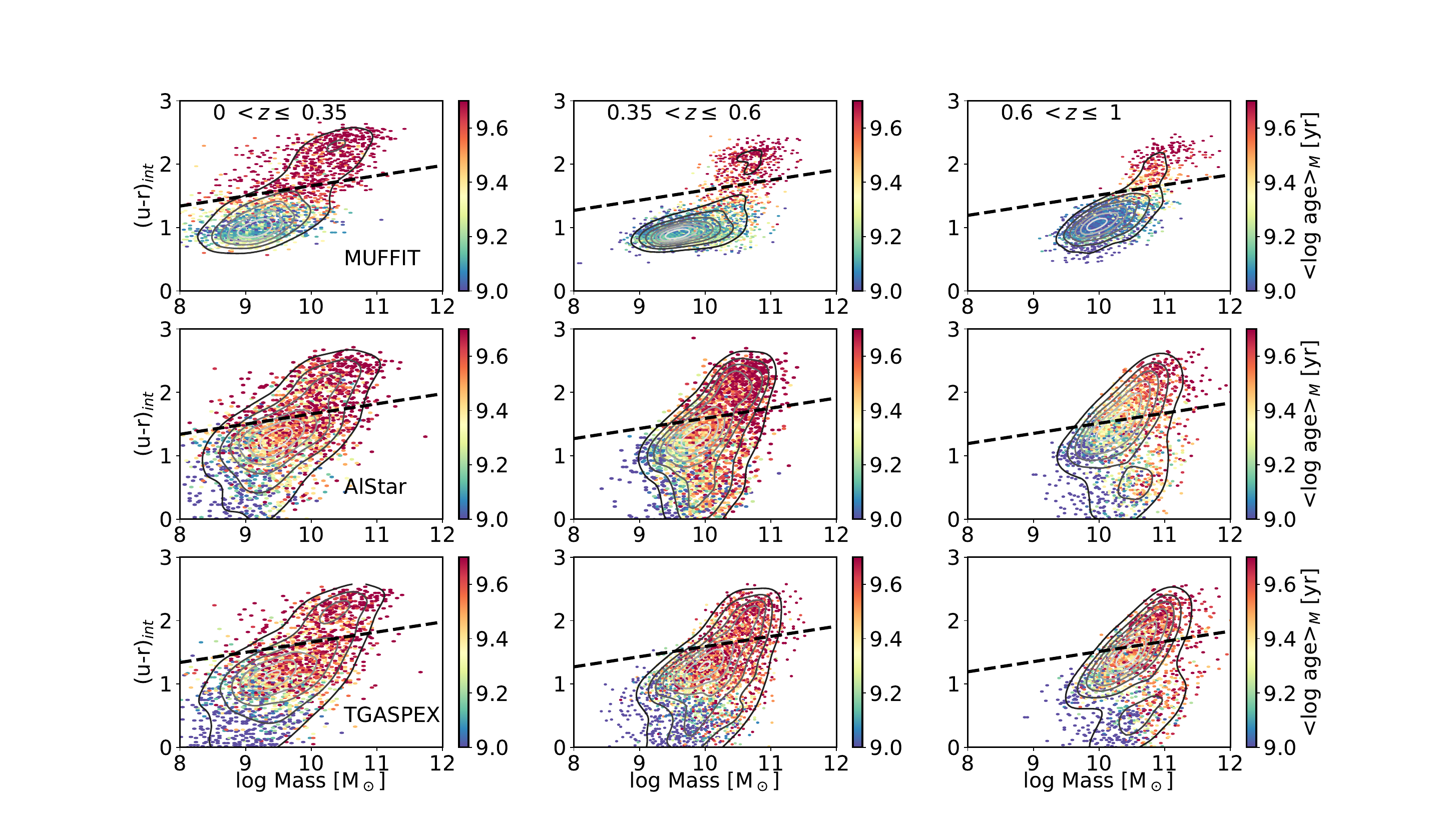}
\caption{Intrinsic colour $(u-r)$ vs. stellar mass for the redshift bins $z \leq 0.35$, $0.35 < z \leq 0.6$, and $0.6 < z \leq 1$ (\textit{from left to right}) for the results determined by  \muff , \alstar , and \tgas . The dashed line shows the $(u-r)_\mathrm{int}^\mathrm{lim}$ for the mean redshift in each bin (details in the text).
}
\label{fig:MassColorEvolution_MUFFIT_ALSTAR_TGASPEX}
\end{figure*}

\begin{figure*}
\centering
\includegraphics[width=\textwidth]{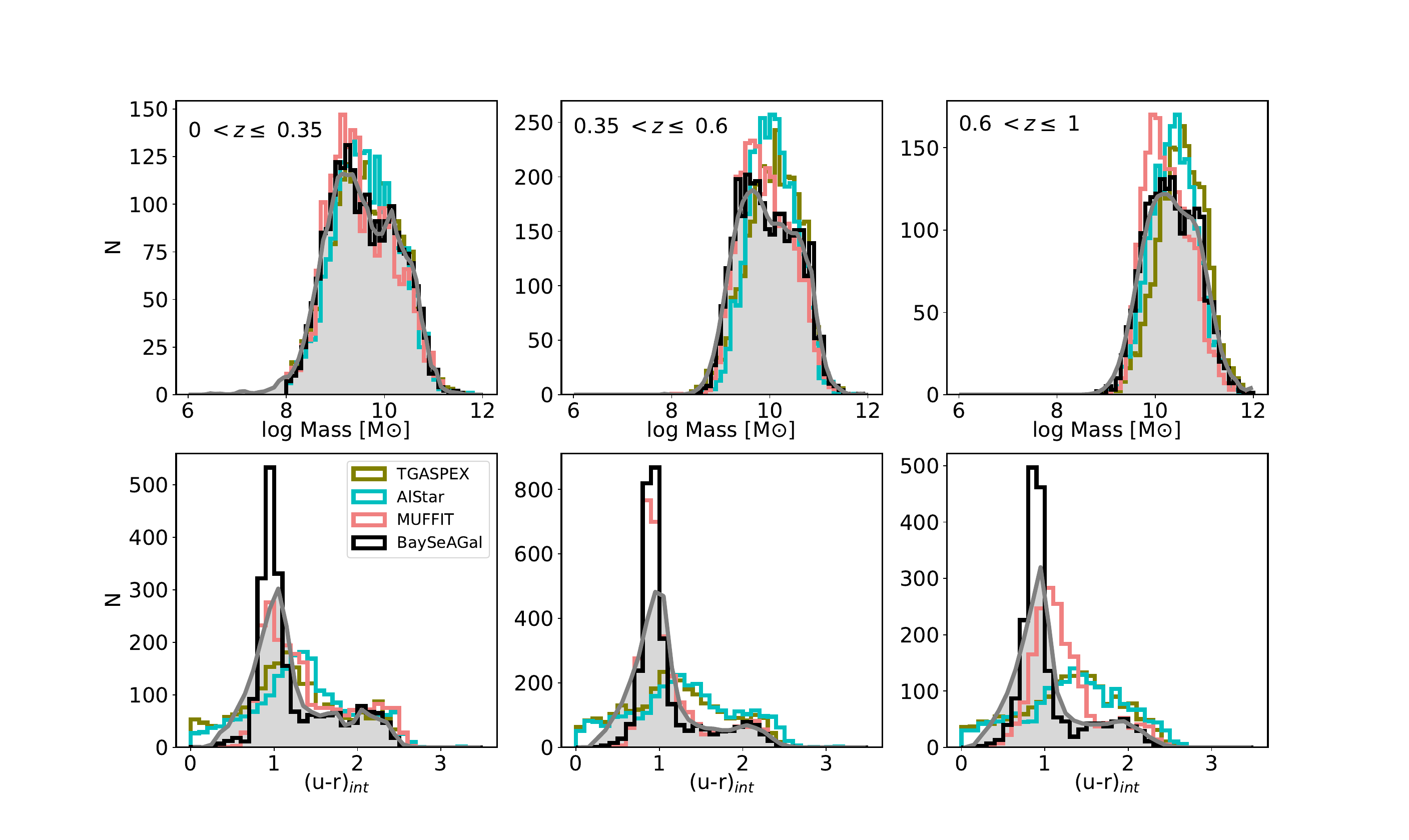}
\caption{Distributions of stellar mass and $(u-r)$ colour, corrected for extinction (top and bottom panels, respectively), obtained by our SED-fitting codes \baysea , \muff , \alstar , and \tgas\ for the \mjp\ galaxies at different redshift bins (see panels). The grey curves illustrate the posterior PDFs obtained by \baysea\ assuming a delayed-$\tau$ SFH.}
\label{fig:hist_SP_methods_evolution}
\end{figure*}

We have discussed the differences in the distribution of galaxy populations when the dust-corrected colour $(u-r)_\mathrm{int}$ is considered instead of the $(u-r)_\mathrm{res}$. Here, we present first the mass-$(u-r)_\mathrm{int}$ diagrams obtained from \muff, \alstar, and \tgas \ (Fig.~\ref{fig:MassColorEvolution_MUFFIT_ALSTAR_TGASPEX}); then we discuss the similarities and differences in the distributions of \logM, $(u-r)_\mathrm{res}$, and $(u-r)_\mathrm{int}$ for three redshift bins: $z \leq 0.35$, $0.35 < z \leq 0.6$, and $0.6 < z \leq 1$. It should be noticed that the distribution of galaxy populations in the mass--$(u-r)_\mathrm{int}$ plane from \muff \ is very similar to the results from \baysea; they both show clear differences with respect to the results from \alstar \ and \tgas. In particular, the line traced by $(u-r)_\mathrm{int}^\mathrm{lim}$ shows a  sharper and clearer division between red and blue galaxies in the plane derived by \baysea \ and \muff \ and than that derived by \alstar \ and \tgas ; thus, as we have already discussed, it gives a larger number of red galaxies from \alstar \ with respect to \baysea\ and \muff.

In general, the distributions of stellar mass present few differences between the SED-fitting codes at the different cosmic epochs explored in this work. As expected for flux-limited samples, the four codes retrieved a higher number of massive galaxies at increasing redshift owing to both the survey detection limit, or depth, and the higher volume observed at higher redshifts.
However, there are still some discrepancies between the results of the four codes. At $z > 0.35$ (see the top-middle and top-right panels of Fig.~\ref{fig:hist_SP_methods_evolution}), there is a larger fraction of massive galaxies derived by \tgas \ and \alstar \ than by \muff , while the same distribution for \baysea \ is a bit broader than the others, with values in between the distributions of \muff \ and \alstar. However, these differences are not so significant as to indicate that there are problems in determining the stellar mass distribution of the \mjp \ galaxies; rather, there are inherent discrepancies in the four methodologies that yield these kinds of differences in the distributions. 

The distributions of  $(u-r)_\mathrm{int}$ from the four codes show larger differences. At low redshift, the distributions are very similar. 
At intermediate redshift, \baysea \ and \muff \ distributions are equal, and they are shifted to bluer colours than the distributions from \alstar \ and \tgas. In the highest redshift bin, $0.6 \le z \le 1$, there are larger discrepancies between \baysea \ and the results from the non-parametric codes. 
The $(u-r)_\mathrm{int}$ distributions from \alstar \ and \tgas \ are similar and more widespread in the colour range of the SSPs than the \muff \  and \baysea \ results. These differences in the  $(u-r)_\mathrm{int}$ distributions are more remarkable in the blue cloud galaxies and are mainly related to differences in the values of extinction, which in turn concern the SFH assumptions adopted in each of the codes. A similar conclusion can be extended to the differences between the distributions of \ageM\ and \ageL\ from \baysea\ and \muff\ with respect to \alstar\ and \tgas;  the last two codes produce older ages for the \mjp\ galaxies (i.e.~redder colours).  

\end{appendix}


\end{document}